\documentclass[12pt]{article}
\usepackage[a4paper]{geometry} 
\usepackage{graphicx}
\usepackage{amsfonts,amsmath,amssymb}
\usepackage{latexsym}
\usepackage{MnSymbol}
\usepackage{epsfig}
\usepackage{hyperref}
\usepackage[final]{showlabels}
\usepackage{xcolor}
\usepackage{appendix}
\usepackage[normalem]{ulem}

\DeclareMathAlphabet{\mathbbold}{U}{bbold}{m}{n}

\usepackage[T1]{fontenc} 
\usepackage[utf8]{inputenc} 
\usepackage[icelandic,english]{babel}

\newcommand{\mathi}{\mathrm{i}}
\graphicspath{{./Plots-Els4/}{./Plots-Egt4/}{./Plots-Ploc/}}

\begin{document}

\def\void{}
\def\labelmark{}

\newenvironment{formula}[1]{\def\labelname{#1}
\ifx\void\labelname\def\junk{\begin{displaymath}}
\else\def\junk{\begin{equation}\label{\labelname}}\fi\junk}%
{\ifx\void\labelname\def\junk{\end{displaymath}}
\else\def\junk{\end{equation}}\fi\junk\labelmark\def\labelname{}}

\newenvironment{formulas}[1]{\def\labelname{#1}
\ifx\void\labelname\def\junk{\begin{displaymath}\begin{array}{lll}}
\else\def\junk{\begin{equation}\label{\labelname}\left.
\begin{array}{lll}}\fi\def\arraystretch{1.5}\junk}%
{\ifx\void\labelname\def\junk{\end{array}\end{displaymath}}
\else\def\junk{\end{array}\right.\end{equation}}
\fi\junk\labelmark\def\labelname{}\def\junk{}
\def\arraystretch{1}}

\newcommand{\beq}{\begin{formula}}
\newcommand{\eeq}{\end{formula}}
\newcommand{\beqv}{\begin{formula}{}}

\newcommand{\rf}[1]{(\ref{#1})}
\newcommand{\oh}{\frac{1}{2}}
\newcommand{\oq}{\frac{1}{4}}
\newcommand{\bea}{\begin{eqnarray}}
\newcommand{\eea}{\end{eqnarray}}
\newcommand{\beas}{\begin{eqnarray*}}
\newcommand{\eeas}{\end{eqnarray*}}
\newcommand{\beqs}{\begin{displaymath}}
\newcommand{\eeqs}{\end{displaymath}}
\newcommand{\ra}{\rightarrow}
\newcommand{\fslu}{\not\mbox{\hspace{-1.5mm}}}   
\newcommand{\fsll}{\not\mbox{\hspace{-.5mm}}}   
\newcommand{\br}{\langle}
\newcommand{\kt}{\rangle}
\newcommand{\bra}[1]{\langle {#1}|}
\newcommand{\ket}[1]{|{#1}\rangle}
\newcommand{\ul}{\underline}
\newcommand{\dg}{\dagger}
\newcommand{\5}{5 \! \times \! 5}
\newcommand{\nn}{\nonumber \\}
\newcommand{\ab}{abelian}
\newcommand{\ch}{chirality}
\newcommand{\Cou}{Coulomb}
\newcommand{\Gf}{Green functions}
\newcommand{\Fm}{Feynman}
\newcommand{\ham}{hamiltonian}
\newcommand{\lag}{lagrangian}
\newcommand{\Lo}{Lorentz}
\newcommand{\Mi}{Minkowski}
\newcommand{\mi}{minkowskian}
\newcommand{\Eu}{Euclid}
\newcommand{\eu}{euclidean}
\newcommand{\WT}{Ward-Takahashi}
\newcommand{\sm}{standard model}
\renewcommand{\a}{\alpha}
\renewcommand{\b}{\beta}
\newcommand{\g}{\gamma}
\newcommand{\G}{\Gamma}
\newcommand{\del}{\delta}
\newcommand{\Del}{\Delta}
\newcommand{\ep}{\varepsilon}
\newcommand{\e}{\epsilon}
\newcommand{\ka}{\kappa}
\newcommand{\kp}{\kappa}
\newcommand{\bx}{\bf x}
\newcommand{\m}{\mu}
\newcommand{\vmu}{\vec{\mu}}
\newcommand{\n}{\nu}
\newcommand{\mn}{\mu\nu} 
\newcommand{\om}{\omega}
\newcommand{\vom}{\vec{\omega}}
\newcommand{\tphi}{\tilde{\phi}}
\newcommand{\phid}{\phi^{\dagger}}
\newcommand{\ff}{\phi^3}
\newcommand{\fF}{\phi^4}
\newcommand{\fn}{\phi^n}
\newcommand{\bpsi}{\bar{\psi}}
\newcommand{\rh}{\rho}
\newcommand{\sg}{\sigma}
\newcommand{\vsig}{\vec{\sigma}}
\newcommand{\Sg}{\Sigma}
\newcommand{\vp}{\varphi}
\newcommand{\vB}{\vec{B}}

\newcommand{\cH}{{\cal H}}

\newcommand{\uN}{\underline{N}}

\newcommand{\be}{\bar{e}}
\newcommand{\bl}{\bar{\l}}
\newcommand{\cL}{{\cal L}}
\newcommand{\vL}{\vec{L}}
\newcommand{\cS}{{\cal S}}
\newcommand{\cD}{{\cal D}}
\newcommand{\cC}{{\cal C}}

\newcommand{\cG}{{\cal G}}
\newcommand{\cM}{{\cal M}}
\newcommand{\cP}{{\cal P}}   
\newcommand{\cB}{{\cal B}}
\newcommand{\cT}{{\cal T} }
\newcommand{\cO}{{\cal O}}
\newcommand{\vS}{\vec{S}}
\newcommand{\cV}{{\cal V}}
\newcommand{\hk}{\hat{k}}
\newcommand{\vk}{\vec{k}}
\newcommand{\vek}{\vec{k}\mbox{\hspace{1mm}}}
\newcommand{\hq}{\hat{q}}
\newcommand{\vx}{\vec{x}}
\newcommand{\vex}{\vec{x}\mbox{\hspace{1mm}}}
\newcommand{\hr}{\hat{r}}
\newcommand{\Tr}{{\rm Tr}\;}
\newcommand{\ben}{\begin{equation}}
\newcommand{\een}{\end{equation}}

\newcommand{\bdm}{\begin{displaymath}}
\newcommand{\edm}{\end{displaymath}}
\newcommand{\hf}{\frac{1}{2}}
\newcommand{\vn}{\vec{\nabla}}
\newcommand{\ve}{\vec{e}}
\newcommand{\vep}{\vec{e}\mbox{\hspace{1mm}}'}
\newcommand{\La}{{\cal L}}
\newcommand{\Ha}{{\cal H}}
\newcommand{\pa}{\partial}
\newcommand{\eps}{\epsilon}
\newcommand{\al}{\alpha}
\newcommand{\D}{{\cal D}}
\newcommand{\N}{{\cal N}}
\newcommand{\tb}{\overline{t}}
\newcommand{\qh}{\hat{q}}
\newcommand{\phih}{\hat{\phi}}
\newcommand{\gb}{\overline{\gamma}}
\newcommand{\zt}{\zeta}
\newcommand{\ztb}{\overline{\zeta}}
\newcommand{\tq}{\tilde{q}}
\newcommand{\Asf}{{\sf A}}
\newcommand{\Bsf}{{\sf B}}
\newcommand{\Csf}{{\sf C}}
\newcommand{\ps}{\psi}
\newcommand{\psb}{\overline{\psi}}
\newcommand{\etb}{\overline{\eta}}
\newcommand{\Phc}{\Phi^{Cl}}
\renewcommand{\d}{\delta}
\newcommand{\cJ}{{\cal J}}
\newcommand{\vph}{\varphi}

\renewcommand{\dag}{\dagger}
\newcommand{\tr}{{\rm tr}\,}
\newcommand{\lla}{\left\langle}
\newcommand{\rra}{\right\rangle}
\newcommand{\DS}{\!\!\!\! /}
\newcommand{\ds}{\!\!\!\! \hspace{1mm} /}

\newcommand{\bbN}{\mathbb N}
\newcommand{\bbZ}{{\mathbb Z}}
\newcommand{\bbR}{{\mathbb R}}
\newcommand{\bbC}{{\mathbb C}}
\newcommand{\bbQ}{{\mathbb Q}}
\newcommand{\bon}{{\bf n}}

\hfill

\vspace{1cm}

\begin{center}

{\Large \bf Quantum walk on a random comb \\
}

\medskip
\vspace{1.5 truecm} 



{\bf Fran\c cois David}

\vspace{0.4 truecm}

Institut de Physique Th\'eorique,\\
Université Paris-Saclay, CNRS, CEA,\\
91191, Gif-sur-Yvette, France

\vspace{1.3 truecm}

{\bf Thordur Jonsson}
\vspace{0.4 truecm}

\vspace{0.4 truecm}

Division of Mathematics

The Science Institute, University of Iceland

Dunhaga 3, 107 Reykjavik, Iceland

\end{center}
 \textwidth 150mm
  \textheight 215mm
   \setlength{\unitlength}{0.01in}
    \def\sepand{\rule{14cm}{0pt}\and}

\vspace{1.3 truecm}

\noindent {\bf Abstract.}  We study continuous time quantum walk on a random comb graph with infinite teeth.
Due to localization effects along the spine, the walk cannot go to infinity in the spine direction, while it can escape to infinity along the teeth of the comb. Starting from an initial vertex, the walk has a nonzero probability to stay trapped in a finite region. 
These results are obtained by studying the spectrum and eigenstates of the random Hamiltonian for the graph
and analysing its properties. We use both analytic and numerical methods, many of which come from the theory 
of Anderson localization in one dimension.

 \topmargin 0pt
 \oddsidemargin 5mm
 \headheight 0pt
 \topskip 0mm

\addtolength{\baselineskip}{0.5\baselineskip}

\pagestyle{empty}
\vfill
\hfill 02/07/2026

 \newpage
 \pagestyle{plain}
 
 \tableofcontents

\section{Introduction}
Quantum walks on graphs have been studied for more than 25 years, the main motivation coming from quantum computation and
the search for efficient algorithms, see \cite{Kempe2003, Venegas-Andraca:2012aa} and references therein. 
The behaviour of classical random walks on graphs reflects the structure of the underlying graph
and the same is the case for quantum walks.  In general the behaviour of quantum walks is quite different from that of classical random walk.
The quantum walks 
move much faster than classical random walks and at the same time the quantum walk can be trapped in a finite subgraph with a finite 
probability as the time goes to infinity, see e.g., \cite{PhysRevA.99.042329,JoyeMerkli2010,Childs:2002aa,Childs_2003}.

In \cite{David_2022} we studied continuous time quantum walk on a regular comb with infinite teeth.  This is the linear graph $\mathbb{Z}$ (the spine) 
with a discrete 
half line $\mathbb{Z}_+$ (a tooth) attached at each vertex.  In this paper we study quantum walk on a random comb which is the random graph obtained by
attaching a tooth to each vertex on the spine with a fixed probability $1-p$, $0<p<1$.  
Classical random walks on random 
combs, trees and surfaces have been studied
extensively, see e.g.,\cite{DurhuusJonssonWheater2010,DurhuusJonssonWheater2007,Durhuus_2006,CurienHutchcroftNachmias2020}.

For the reader's convenience, we briefly outline the main results of \cite{David_2022} before 
describing the properties of quantum walk on the random comb.
There are two kinds of eigenfunctions of the Hamiltonian on the regular comb, functions which are oscillatory along the spine and also in the teeth
and functions which decay exponentially in the teeth but are oscillatory along the spine.
It follows that a quantum walk which starts somewhere on the spine can escape to infinity
either along the spine or along the teeth and the probabilities of these escapes (which sum to 1) can be calculated.

On the random comb the eigenfunctions of the Hamiltonian behave quite differently.  
The random teeth effectively play the role of a 
random potential and the behaviour of the eigenfunctions of the Hamiltonian along the spine can 
be described by the Anderson tight
binding model which in one dimension has only exponentially decaying eigenfunctions.  
Along the teeth the eigenfunctions are oscillatory or
exponentially decaying (depending on the energy) so the walk can either escape to infinity along the 
teeth or it is trapped in a finite region.
This means that there are finite regions $R$ such that the probability $P(t)$ that the 
walk is located inside $R$ at time $t$ converges to a positive constant as $t\to\infty$.

In the next section, we introduce our notation and define the quantum walk model.  Then we 
study quantum walk on a finite random comb before
discussing the walk on an infinite random comb where 1-dimensional Anderson localization plays
a major role.  We then describe the results of numerical simulations and study quantum
diffusion on a random comb in some detail.  A few technical points are relegated the appendices.


\section{The model}
\label{sTheModel}

The random comb ${\mathcal C}$ is the linear graph $\mathbb{Z}$ with vertices labelled by the integers, 
with an integer half-line $\bbZ_+$ 
attached at the vertices with a probability $1-p$, so there is no half line attached with probability $p$.
When $p=0$ this is the regular infinite comb model studied in \cite{David_2022}.

Let $n$ be the coordinate along the spine.  If a tooth is attached at $n$, then
we denote the vertices in the tooth by $(n,j)$, $j\in\bbZ_+$, $j$ being the distance from the spine.  
Vertices on the spine with no tooth attached will often be referred to as \emph{holes}.
If no tooth is attached at $n$ we denote that vertex by $(n,0)$.

\begin{figure}[h]
\begin{center}
\includegraphics[width=2.85in]{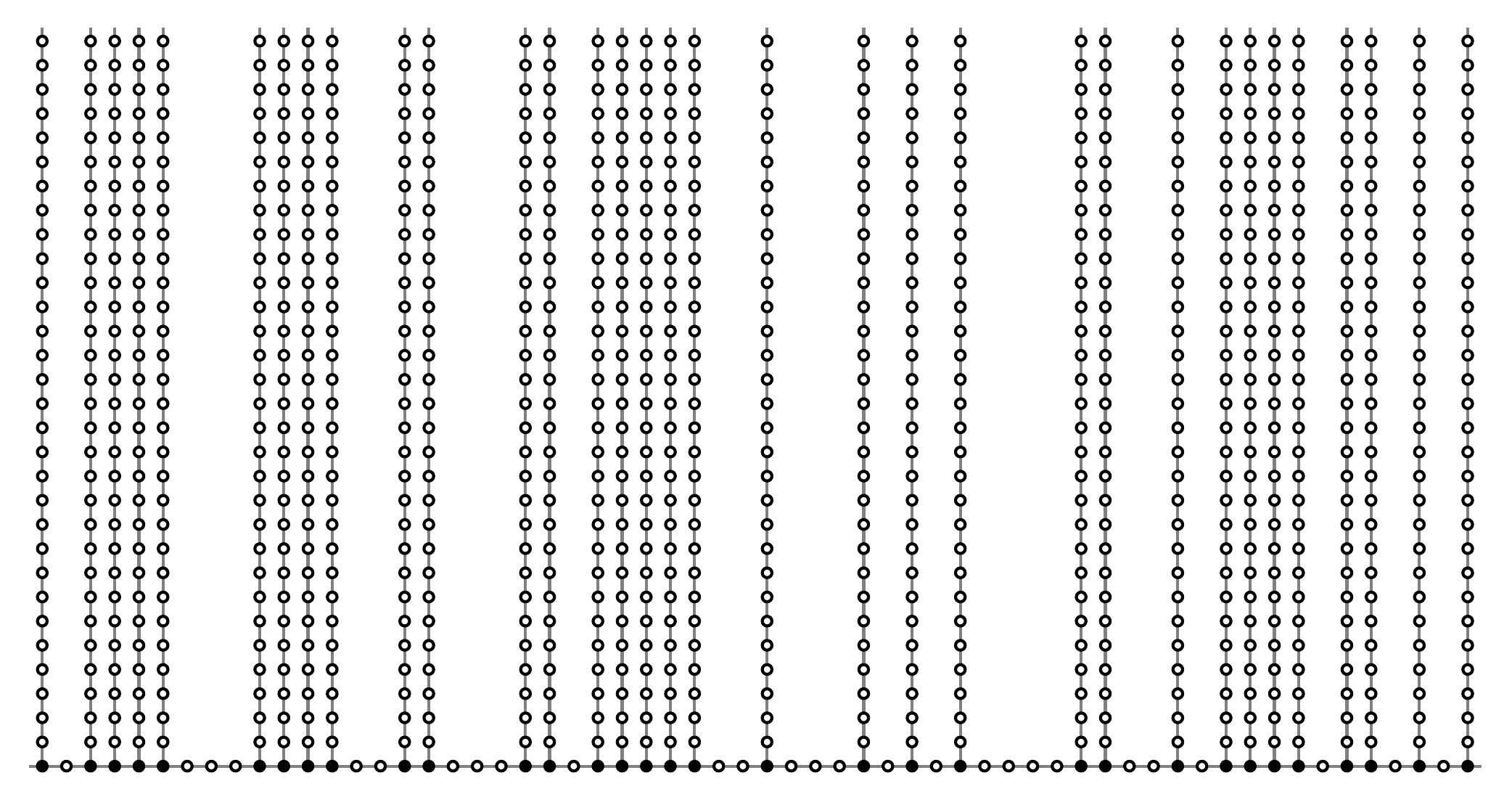}
\includegraphics[width=2.85in]{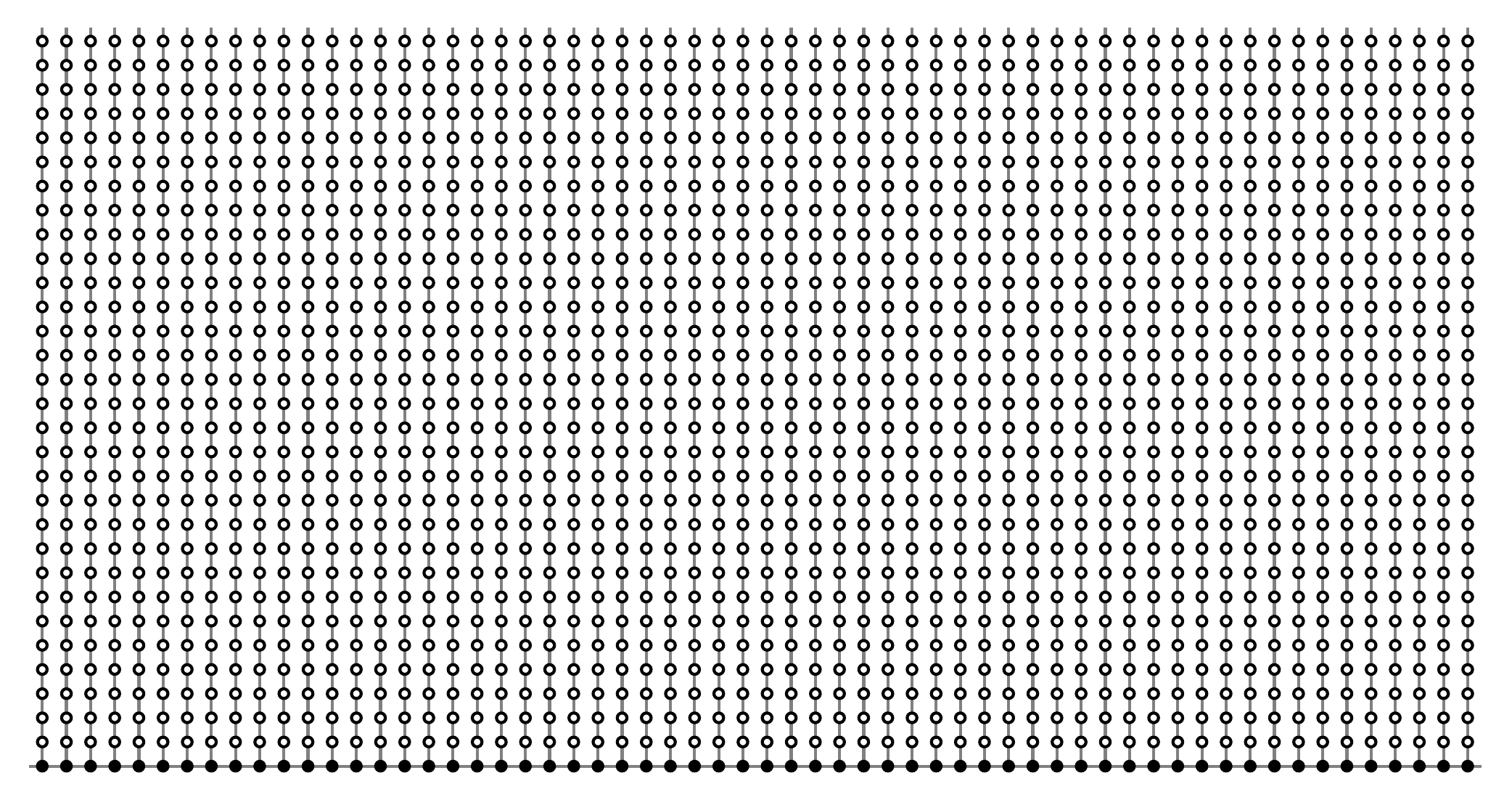}
\caption{A random comb ($p=.33$) vs. a regular comb ($p=0$)}
\label{ }
\end{center}
\end{figure}

We denote by $\cH$ the Hilbert space of square summable functions on the vertices of the graph $\mathcal C$ with the 
standard inner product.  At a fixed time, the state of a quantum walk on $\mathcal C$ is given by a normalized function in $\cH$.
The time development of the quantum walk on $\mathcal C$ is described by the 
unitary time evolution operator 

\begin{equation}
\label{theU}
U(t)=e^{-\mathi t H},
\end{equation}
where the Hamiltonian $H$ is minus the Laplacian on $\mathcal C$, defined on $\cH$ as 
\begin{equation}
\label{theH}
H\phi (n,j)= \sum_v (\phi(n,j) - \phi(v)),
\end{equation}
where the sum runs over the nearest neighbours of the vertex $(n,j)$.  
We will sometimes use the Dirac notation and 
denote by $|n,j\kt$ the function which takes the value $1$ at the vertex $(n,j)$ and is zero elsewhere.  These functions 
form an orthonormal basis for $\cH$ and we write $\phi(n,j)=\br n,j|\phi\kt$ for functions in $\cH$.  The inner product is
written
$\br \phi|\psi\kt=\sum_{(n,j)}\bar{\phi}(n,j)\psi (n,j)$.

If a quantum walker is located at the vertex $(n_1,j_1)$ at time $t=0$ (i.e.\ its state at $t=0$ is $|n_1,j_1\kt$)
then the probability amplitude that the walker
is located at $(n_2,j_2)$ at time $t$ is given by
\begin{equation}
\label{theAt}
A_t(n_1,j_1;n_2,j_2)=\br n_2,j_2|e^{-\mathi tH} |n_1,j_1\kt.
\end{equation}
The probability that the walker is at $(n_2,j_2)$ at time $t$, given that he is at $(n_1,j_1)$ at time $t=0$, is
\begin{equation}
\label{thePt}
P_t(n_1,j_1;n_2,j_2)=| A_t(n_1,j_1;n_2,j_2)|^2.
\end{equation}

As for the regular infinite comb, we are interested in the nature of the energy eigenstates, 
and especially the large time behaviour of the wave functions, when starting from a state localized on a single vertex (on the spine) at 
the initial time $t=0$.
We recall that in  \cite{David_2022} it was shown that for the regular comb the energy eigenstates $H\phi = E \phi$ 
can be separated into two classes:
The $0\le E\le 4$ states have the form
\begin{equation}
\label{phiE<4}
\phi (n,j)=A_ne^{\mathi\alpha n+\mathi\theta j}+B_ne^{\mathi\alpha n-\mathi\theta j}
\ ,\quad \theta\in [0,\pi],\quad \alpha\in [0,2\pi)
\ ,\quad E = 2 -2 \cos\theta
\end{equation}
and can propagate along the teeth.
The $E>4$ states have the form
\begin{equation}
\label{phiE>4}
\psi(n,j)=C_n(-1)^je^{\mathi\alpha n-\sigma j},\quad \sigma >0\ ,\quad E=2+2\cosh\sigma
\end{equation}
and do not propagate along the teeth, but are localized in the (vicinity of the) spine.


\section{The finite random comb}
\label{sFiniteComb}
\subsection{The geometry}
\label{ssGeometry}
If we replace the spine $\bbZ$ of the infinite comb by a finite linear graph but attach infinite teeth
in the same way as before to the vertices on the (now finite) spine we get what will be called a finite
random comb.
Two spine geometries will be considered, the open one and the closed (periodic) one.

For the open geometry, the spine is a linear chain of length $L$, with $L+1$ vertices labelled $n=0,1,\cdots, L$. 
No teeth are attached to the endpoints $n=0$ and $n=L$, and Dirichlet boundary conditions are imposed at the 
endpoints, so for the eigenfunctions $\phi$ of the Hamiltonian we require $\phi(0)=0$ and $\phi(L)=0$.
The number of active sites with a non-trivial wave function is $N=L-1$, and if $N_t$ and $N_h$ denote the number of 
teeth and holes, respectively, we have $N_t+N_h=N$.

For the closed (periodic) geometry, the spine is closed chain of length $N$, with $L$ vertices labelled $n=1,\cdots, L$. 
The vertex $L+1$ is identified with the vertex $1$, i.e., the eigenfunctions satisfy $\phi(n+L)=\phi(n)$. The number 
of active sites is $N=L$, and  $N_t+N_h=N$.

The open case is a bit easier to study mathematically, and the periodic case better for numerical calculations, 
but they are expected to be equivalent in the infinite comb limit $L\to\infty$.
Note that a variant of the open comb consists in replacing the two endpoints at $n=0$ and $n=L+1$ by infinite half lines with no teeth. In other words, it corresponds to an infinite comb ($n\in\mathbb{Z}$) with no teeth for $n<0$ and $n >L$.

\subsection{The eigenvalue equation}
\label{ssEvEquation}

We are interested in finding the eigenfunctions of the Hamiltonian $H$ on the finite comb.
Along the teeth, the energy eigenstates $\phi$ are still of the form \rf{phiE<4} and \rf{phiE>4} when $E<4$ or $E>4$, respectively.
We denote by $C_n$ the values of the eigenfunctions along the spine
\begin{equation}
\label{Cdef}
\phi(n,0)=C_n.
\end{equation}
The eigenfunctions with $0\le E<4$ energy can therefore be written as
\begin{equation}
\label{phiEle4}
\begin{split}
\phi (n,j)&=A_n\,e^{\mathi\theta j}+B_n\,e^{-\mathi\theta j}\quad\text{with\ } A_n+B_n=C_n,
\quad\text{if $n$ is a tooth}\\
&=C_n,\quad\text{if $n$ is a hole},
\end{split}
\end{equation} 
where again $$E =E(\theta)= 2 -2 \cos\theta\ ,\quad\theta\in [0,\pi]\ .$$
Similarily, the $E>4$ eigenstates can be expressed as
\begin{equation}
\label{phiEgt4}
\begin{split}
\phi(n,j)&=C_n(-1)^je^{-\sigma j},\quad\text{if $n$ is a tooth}\\
&=C_n,\quad\text{if $n$ is a hole},
\end{split}
\end{equation}
where again $E =2+2\cosh\sigma$.

The eigenvalue equation along the spine is
\begin{equation}
\label{WEt}
3\phi(n,0)-\phi(n-1,0)-\phi(n+1,0)-\phi(n,1)= E\, \phi(n,0),
\quad\text{if $n$ is a tooth}
\end{equation}
and
\begin{equation}
\label{WEh}
2\phi(n,0)-\phi(n-1,0)-\phi(n+1,0)= E\, \phi(n,0),
\quad\text{if $n$ is a hole}.
\end{equation}
For the $E<4$ states, \rf{WEt} and \rf{WEh}  lead to
\begin{equation}
\label{E<4tooth}
-C_{n-1}-C_{n+1}+A_n(3-e^{\mathi \theta})+ B_n(3-e^{-\mathi \theta})=(A_n+B_n)(2-2\cos\theta),
\ \text{if $n$ is a tooth}
\end{equation}
and
\begin{equation}
\label{E<4hole}
-C_{n-1}-C_{n+1}+2C_n=C_n(2-2\cos\theta),
\ \text{if $n$ is a hole}.
\end{equation}
Together with $C_n=A_n+B_n$ if $n$ is a tooth, this gives a set of $N=N_t+N_h$ equations for the 
$2N_t+N_h$ coefficients (the $A_n$'s and $B_n$'s for the teeth and the $C_n$'s for the holes).
Thus for a given $\theta$ (i.e.\ given $E$) the $A_n$'s and the $C_n$'s are determined 
(up to a normalization) by the $B_n$'s 
and the boundary conditions at the ends of the finite comb.  

For the $E>4$ states , \rf{WEt} and \rf{WEh}  lead to
\begin{equation}
\label{E>4tooth}
-C_{n-1}-C_{n+1}+C_n(3+e^{-\sigma})=C_n(2+2\cosh\sigma),
\quad \text{if $n$ is a tooth}
\end{equation}
and
\begin{equation}
\label{E>4hole}
-C_{n-1}-C_{n+1}+2C_n=C_n(2+2\cosh\sigma),
\quad \text{if $n$ is a hole}.
\end{equation}
We have $N=N_t+N_h$ equations for the $N$ coefficients $C_n$, so they are determined (up to a normalization) by $\sigma$ and the 
boundary conditions.

\subsection{$E<4$ states and the $S$-matrix}
\label{ssSmatrix}
\subsubsection{The $S$-matrix}
\label{sssSmatrix}
Let us first discuss the energy eigenstates with $E<4$, which propagate in the teeth.
As for the regular comb, the spectrum of $H$ for $0\le E\le 4$ is continuous. Each energy eigenspace 
$\mathcal{H}_\theta$ of states $\phi$ 
such that $H\phi=E(\theta)\phi$ is finite dimensional with dimension $N_t$ (the number of teeth).

In order to study the continuous spectrum eigenfunctions, we consider the $S$-matrix (or reflection matrix), 
defined as follows (see, e.g., \cite{PhysRevA.80.052330}).
First, in order to simplify the discussion, let us add additional variables $A_n$ and $B_n$ when the site $n$ is a hole, 
with the constraint
$A_n+B_n=C_n$. The general form of the energy eigenfuction \ref{phiEle4} becomes
\begin{equation}
\label{phiEle4bis}
\begin{split}
\phi (n,j)&=A_n\,e^{\mathi\theta j}+B_n\,e^{-\mathi\theta j}
\quad\text{if $n$ is a tooth,}\\
\phi (n,0)&=A_n+B_n\quad\text{if $n$ is a hole}.
\end{split}
\end{equation} 
The eigenvalue equations \rf{WEt} \rf{WEh} now read
\begin{equation}
\label{E<4tfull}
\begin{split}
-A_{n-1}-A_{n+1}+A_n(1+e^{-\mathi\theta})&=B_{n-1}+B_{n+1}- B_n(1+e^{\mathi\theta})
\ \text{if $n$ is a tooth}\\
-A_{n-1}-A_{n+1}+A_n(e^{\mathi\theta}+e^{-\mathi\theta})&= B_{n-1}+B_{n+1}-B_n(e^{\mathi\theta}+e^{-\mathi \theta})\ \text{if $n$ is a hole}.
\end{split}
\end{equation}
Let $A$ and $B$ denote
the $N$-component column vectors $A={(A_n)}_{\scriptscriptstyle{n\in\mathrm{spine}}}$ and $B={(B_n)}_{\scriptscriptstyle{n\in\mathrm{spine}}}$.
Then \rf{E<4tfull} can be written 
\begin{equation}
\label{XAYBeq}
\mathbb{X}\cdot A+ \mathbb{Y}\cdot B=0,
\end{equation}
where $\mathbb{X}$ and $\mathbb{Y}= \mathbb{\bar X}$ (its complex conjugate) are $N\times N$ matrices (which depend on 
$\theta$ and the configuration of teeth). 

Let us begin by assuming that the matrix $\mathbb{X}$ is invertible, or equivalently that $\mathbb{Y}$ is invertible.
Eq.~\rf{XAYBeq} gives a mapping $B\mapsto A$, to which we  associate an $N\times N$  matrix $\mathbb{S}$ 
\begin{equation}
\label{BigSdef}
\mathbb{S}=-\mathbb{X}^{-1} \mathbb{Y}\quad\text{such that}\quad A=\mathbb{S} B.
\end{equation}

In the decomposition \ref{phiEle4} of the energy eigenstates, for any tooth $n$, the component $B_ne^{-\mathi\theta j}$ 
corresponds to an incoming wave of energy 
$E$ propagating toward the spine along the tooth $n$, while the component $A_ne^{\mathi\theta j}$ corresponds to an outgoing 
wave of energy $E$ propagating away from the spine along the tooth $n$. 
Moreover, for any hole $m$, the variable $C_m$ depends only on the $N_t$ variables 
$\{B_n:\,n$ tooth$\}$, not of the auxiliary $N_h$ variables $\{B_n:\,n$ hole$\}$. Therefore, the  
$\mathtt{hole}\to\mathtt{tooth}$ matrix elements of the  matrix $\mathbb{S}$ must be zero
\begin{equation}
\label{Sdef}
\mathbb{S}_{\mathtt{tooth}\,\mathtt{hole}}\ =\ 0
\end{equation}
so that the restriction $S$ of $\mathbb{S}$ to $\mathtt{tooth}\to\mathtt{tooth}$ elements is an 
$N_t\times N_t$  matrix which describes the reflection of an incoming wave along the teeth $B_\mathrm{t}$, with energy $E=2-2\cos\theta$ 
into an outgoing wave $A_\mathrm{t}$ with the same energy,
\begin{equation}
\label{ }
 B_{\mathrm{t}}\ \mapsto\  A_{\mathrm{t}}= S  B_{\mathrm{t}},
\end{equation}
where 
$A_\mathrm{t}$ and $B_\mathrm{t}$ are $A_\mathrm{t}={(A_n)}_{\scriptscriptstyle{n:\mathrm{teeth}}}$ and $B_\mathrm{t}={(B_n)}_{\scriptscriptstyle{n:\mathrm{teeth}}}$.
We will show in Appendix \ref{App6} that the matrix $S$ is unitary\footnote{The $N\times N$ 
matrix  $\mathbb{S}$  is not unitary, see Appendix \ref{App6}}
with respect to the standard scalar product on the $N_t$-dimensional Hilbert space $\mathcal{H}_{\mathrm{teeth}}=\mathop{\oplus}\limits_{n:\mathrm{tooth}} \mathbb{C}=\mathbb{C}^{N_t}$
which is given by
\begin{equation}
\label{SPteethDef}
 {(X|Y)}=\sum_{n:\mathrm{tooth}} \overline X_n Y_n 
\end{equation}
for $X,Y\in \mathcal{H}_{\mathrm{teeth}}$.
The special case $\theta=0$ or $\pi$ is discussed in Appendix~\ref{App5b}, where 
we show that the S-matrix reduces to $S=-1$ 
(this is a bit formal since there are no propagating in and out states along the teeth). 

The above definitions of the matrices $\mathbb{S}$ and $S$ are valid if $\mathbb{X}$ is invertible.
In Appendix~\ref{App6} we also show that this is the case for generic values of the variable $\theta$, 
and that for only a finite number of isolated values of $\theta$ (this number being bounded by $2N$), 
the matrix $\mathbb{X}$ is not invertible.  In this case 
the matrix $\mathbb{S}$ is ill-defined. However, by analytic continuation in the 
variable $z=e^{i\theta}$, we show that $\mathbb{S}$  can be defined as a meromorphic 
(matrix valued) function of $z$, and that the $N_t\times N_t$ S-matrix $S$ is in fact well defined 
as a regular function of $\theta$, and unitary, in the whole range $0<\theta<\pi$, even at the isolated 
values where the $N\times N$ matrix $\mathbb{S}$ is ill-defined (which correspond to poles in the $z$ variable).

\subsubsection{The full scalar product}
\label{sssScalProd}

We define $\mathcal{H}_{E<4}\subset \mathcal{H}$ to be the subspace spanned by states of energy
$E<4$. Let $|\Phi_\theta\rangle$ and $|\Phi'_{\theta'}\rangle$ be two energy eigenstates 
with energies $E=2-2\cos\theta$ 
and $E'=2-2\cos\theta'$, respectively.  
They have wave functions of the form \rf{phiEle4}
\begin{equation}
\label{ }
\phi_\theta (n,j)=A_n\,e^{\mathi\theta j}+B_n\,e^{-\mathi\theta j}
\ ,\quad
\phi'_{\theta'} (n,j)=A'_n\,e^{\mathi\theta' j}+B'_n\,e^{-\mathi\theta' j}
\end{equation}
with $A,B$ and $A',B'$ satisfying \rf{XAYBeq}.
The scalar product in $\mathcal{H}_{E<4}$
\begin{equation}
\label{standSP}
\langle\Phi_\theta  | \Phi'_{\theta'} \rangle = \sum_{n,j} \bar\phi_\theta (n,j) \phi'_{\theta'} (n,j)
\end{equation}
is calculated in Appendix~\ref{Appendix7}. 
It is found to be
\begin{equation}
\label{SPPhiAB}
\langle\Phi_\theta  | \Phi'_{\theta'} \rangle = 2\pi\,\delta(\theta-\theta')\,{(A | A')} = 2\pi\,\delta(\theta-\theta')\,{(B | B')} .
\end{equation}
We obtain a partition of unity in $\mathcal{H}_{E<4}$ by choosing an orthonormal 
basis in each subspace $\mathcal{H}_\theta$ spanned by eigenstates of energy
$E=2-2\cos\theta$.
Two such bases are interesting and will be considered below.

\subsubsection{The $S$-matrix eigenvector basis}
\label{sssSegv}

We consider the eigenvectors of the matrix $S$.
 The eigenvalues of $S$ are pure phases $\zeta$ of the form
\begin{equation}
\label{ }
\zeta=e^{\mathi\delta}\ ,\quad -\pi<\delta\le \pi ,
\end{equation}
with $\delta$ the phase shift between the in-wave (the $B$'s) and the out-wave (the $A$'s).
The corresponding eigenvector of $S$ is a solution of \rf{XAYBeq} such that
\begin{equation}
\label{ASB}
A_n=e^{\mathi\delta}B_n\ ,\quad \text{$n$ tooth}.
\end{equation}
Any eigenvector of $S$ can be extended to an eigenvector of $\mathbb{S}$ by 
associating the $C_n$'s on the holes 
to the pairs $A_n,\,B_n$ such that 

\begin{equation}
\label{CABd}
C_n=A_n+B_n=(1+e^{\mathi\delta})B_n.
\end{equation}
This makes sense as long as the phase shift $\delta$ is different from $\pi$. 
The special case $\delta=\pi$ 
corresponds to  solutions such that $C_n=0$ and will be discussed in Appendix~\ref{App5b}.

We let  $\phi_{\theta,\delta}$ denote the wavefunction of  a state $|\Phi_{\theta,\delta}\rangle$ which is 
both an eigenfunction of the hamiltonian $H$  with eigenvalue $E=2-2 \cos\theta<4$ and an 
eigenfunction of  the S-matrix $S$ 
with phase shift $\delta$.
The $B_n$'s and the $A_n$'s  satisfy equations
\rf{E<4tooth} and \rf{E<4hole} together with \rf{CABd}. The $A_n$'s can be eliminated through \ref{ASB}, and the eigenvalue equation 
can be rewritten as a set of $N$ equations for the $B_n$'s:
\begin{equation}
\begin{split}
\label{BthEq}
2\,B_n-B_{n-1}-B_{n+1} +V(E,\delta)\,B_n&=E\, B_n,\quad\text{if $n$ is a tooth}\\
2\,B_n-B_{n-1}-B_{n+1} &=E\, B_n,\quad\text{if $n$ is a hole},
\end{split}
\end{equation}
where
\begin{equation}
\label{VEd}
V(E,\delta)=1-\cos\theta + \sin\theta \tan(\delta/2)=E/2+\sqrt{E(1-E/4)}\,\tan(\delta/2).
\end{equation}

\subsubsection{The $S$-matrix column vector basis}
\label{sssScvec}
For each value of the energy $E$, another orthonormal basis for the 
subspace $\mathcal{H}_\theta$ is made up of the column vectors of the matrix $S=S(\theta)$. 
This basis is defined as follows.
To each tooth $t$ of the chain $\mathcal{C}$, we associate the state $|\Upsilon_{\theta,t}\rangle$, 
whose wave  function $\phi^{(t)}$ 
is obtained by solving the eigenvalue equation \rf{E<4tfull} 
 with the $B$-coefficients $B_n^{(t)}$ corresponding to an incoming particle localized on the tooth $t$, 
i.e.,
\begin{equation}
\label{column}
B^{(t)}_n=\delta_{n,t}.
\end{equation}
Note that here, since $t$ is a tooth, the $B_n$'s are set to zero if $n$ is a hole.
From \rf{E<4tfull}, the $A^{(t)}_n$ are solutions of 
\begin{equation}
\label{E<4column}
\begin{split}
-A_{n-1}-A_{n+1}+A_n(1{+}e^{-\mathi\theta})&=\delta_{n,t+1}+\delta_{n,t-1}-\delta_{n,t}(1{+}e^{\mathi\theta})
\quad \text{if $n$ is a tooth},\\
-A_{n-1}-A_{n+1}+A_n(e^{\mathi\theta}{+}e^{-\mathi\theta})&= \delta_{n,t+1}+\delta_{n,t-1}- 
\delta_{n,t}(e^{\mathi\theta}{+}e^{-\mathi\theta})\ \text{if $n$ is a hole}.
\end{split}
\end{equation}
From the unitarity of $S$, 
$$
\sum_{n:\mathrm{teeth}} \bar A^{(t)}_n A^{(t')}_n=\delta_{t,t'},
$$
and Eq.\ \rf{SPPhiAB}, we see  that
the $|\Upsilon_{\theta,t}\rangle$  form an orthonormal (up to a factor of $2\pi$) basis for $\mathcal{H}_{\scriptscriptstyle{E{<}4}}$:
\begin{equation}
\label{OrthNUps}
\langle \Upsilon_{\theta,t} | \Upsilon_{\theta',t'}\rangle\ =\ 2\pi\,\delta(\theta-\theta')\,\delta_{t,t'}.
\end{equation}
Note also that if we define
\begin{equation}
\label{ A2Atild}
\tilde A_n=A_n+\delta_{n,t}
\end{equation}
then \rf{E<4column} can be rewritten
\begin{equation}
\label{E<4columnA}
\begin{split}
-\tilde A_{n-1}-\tilde A_{n+1}+\tilde A_n(1+e^{-\mathi\theta})&= -\delta_{n,t}(e^{i\mathi\theta}-e^{-\mathi\theta}),
\quad \text{if $n$ is a tooth}\\
-\tilde A_{n-1}-\tilde A_{n+1}+\tilde A_n(e^{\mathi\theta}+e^{-\mathi\theta})&= 0,\quad \text{if $n$ is a hole}.
\end{split}
\end{equation}

\begin{figure}
\begin{center}
\includegraphics[width=3.in]{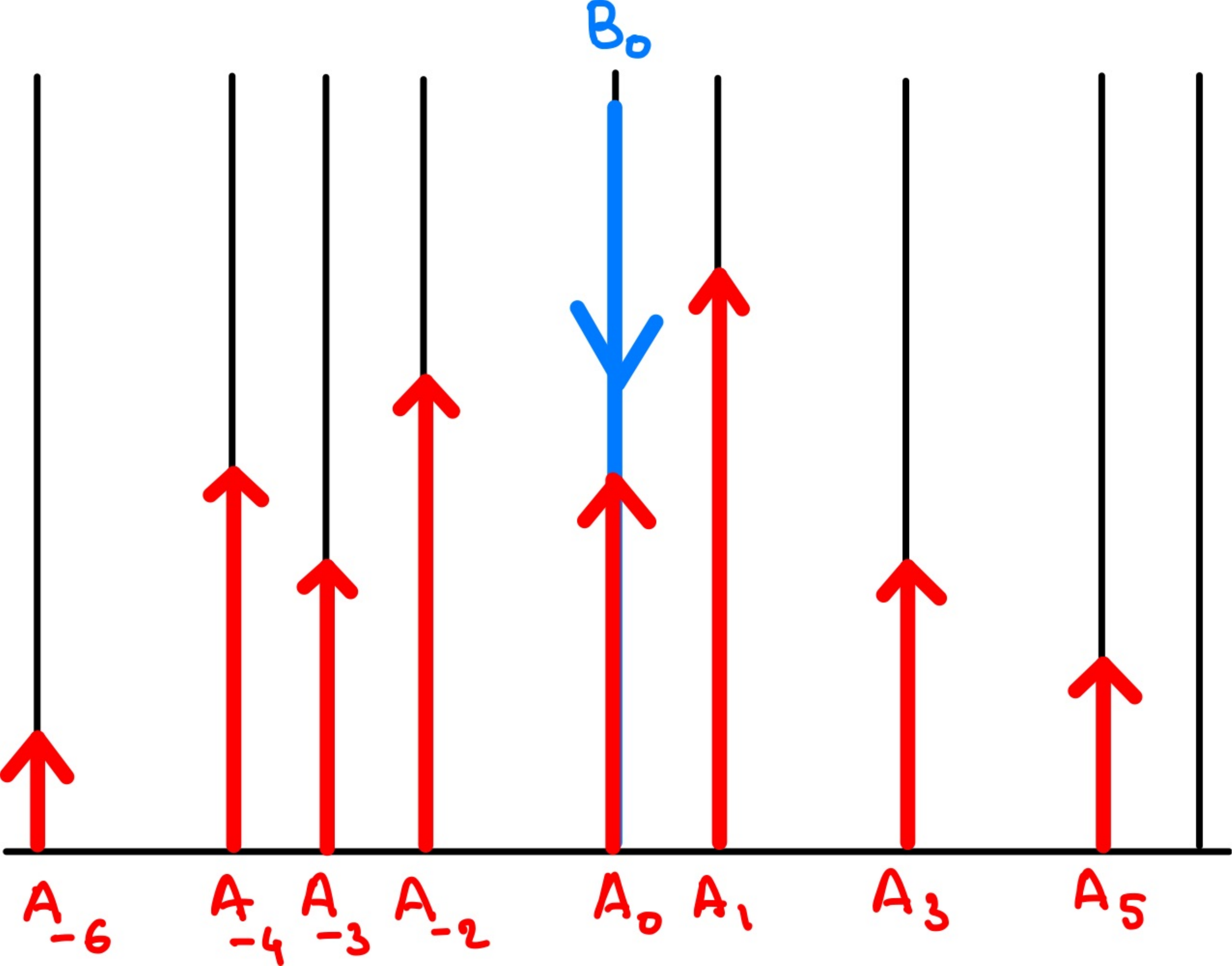}
\caption{In and Out states}
\label{ BtoAS}
\end{center}
\end{figure}

\subsection{The $E>4$ states}
The $E>4$ eigenstates of $H$ are of the form
given in \rf{phiEgt4}.
The $C_n$'s obey
\begin{equation}
\label{E>4equ}
\begin{split}
-C_{n-1}-C_{n+1}+C_n(3+e^{-\sigma})&=C_n(2+2\cosh\sigma),
\quad \text{if $n$ is a tooth}
\\
-C_{n-1}-C_{n+1}+2C_n&=C_n(2+2\cosh\sigma),
\quad \text{if $n$ is a hole}.
\end{split}
\end{equation}
Finding the eigenstates of $H$, namely the values of $\sigma$ where \rf{E>4equ} has 
a solution on the finite chain, is not a standard linear eigenvalue problem.
It has a finite number of solutions $N_{E>4} $ with $\sigma>0$
which satisfies the bounds
\begin{equation}
\label{NEgt4bound}
N_t/2 \le N_{E>4} \le N_t 
\end{equation}
as we show in Appendix A.3 where a formula for $N_{E>4}$ is derived.

However, this problem can be mapped onto a standard linear problem by the following trick.
To every $C_n$ let us associate a $\tilde C_n$ defined as
\begin{equation}
\label{CCtilde}
\tilde C_n=e^{-\sigma}C_n .
\end{equation}
Eq.\ \rf{E>4equ} can then be rewritten as
\begin{equation}
\label{E>4equ2}
\begin{split}
-C_{n-1}-C_{n+1}+C_n&=e^\sigma C_n,
\quad \text{if $n$ is a tooth}
\\
-C_{n-1}-C_{n+1}-\tilde C_n&=e^\sigma C_n,
\quad \text{if $n$ is a hole}.
\end{split}
\end{equation}
Regrouping the $C$'s and $\tilde C's$ into a $2N$ component vector 
$\mathbf{C}=(C,\tilde C)$, \rf{CCtilde} and \rf{E>4equ2} take the form of a linear
eigenvalue equation
\begin{equation}
\label{ }
\mathbb{M}\mathbf{C}=e^{\sigma}\mathbf{C}
\quad\text{with}\qquad \mathbf{C}=(C,\tilde C),
\end{equation}
where $\mathbb{M}$ is a $2N\times 2N$ matrix, which depends on the configuration of holes and teeth on the 
chain, and all the matrix elements are $0$ or $\pm 1$.

The matrix $\mathbb{M}$ is real, but not symmetric. A priori it has $2N$ eigenvalues, 
real ones or pairs of 
complex conjugate ones. Only the eigenvectors associated to eigenvalues $\lambda$ of $\mathbb{M}$ which are real 
and lie in the interval $(1,3)$ (see the next paragraph) 
correspond to physical solutions of \rf{E>4equ} and $E>4$ eigenstates of $H$.
This formulation allows an easy numerical study of the $E>4$ solutions.

An alternative approach to the problem is to observe that the eigenvalue problem can be written
\beq{alternative}
(H_0+W(n))C_n=(2+2\cosh\sigma)C_n
\eeq
where $H_0$ is the Laplacian on the discrete line and $W(n)=(1+e^{-\sigma })\chi_n$ 
where $\chi_n$ is the characteristic function for the teeth,
\begin{equation}
\label{characteristic}
\chi_n\ =\ \begin{cases}
     1 & \text{if $n$ is a tooth}, \\
     0 & \text{if $n$ is a hole}.
\end{cases}
\end{equation}

The spectrum of $H_0$ lies in $[0,4]$ so the eigenvalues $f_i$ of $H_0+W(n)$ lie in 
the interval $[0,5+e^{-\sigma}]$. 
Varying $\sigma$ we obtain a solution to the original problem exactly when one of 
the $f_i's$ equals $2+2\cosh\sigma$.  
A necessary condition for this is
that $e^\sigma <3$ which implies $2+2\cosh\sigma <16/3$.

\subsection{Mapping on the Anderson model}
\label{Anderson}
Both the $E>4$ and the $E<4$ eigenvalue equation on the spine can be mapped 
onto a well-known one-dimensional model, the tight-binding binary chain model 
\cite{LuckBook1992}
\cite{PhysRev.105.425}
\cite{Nieuwenhuizen:1985aa}, which is a special case of the Anderson model \cite{PhysRev.109.1492}
\cite{LiebMattis66}
\cite{Lifschitzetal1988}.
The tight binding chain model is a disordered quantum model on a linear chain with a nearest neighbour 
hopping term and a random potential $\mathcal{V}(n)$ on each site $n$. The energy $\mathcal{E}$ eigenfunctions 
obey the equation
\begin{equation}
\label{bcgeneric}
-\varphi({n+1})-\varphi({n-1}) + \mathcal{V}(n)\varphi(n)=\mathcal{E}\,\varphi(n).
\end{equation}
The binary chain is the case where the potential $\mathcal{V}(n)$ takes only two values
(Bernoulli potential).

For a random comb with distribution of teeth given by a function ${\chi}_n$, the eigenvalue equation on the spine 
can be written as the equation of the Anderson model \rf{bcgeneric}, with
\begin{equation}
\label{VnVchi}
\varphi(n)=C_n\ ,\quad \mathcal{V}(n) = \mathcal{V}\,\chi_n\ ,\quad \mathcal{E}=E-2.
\end{equation}
For the $E>4$ comb eigenstates, localized on the spine, we find that 
\begin{equation}
\label{V&E>4}
\mathcal{V}= 1+e^{-\sigma}\quad,\quad \mathcal{E}=2\cosh\sigma.
\end{equation}
In this case the parameters $\mathcal{E}$ and $\mathcal{V}$ are not independent but related by
\begin{equation}
\label{A1}
\mathcal{E}=\mathcal{V}-1 +{1\over \mathcal{V}-1}.
\end{equation}
For the  $E<4$ energy and phase-shift eigenstates, $\mathcal{V}$ and $\mathcal{E}$ are given by
\begin{equation}
\label{A2}
\mathcal{E}= E-2=-2\cos\theta\ ,\quad \mathcal{V}= 1+{\mathcal{E}/ 2}+\sqrt{1-\mathcal{E}^2/4}\,\tan(\delta/2),
\end{equation}
cf.\ \rf{VEd}.
In this case $\mathcal{E}$ and $\mathcal{V}$ can be viewed as independent parameters. 

\begin{figure}[h!]
\begin{center}
\includegraphics[width=4in]{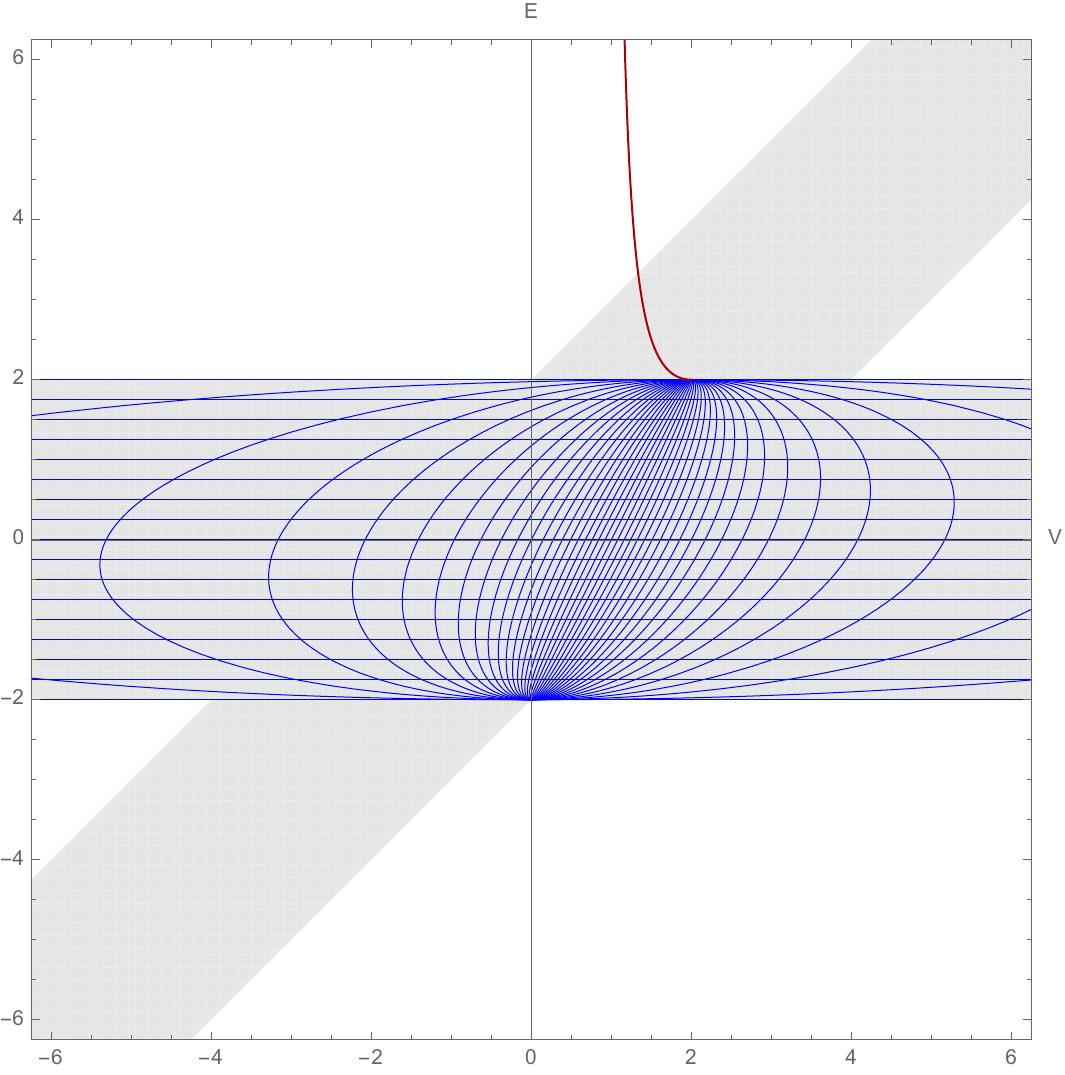}
\caption{The ellipses describe the location of the $|\mathcal{E}|<2$ eigenstates in the 
$\mathcal{E},\mathcal{V}$ plane for a fixed value of the 
phase shift $\delta$ 
and the red curve corresponds to the $|\mathcal{E}|>2$ states, cf.\ \rf{A1} and \rf{A2}} \label{ VEcurves}
\end{center}
\end{figure}

\subsection{The spectrum of the binary chain}
\label{ssSpecBinC}
\subsubsection{The spectral flow}
The random binary chain described in the previous subsection 
is a prototypical model for disordered systems, and there is an extensive literature 
in physics and in mathematics, see e.g.,\ \cite{LuckBook1992} \cite{Comtet_2013}.  We will denote the energy and disorder strength by
$E$ and $V$, respectively, so when we connect this to the random comb we have $E=-2\cos\theta$ or 
$E=2\cosh\sigma$ which amounts to shifting the energy by 2.  This convention will be used in the rest of 
this section.

Let us recall some well known (and maybe not so well known) properties of this model and its spectrum, 
as a function of the  parameters $E$ and $V$, first for a finite chain and then in the infinite length limit.
We start from the eigenvalue equation
\begin{equation}
\label{bcgeneric1}
\begin{split}
H\phi\,(n)=&-\phi({n+1})-\phi({n-1}) + V(n)\,\phi(n)=E\,\phi(n)
\\ 
&\ 
V(n) = V\,\chi_n,\quad
\chi_n= 0, 1
\end{split}
\end{equation}
on a chain with length $L$, where we have either periodic boundary 
conditions, $\phi(n)=\phi({n+L})$, or open (Dirichlet) boundary conditions, $\phi(n)=0$ if $n\le 0$ or $n\ge L+1$.
For the random binary chain, the $\chi_n$ are i.i.d.\ random variables with $P(\chi_n{=}0)=p$ and $P(\chi_n{=}1)=1-p$.

Since $H$ is symmetric, its eigenvalues are real and we can take the corresponding eigenfunctions real too.
The spectrum of $H$ consists of $L$ eigenvalues $E_\alpha$, which are contained in the set 
$[-2,2]\cup [V-2,V+2]$ (and with probability one in the $L\to\infty$ limit for the random chain)
\cite{CylonFroeseFroeseSimonBook1987}.  For Dirichlet 
boundary conditions, the spectrum is simple but degeneracy can occur
on the periodic comb.
We are interested in the variation of the spectrum with the spectral parameter $V$.
An example of such a spectral flow is shown on Fig.~\ref{fEflow1}

\begin{figure}[h]
\begin{center}
\includegraphics[width=4in]{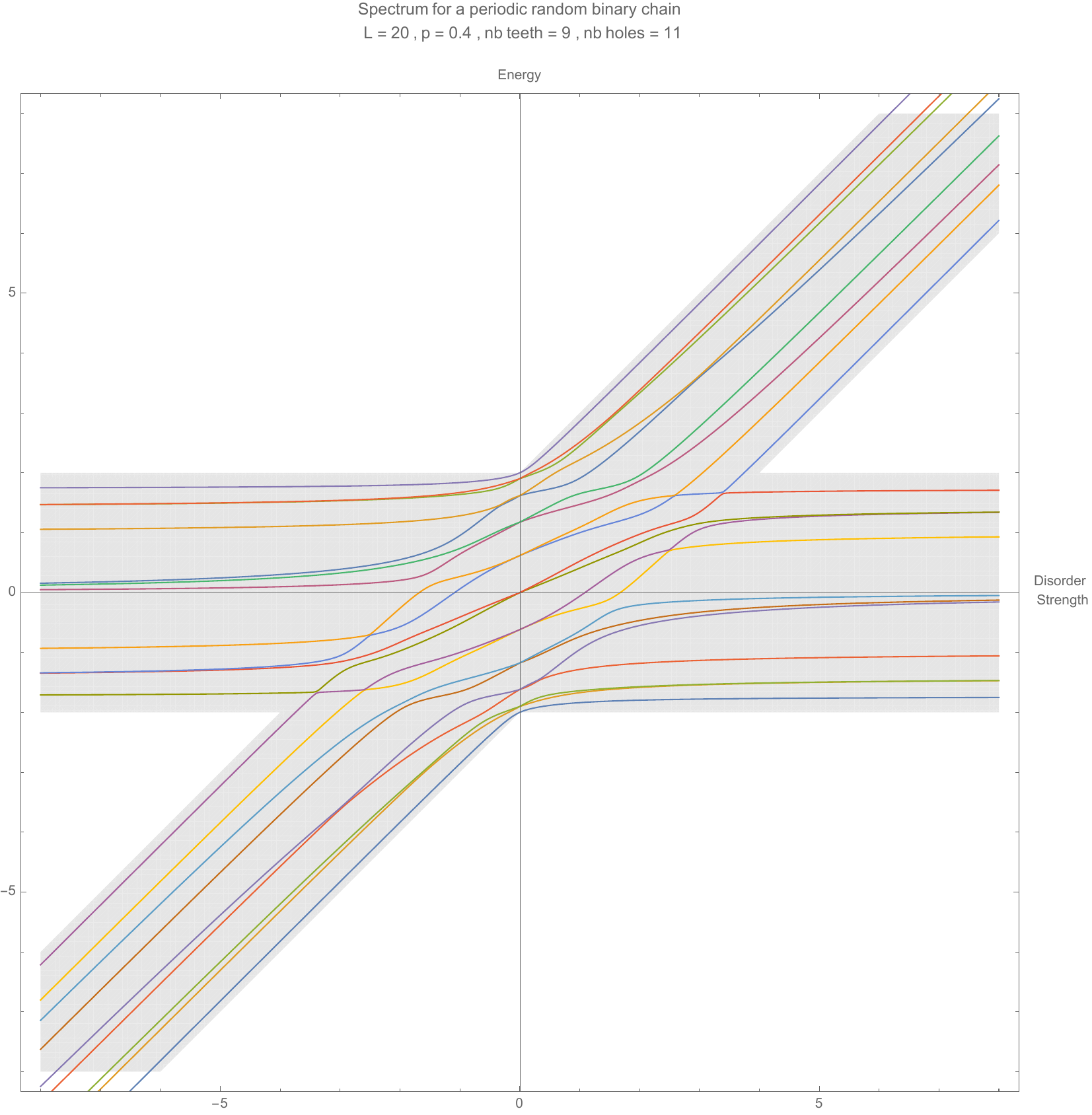}
\caption{The flow of eigenvalues $E$ as a function of the disorder $V$ for a  random 
comb $\mathcal{C}$ of length $L=20$ (here with $N_t=9$ teeth and $N_h=11$ holes) and with periodic
boundary conditions.}
\label{fEflow1}
\end{center}
\end{figure}

\subsubsection{Symmetries of the spectrum}
We first note that if $E$ is an eigenvalue for a given $V$ and a fixed distribution of teeth 
$\chi_n$, then $-E$ is an
eigenvalue for $-V$ and the same $\chi_n$: 
\begin{equation}
\label{ParitySymmetryFlow}
E\to -E\ ,\quad V\to -V.
\end{equation}
This we prove by setting $\psi(n)=(-1)^n\phi(n)$
where $\phi$ is an eigenfunction with eigenvalue $E$.
Then $\psi$ satisfies the equation
\beq{b2}
\psi(n-1)+\psi(n+1)+V(n)\psi(n)=E\psi(n),
\eeq
which proves the claim.

If $E$ is an eigenvalue for a given $V$ and $\chi_n$, then $V-E$ is an eigenvalue for the same $V$
and the chain characterized by a new tooth distribution $\chi_n'=1-\chi_n$ 
(or, equivalently, by a new potential $V'$ such that $V'(n)=V-V(n)$): 
\begin{equation}
\label{ToothHoleSymmetry}
E\to V-E\ ,\quad\chi_n \to 1-\chi_n\ ,\quad\text{$V$ fixed}.
\end{equation} 
The eigenvalue equation can be
rewritten
\beq{b3}
-\phi(n+1)-\phi(n-1)+V(\chi_n-1)\phi (n)=(E-V)\phi(n),
\eeq
and using the previous result proves the claim.

\subsubsection{Monotonicity}
For a given $\chi_n$, except at some special values of $V$ where level crossing may occur, 
the spectrum is non-degenerate and we can follow a given eigenvalue $E$ and the corresponding 
eigenfunction $\phi$ as functions of $V$.
Then we have
\beq{bb4}
0\le{dE\over dV}\le 1.
\eeq
In order to prove this let us normalize $\br\phi |\phi\kt=\sum_n|\phi(n)|^2=1$.  Then
$
H\phi =E\phi
$
where $H=H_0+V\,T$, $T$ is the diagonal matrix with entries $\chi_n$ and $H_0$ is given by
$$
H_0\,\phi(n)=-\phi(n-1)-\phi(n+1).
$$
It follows that $E=\br\phi |H|\phi\kt$ and 
\bea
{dE\over dV} & = & \br\phi |T|\phi\kt +\br{d\phi\over dV}|H|\phi\kt +\br\phi |H|{d\phi\over dV}\kt
= \br\phi |T|\phi\kt
=\sum_{k:\chi_k=1}|\phi(k)|^2
\eea
where we have used the normalization condition.  The claim \rf{bb4} follows.

In fact, one has the stronger bounds
\begin{equation}
\label{bb4gen}
0<{dE\over dV}<1
\end{equation}
unless the eigenfunction $\phi$ vanishes at all the vertices where there is a tooth, 
or it vanishes at the $\chi_n=0$ vertices (the holes), respectively.   
This happens only for special values of $E$, and for
very special configurations, where the teeth are distributed in a periodic way, which 
has probability $0$ in the limit $L\to\infty$.

\subsubsection{Non-crossing}
For the open chain (Dirichlet b.c.), there is no level crossing and the spectrum of $H$ is 
non-degenerate for any finite $V$.  
For completeness we provide a proof.  Suppose 
we have two eigenfunctions of $H$, $\phi_1$ and $\phi_2$, with the same eigenvalue $E$. 
Multiplying the equation for $\phi_1$
by $\phi_2$ and the equation for $\phi_2$ by $\phi_1$ and subtracting we get
$$
\phi_1(n)(\phi_2(n+1)+\phi_2(n-1))-\phi_2(n)(\phi_1(n+1)+\phi_1(n-1))=0
$$
which can be written $W(n+1)-W(n)=0$, with the discrete Wronskian
$$
W(k)=\phi_1(k)\phi_2(k-1)-\phi_2(k)\phi_1(k-1),
$$
which is therefore a constant $W$ along the chain.
By the Dirichlet boundary conditions it follows that $W=0$.  Hence,
$$
\phi_1(n)\phi_2(n-1)=\phi_2(n)\phi_1(n-1)
$$
which implies that $\phi_2$ and $\phi_1$ are proportional.

For periodic boundary conditions, the above argument does not apply. For example, if 
$V=0$ the spectrum is degenerate, with degeneracy two for all eigenvalues, except for 
the smallest one
(and the largest one when $L$ is even). When $V\neq 0$ and we have a 
generic random configuration of teeth 
we expect the spectrum to be non-degenerate, but one can find counterexamples for very 
specific chain configurations 
and special values of $V$ and $E$ (again with periodic distributions of teeth).

\subsubsection{{The spectrum for $|V|>4$}}

Let us consider the spectrum of $H$ for large $|V|$, and more specifically when
$V<-4$ or $V>4$.  
It consists of two parts, one which lies in the interval $[-2,2]$ and a disjoint one in $[V-2,V+2]$.  
We claim that the number of eigenvalues in the second component
is equal to the number of teeth ($\chi_n=1$ sites).  An intuitive argument is that 
(by symmetry of the spectrum) we can consider the
limit $V\to-\infty$.  Then the eigenfunctions with the low energy will be concentrated on the tooth 
vertices and therefore
the space they span has dimension equal to the number of teeth.
\begin{figure}[h!]
\begin{center}
\includegraphics[width=4in]{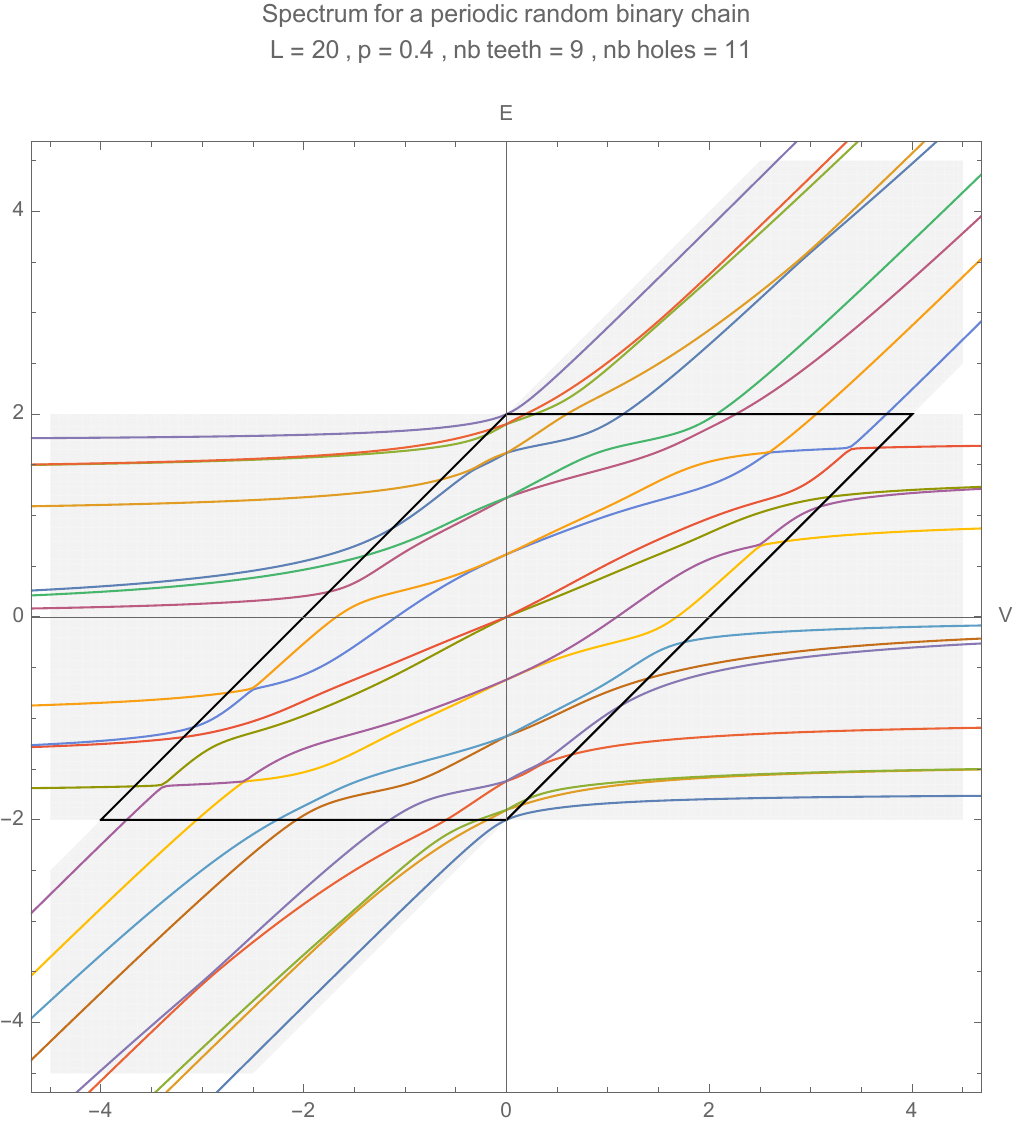}
\caption{Enlarged view of the spectral flow of Fig.~\ref{fEflow1}, with the 4 segments considered in the argument.}
\label{lozenge}
\end{center}
\end{figure}

A more precise argument for this goes as follows.  Consider a comb with $N_t$ teeth and $N_h$ holes.
We first assume the comb is periodic so its total length is $L=N_t+N_h$.
Consider the states with fixed energy $E=-2$ and $-4<V<0$ (see the bottom edge of the 
parallelogram in Fig.~\ref{lozenge}).  In this case the 
eigenvalue equation becomes
\beq{z1}
2\phi(n)-\phi(n-1)-\phi(n+1)=-V\phi(n),
\eeq
if there is a tooth at vertex $n$, and if there is no tooth at $n$ the equation can be written
\beq{z2}
\phi(n)-\phi(n-1)=\phi(n+1)-\phi(n).
\eeq
This implies that along a string of holes between two teeth the eigenfunction $\phi$ is linear.  To be 
concrete, suppose there is a tooth at $n=0$ and the next tooth is at $n=k+1$, $k\geq 0$.  Then 
on the vertices
$n$ between $0$ and $k+1$ we have
\beq{z3}
\phi(n)-\phi(0)=n(\phi(1)-\phi(0))={n\over k+1}(\phi(k+1)-\phi(0) ).
\eeq
This means that we can consider the reduced chain of length $N_t$, where there is a tooth at each vertex, 
with sites labelled by
$a=1,2,\ldots N_t$, but now the edges have a weight $k_a+1\geq 1$, where $k_a\geq 0$ is the number of holes 
on the original comb between $a$ and $a+1$.   We define
\beq{z4}
\tilde{\phi}(a)=\phi(n_a),
\eeq
where $n_a$ is the coordinate of tooth $a$ on the original comb.  The eigenvalue equation on the reduced 
weighted comb
is, by \rf{z1} and \rf{z3},
\beq{z5}
{1\over k_{a-1}+1}(\tilde{\phi}(a)-\tilde{\phi}(a-1))+{1\over k_{a}+1}(\tilde{\phi}(a)-\tilde{\phi}(a+1))
=-V\tilde{\phi}(a).
\eeq
We note that $k_0=k_{N_t}$ by the periodic boundary conditions.
This is an eigenvalue equation for the self-adjoint Hamiltonian
\beq{z6}
\tilde{H}=\sum_{a=1}^{N_t} \left[\left( {1\over k_{a-1}+1}+{1\over k_a+1}\right) |a\kt\br a| -
\left( {|a+1\kt\br a|\over k_a+1}+{|a-1\kt\br a|\over k_{a-1}+1}\right)\right],
\eeq
where the bras and the kets are labelled by the coordinate on the reduced lattice.  This Hamiltonian has 
$N_t$ eigenstates $\tilde{\phi}_\alpha$ with energies $\tilde{E}_\alpha$, $\alpha=1,2,\ldots N_t$.
The eigenvalues lie in the interval $[0,4]$ since 
\bea
\br \tilde{\phi}|\tilde{H}|\tilde{\phi}\kt & = & \sum_a{1\over k_a+1}(\tilde{\phi}(a)-\tilde{\phi}(a+1))^2\nonumber\\    
 & \leq & \sum_a (\tilde{\phi}(a)-\tilde{\phi}(a+1))^2 \nonumber\\
 &\leq & 4\br\tilde{\phi}|\tilde{\phi}\kt.
 \eea
 
 By the symmetry $E\mapsto -E$ and $V\mapsto -V$ we also have $N_t$ states on the interval $E=2$ and $0<V<4$.
 The symmetry $p\mapsto 1-p$ and $E\mapsto E-V$ shows that we have $N_h$ states on the linear segments
 $-2<E<2$, $V=\pm 2+E$.
 
 In the case of Dirichlet boundary conditions, the Hamiltonian on the reduced lattice 
 consisting of sites with a tooth is again
 given by \rf{z6} where now $k_0$ is the number of holes between $-N$ and the location 
 of the first tooth and $k_{N_t}$ is the number of holes
 between the last tooth and $N$.
 
 \subsubsection{The $V\to\infty$ limit of the spectrum}
 \label{sssVtoinf}
Here we show that when the disorder strength ${V\to\pm\infty}$, the 
energy spectrum has a limit, which is the same when $V\to+\infty$ and when $V\to-\infty$. 
In this limit the spectrum is in general degenerate.

For simplicity let us consider how the spectrum in the band $-2<E<2$ behaves in the limit $V\to +\infty$.
Heuristically, for such eigenstates, the teeth become walls with infinite height, where the wave function 
$\phi$ vanishes. Thus eigenfunctions are confined in strings of neighbouring holes where $\chi_n=0$, 
and are solutions of the free equation $-\phi(n+1) -\phi(n-1)=E\,\phi(n)$, with Dirichlet boundary conditions 
$\phi=0$ at the endpoints of the string where $V=+\infty$.
Let us consider eigenfunctions which are supported in one
such string $\mathcal{S}$ with $\ell_\mathcal{S}$ holes, and endpoints 
$n_0,n_1=n_0+\ell_\mathcal{S}+1$.
The energy eigenvalues are $E_{\ell_\mathcal{S},k}=-2\cos (\pi k/(\ell_\mathcal{S}+1))$, 
$k=1,\cdots \ell_\mathcal{S}$, with eigenfunction 
$\phi_{\ell_\mathcal{S},k}(n)=\sin(\pi (n-n_0) k/(\ell_\mathcal{S}+1))$ 
for $n_0< n< n_1$, and zero elsewhere.

The $V=+\infty$ spectrum for a configuration $\mathcal{C}$ is obviously the union of the 
$E_{\ell,k}$'s for all the 
lengths $\ell$ of strings of holes in the configuration $\mathcal{C}$. The spectrum 
is in general highly degenerate, 
since an energy level $E=-2\cos(\pi r/s)$, $r<s$, can be obtained from every string of holes 
of length $\ell=s m-1$, $m\in {\mathbb{Z}}^+$ in 
the configuration.

Using the standard methods of degenerate perturbation theory, it is easy to show that for 
large $|V|$, the eigenvalues and the eigenstates in the $-2\le E\le 2$ band can be constructed 
out of those at 
$V=\pm\infty$ by a perturbation expansion in $1/V$. Therefore, the spectral flow is 
continuous at $V=\pm\infty$, i.e., at $V^{-1}=0$, 
with energy levels merging. The details are left to the reader, but we illustrate this 
result in Fig.~\ref{spectruminfinity}.
\begin{figure}
\begin{center}
\includegraphics[width=5in]{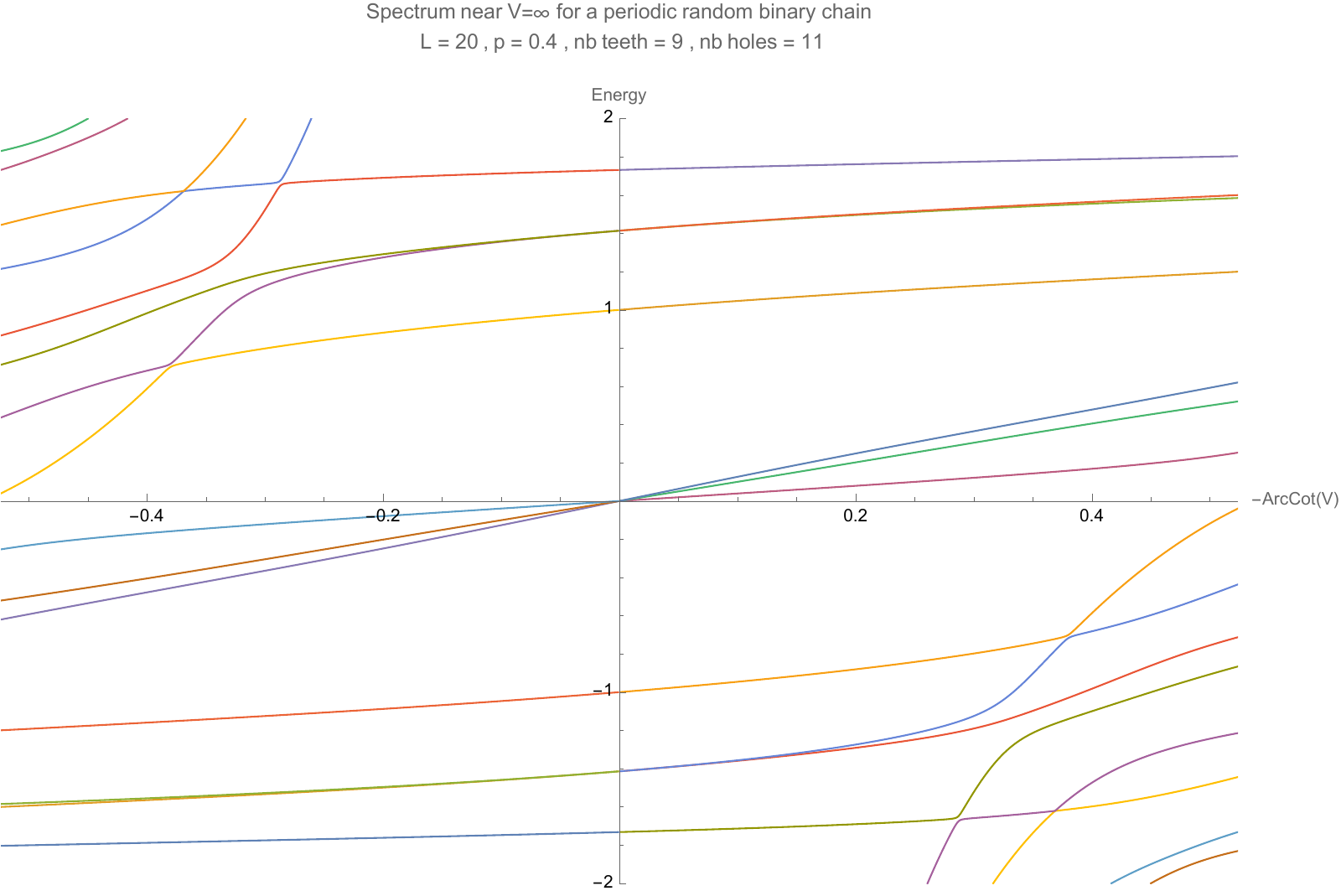}
\caption{The spectral flow  $E(V)$ of Fig.~\ref{fEflow1}  at $V=\infty$, $E$ is plotted 
here as a function of $U=-\mathrm{arccot}(V)$}
\label{spectruminfinity}
\end{center}
\end{figure}
The same results apply for the spectrum in the limit $V-E$ finite, $V\to\pm\infty$.

\subsubsection{Analyticity and periodicity of the spectral flow}

Using  the parity symmetry \rf{ParitySymmetryFlow} of the spectral flow, 
its $\mathtt{holes}\,\leftrightarrow\,\mathtt{teeth}$ symmetry \rf{ToothHoleSymmetry} and 
its continuity and analyticity as $V\to\pm\infty$ as well as $V,E\to\pm\infty$, 
with $V-E$ finite, one can represent the spectral flow by "Penrose-like" coordinates
\begin{equation}
\label{vevar}
\mathbf{v}=2\,\arctan({(V-E)/2})
\quad,\qquad
\mathbf{e}=2\,\arctan({E/2})
\end{equation}
which map the $(V,E)$ plane into the square $[-\pi,\pi]^2$.
The eigenvalues of $H$ corresponds to the zeroes of the determinant of $H-E$, 
\begin{equation}
\label{ detH-E=0}
\det(H-E)=\mathcal{R}(\mathbf{x},\mathbf{y})=0
\end{equation}
which is a rational function $\mathcal{R}$ (with integer real coefficients) in the variables
\begin{equation}
\label{xyvar}
\mathbf{x}=\mathrm{e}^{\mathi\mathbf{v}}\ \text{and}\quad \mathbf{y}=\mathrm{e}^{\mathi\mathbf{e}}
\end{equation}
since $E=-\mathi(\mathbf{y}-1)/(\mathbf{y}+1)$ and $E-V=-\mathi(\mathbf{x}+1)/(\mathbf{x}-1)$.
\begin{figure}[h]
\begin{center}
\includegraphics[width=4.in]{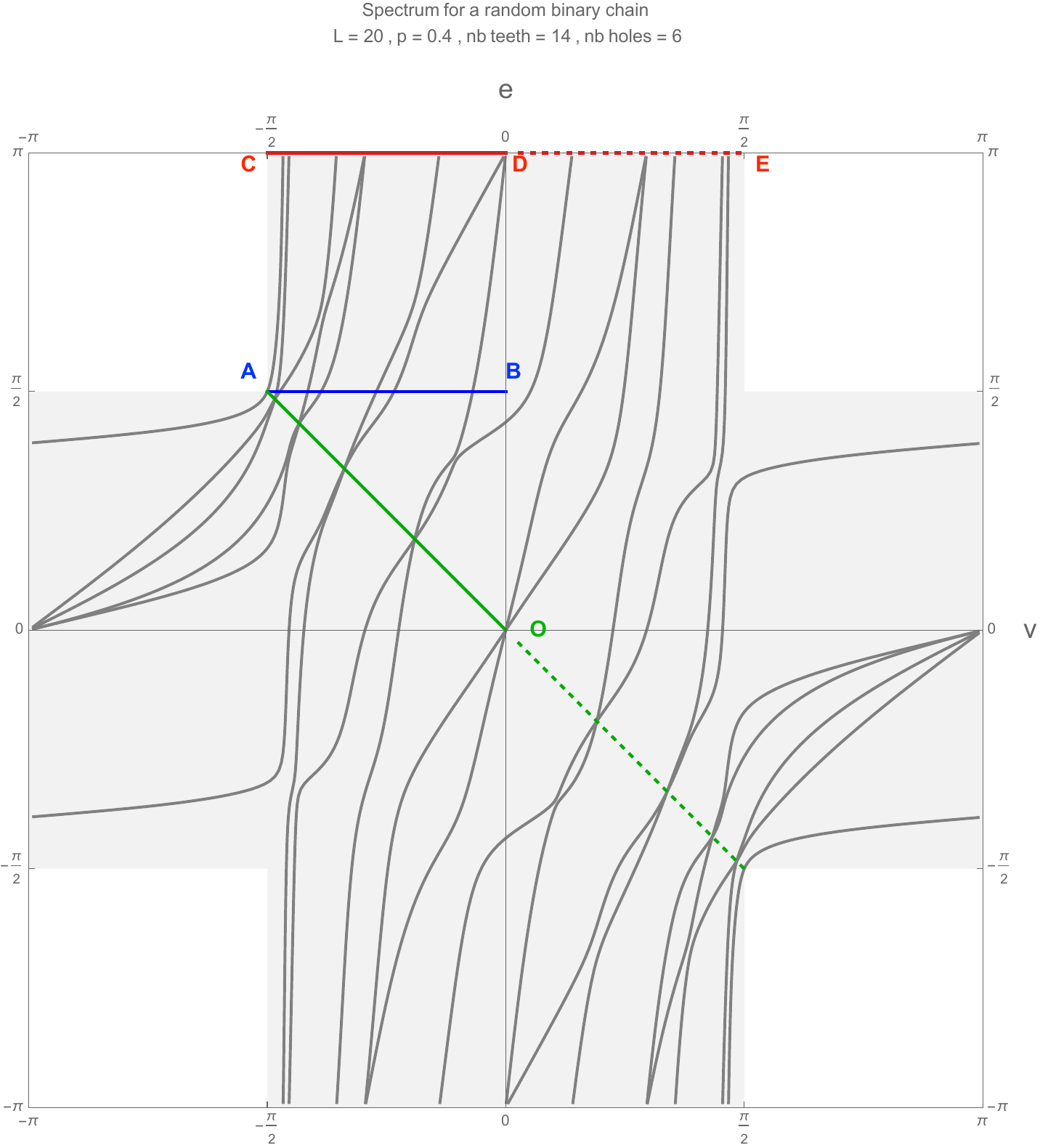}
\caption{Examples of a spectral flow in the toric $\mathbf{v},\mathbf{e}$ coordinates.}
\label{Fveflows}
\end{center}
\end{figure}
Therefore the spectrum is the restriction to the real torus $|\mathbf{x}|=|\mathbf{y}|=1$ of the complex algebraic curve $\mathcal{R}(\mathbf{x},\mathbf{y})=0$, and is a periodic real analytic curve in the variables $\mathbf{v},\mathbf{e}$.
We will use this picture of the spectral flow to derive an explicit formula for $N_{E>4}$ in Appendix A.3.


\section{The infinite comb}
\label{sInfiniteC}
The infinite random comb is obtained by letting the length $L$ of a finite random comb tend to infinity.
The behaviour of the eigenfunctions of the Hamiltonian in this limit is the same in the teeth as for 
the finite random combs.   
However, the behaviour of these functions on the spine is given by the 
eigenfunctions of the one-dimensional Anderson model with a Bernoulli potential as explained 
in Section 3.5.  We begin by reviewing briefly the main results about the one-dimensional Anderson 
model on the discrete line.  The most important feature is that the eigenfunctions decay exponentially 
along the spine with probability 1.  This can, of course, be seen in simulations of finite random combs
if they are large enough so that $L\gg \gamma^{-1}$ where $\gamma$ is the Lyapunov 
exponent defined below.

\subsection{The Anderson model on the line}
\label{ssAnderson}

The results we describe here are discussed in the books \cite{LuckBook1992} \cite{Comtet_2013} and the details needed
for the Bernoulli potential are established in \cite{CarmonaKleinMartinelli}.
The Anderson model on $\bbZ$ is described by \rf{bcgeneric}
\begin{equation}
\label{anderson }
-\phi(n-1) -\phi(n+1) + V(n)\,\phi(n) = E\,\phi(n),
\end{equation} 
where the local potentials $V(n)$ are given by i.i.d.\ random variables 
with a probability density distribution $\mu(V)$.  
Introducing the matrix
\beq{transfermatrix}
T_{n,E} = \left(\begin{array}{cc} V(n)-E & -1\\ 1 & 0\end{array}\right)
\eeq
the eigenvalue equation can be written
\beq{anderson2}
\left(\begin{array}{c}\phi(n+1)\\ \phi(n)\end{array}\right)= T_{n,E}\left(\begin{array}{c}
\phi(n)\\\phi(n-1)\end{array}\right)
\eeq
so the solution can be expressed in terms of initial conditions by
\beq{solution}
\left(\begin{array}{c}\phi(n+1)\\\phi(n)\end{array}\right)=\prod_{j=1}^nT_{j,E}\left(\begin{array}{c}
\phi(1)\\\phi(0)\end{array}\right)
\eeq
and similarly for $n<0$.  The limit
\beq{lyapunov}
\lim_{n\to\pm\infty}{1\over |n|}\log\| \prod_{j=1}^nT_{j,E}\|
\eeq
exists with probability 1 and defines a non-random number $\gamma (E)$ 
called the Lyapunov exponent.
If $\mu$ is not concentrated on a single point, $\gamma (E)>0$.
Furthermore, there exists a random unit vector ${\bx}_E\in\bbR^2$ such that
\beq{limit}
\lim_{n\to\pm\infty}{1\over |n|}\log\| \prod_{j=1}^nT_{j,E} {\bx} _E \|
\eeq
equals $-\gamma (E)$ if $\bx$ is proportional to ${\bx}_ E$ but $\gamma (E)$ if not.  
It follows that the square summable 
eigenfunctions satisfy
\beq{decay}
|\phi(n)|\leq C_E\,e^{-\gamma (E)|n|},
\eeq
where $C_E$ is independent of $n$.
The spectrum is a pure point spectrum which, if the potential takes only the values $0$ and $V$, 
is dense in $[-2,2]\cup [-2+V,2+V]$.

The normalized integrated density of states (IDOS) $\eta(E)$, defined as the limit
\begin{equation}
\label{IdosAver}
\eta(E)=\lim_{L\to\infty}\, {1\over L}\,\mathbb{E} \left[ \text{$\#$ of states of the chain 
with length $L$ with energy $E'\le E$} \right],
\end{equation} 
is known to exist and be Hölder continuous.  
The IDOS is related to the Lyapunov exponent through the
Thouless formula:
\beq{Thouless}
\gamma (E)=\int\log |E-E'|\,d\eta(E').
\eeq

Another approach to the Anderson model is to look at the ratios
\begin{equation}
\label{ratio }
\psi(n)=\phi(n)/\phi(n-1)
\end{equation}
which obey the stochastic recursion relation 
\begin{equation}
\label{Ricatti}
\psi(n+1)=-1/\psi(n)+V(n)-E.
\end{equation}
It is known from Furstenberg's theorem \cite{Furstenberg1963} that a.s.\ 
in the $n\to\infty$ limit, the probability distribution for $\psi$ converges to a unique 
invariant stable distribution $W_E(\psi)$ which obeys the integral equation
\begin{equation}
\label{WIntEqu}
W_E(\psi) = \int dV\, \mu (V) {1\over (V-E-\psi)^2}\,W_E\left({1\over V-E-\psi}\right).
\end{equation}
This distribution is real and positive, but singular 
(neither absolutely continuous nor discrete) and fractal.

From this $W_E$ one can extract the Lyapunov exponent $\gamma (E)$  which is given by
\begin{equation}
\label{LyapAver}
\gamma (E)=\int d\psi\, W_E(\psi)\, \log |\psi|.
\end{equation}
The IDOS is related to $W_E$ by
\begin{equation}
\label{ }
\eta(E)=\int_{-\infty}^{\ 0} \hskip -.5 ex d\psi\ W_E(\psi).
\end{equation}
For numerical calculations it is useful to consider the density $W_E(\psi)$ as we
will see below.

\subsection{Localization on the infinite random comb}

The eigenfunctions of the Hamiltonian for the random comb are all exponentially
decaying along the spine due to the Anderson localization, i.e.,
if $\phi$ is an eigenfunction of the random Hamiltonian \rf{theH} with eigenvalue $E$,
then there are constants $C_E$ and $\gamma >0$ such that
\beq{decay}
|\phi(n,j)|\leq C_E e^{-\gamma |n|}
\eeq
with probability 1.   We expect the spectral flow on a finite comb depicted 
in Fig.~\ref{fEflow1} to converge to a limiting flow which is given by continuous
spectral curves dense in the ribbons 
$[-2,2]\cup [-2+V,2+V]$ in the $V,E$ plane. 
The density of states $\rho_V (E)$ (for a fixed $V$),
which is given by the derivative of the IDOS $\eta$,
satisfies 
\beq{density1}
\int_{V-2}^{V+2} \rho_V(E)\,dE = p~~{\rm and}~~ 
\int_{-2}^2\rho_V(E)\,dE =1-p
\eeq
for $V>4$ or $V<-4$ by the results of subsection 3.6.5.


\section{Numerical studies and results}
\label{sNumRes}

\subsection{Methods}
\label{sNumRes1D}
In this section we study by numerical methods the model of a random comb. 
We shall illustrate the statements and the analytical results discussed in the 
previous sections. 
We shall mainly discuss the localization length $\xi$ for the wave functions of 
the random comb in the energy regimes $E<4$ and $E>4$, but we 
also study the associated integrated density of states $\eta$.
 
We adapt to our case the standard method used to study the one-dimensional
Anderson model through the iteration of the stochastic equation \rf{Ricatti} for the 
ratio variables $\psi(n)$ defined by \rf{ratio } (see the Sec.\ 4.1). It leads to the 
Lyapunov exponent $\gamma$ for the stochastic discrete dynamical system, which is the 
inverse of the localization length $\xi$ for the original system. We refer 
to \cite{LuckBook1992} and \cite{Comtet_2013} (among many references) for an introduction 
to these methods, as well as to the seminal works of 
Dyson \cite{PhysRev.92.1331} and Schmidt \cite{PhysRev.105.425} on the random harmonic chain, 
and of Anderson \cite{PhysRev.109.1492} for the propagation on random chains and more general lattices.

For the binary chain, the random variable $V$ takes only two values. For this case, as well as 
for more complicated chain models, numerical estimates of the Lyapunov exponent $\gamma$ and of 
the integrated density of states $\eta$ are obtained by first constructing large samples of the 
stable distribution $W_E(\psi)$, by iterating $\mathcal{N}$ times the recursion relation 
\ref{Ricatti}, starting from some initial value $\psi_0$ for the random variable $\psi$ and 
thus getting a sample of size $\mathcal{N}$ of random $\ps$'s . From these large samples 
$\mathcal{S}$ of $\psi$'s, one obtains estimates for $\gamma$ and $\eta$ by taking the
average over the sample of the quantities \rf{LyapAver} and \rf{IdosAver}
\begin{equation}
\label{Lambda+Eta2}
\overline\gamma=\overline {\log |\psi |}={1\over \mathcal{N}} \sum_{i=1}^\mathcal{N} \log |\psi_i | \quad,\qquad 
\overline\eta=\overline{\theta(-\psi)}={1\over \mathcal{N}} \sum_{i=1}^\mathcal{N}\theta (-\psi_i )
\end{equation}
with $\theta$ the Heaviside function.  The quantities
$\overline\gamma$ and $\overline\eta$ are still random variables for finite $\mathcal{N}$, 
but when $\mathcal{N}$ goes to $\infty$, they converge almost surely to $\gamma$ and $\eta$:
\begin{equation}
\label{Lambda+Eta}
\lim_{\mathcal{N}\to\infty} \overline\gamma \ \mathop{=}\limits_{\mathrm{a.s.}}\ \gamma
\quad,\qquad
\lim_{\mathcal{N}\to\infty} \overline\eta \ \mathop{=}\limits_{\mathrm{a.s.}}\ \eta.
\end{equation}
In the following sections, we apply these methods to our model. In practice it is 
sufficient to use sample sizes $\mathcal{N}$ of order $10^5$ to $10^6$.

\subsection{$E>4$ eigenstates}
\subsubsection{Principle of the calculations}
As discussed in Sec.\ 3, the random comb has energy eigenstates with $E>4$ which 
are localized near the spine and decay exponentially along the teeth.
For the regular comb with no disorder ($p=0$) these states are plane waves along the spine, 
and exist in the energy band
$$4\ <\ E\ \le\, E_{\mathrm{max}}=16/3$$
For the random comb, the $E>4$ eigenstates take the form
\begin{equation}
\label{ }\begin{split}
\psi(n,j)\ =&\ C_n\,(-1)^j\,e^{-\sigma j}\qquad\text{if $n$ is a tooth}\\
\psi(n)\ =&\ C_n \qquad\text{if $n$ is a hole}
\end{split}
\end{equation}
with the energy still given by
\begin{equation}
\label{Esigma}
E=2+2\,\cosh\sigma .
\end{equation}
For a finite chain of length $N$ the energy $E$ is quantized, and there is a finite number
$N_{E>4}$ of such $4<E\le E_{\mathrm{max}}$ states. This number is of order $\mathrm{O}(N)$
and is calculated in Appendix A3.
These states are still localized near the spine, and do not propagate along the teeth. 

For an infinite comb with disorder ($p\in]0,1[$), these eigenstates are now localized along the spine. 
This means that the amplitude along the spine $C_n=\psi(n,0)$ decays exponentially with the 
position $n$ along the spine when $n\to\pm\infty$ as
$$
|C_n|\ \sim\ e^{-|n| / \xi}.
$$
The spine localization length $\xi$ depends on the energy $E$ of the state and 
the disorder, i.e., the hole probability $p$.

The standard methods used to study the disordered linear chain can easily be adapted 
to studying the $E>4$ states of the random comb. 
The eigenvalue equation for the wave function on the spine $\psi(n,0)=C_n$ 
was discussed in Sec.\ 3.4 and is given by Eq.~\rf{E>4equ}.
One defines the Riccati variable $\psi$ as
$$
\psi_n=C_n/C_{n-1}.
$$ 
It obeys the recursion relation
\begin{equation}
\label{RecRelE>4}
\psi_{n+1}=-1/\psi_n+\begin{cases}
     \ e^{\sigma}-1& \text{if $n$ is a tooth (prob. $=1-p$)}, \\
     \ e^{\sigma}+e^{-\sigma}& \text{if $n$ is a hole (prob. $=p$)}
\end{cases}
\end{equation}
with $\sigma$ related to the energy $E$ by \rf{Esigma}.
For a given $E>4$ (a choice of $\sigma$) and a given hole probability $p\in ]0,1[$, 
starting from some initial $\psi_0$ one can iterate \rf{RecRelE>4} and get large 
samples of $\psi$'s.
The relations \rf{Lambda+Eta2} are still valid for the Lyapunov exponent 
$\gamma=\gamma(E,p)$, which is
related to the localization length $\xi$ by
\begin{equation}
\label{xi2gamma2}
\xi=1/ \gamma
\end{equation} and for the integrated density of $E>4$ states  $\eta(E,p)$
defined in \rf{IdosAver}.

\subsubsection{Lyapunov exponent for the $E>4$ states}
\label{sssGammaE>4}
First we show in Fig.~\ref{lambdaE>4}  estimates for the Lyapunov exponent $\gamma$ for 
these states, as obtained from \rf{Lambda+Eta} and \rf{RecRelE>4}. 
Each curve in the figure corresponds to a fixed value of the hole probability $p$,
and $\gamma$ is plotted as a function of the energy $E$. 
The various curves correspond to different values of the hole probability $p$, 
starting from $p=0.1$ and going to $p=.975$ close to $p=1$, which 
corresponds to the linear chain with no teeth.
The curves are obtained using $\mathcal{N}=10^5$ iterations.
\begin{center}
\includegraphics[width=3.8in]{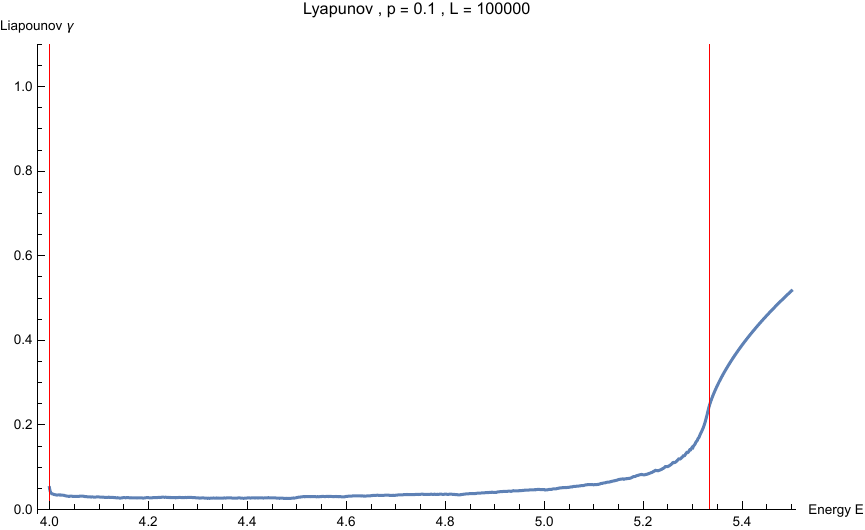}
\\
\includegraphics[width=3.8in]{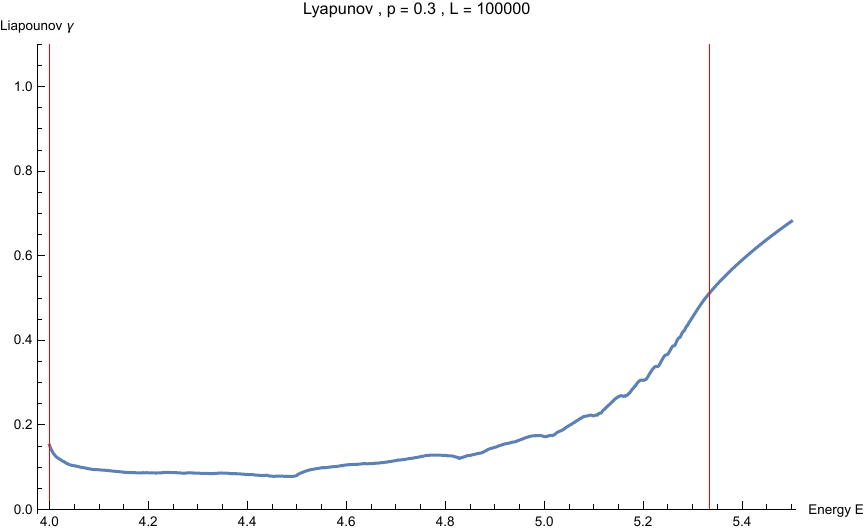}
\\
\includegraphics[width=3.8in]{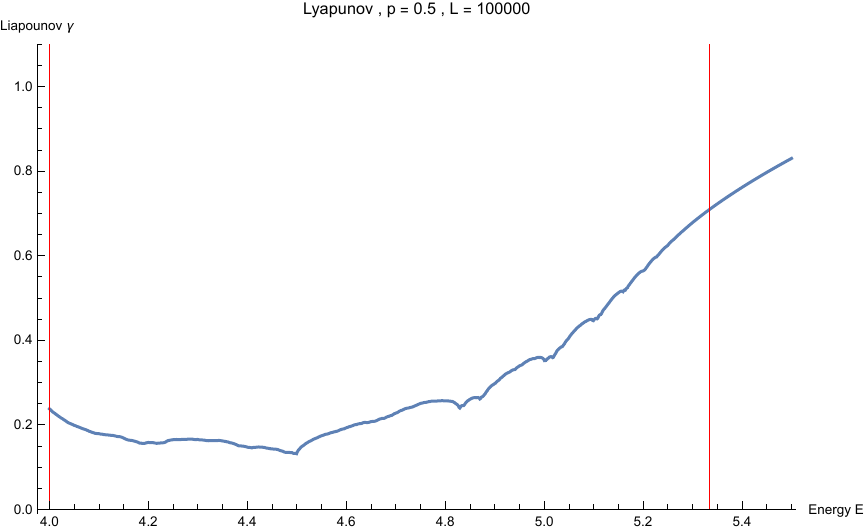}
\\
\includegraphics[width=3.8in]{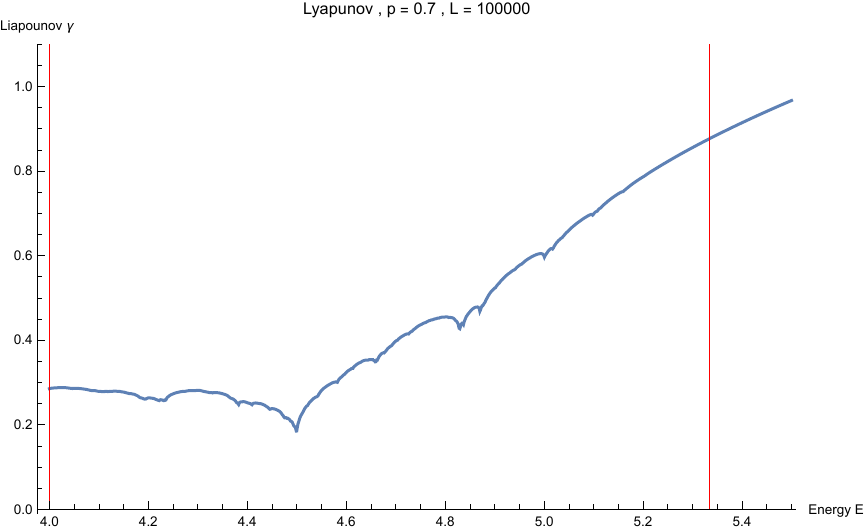}
\\
\includegraphics[width=3.8in]{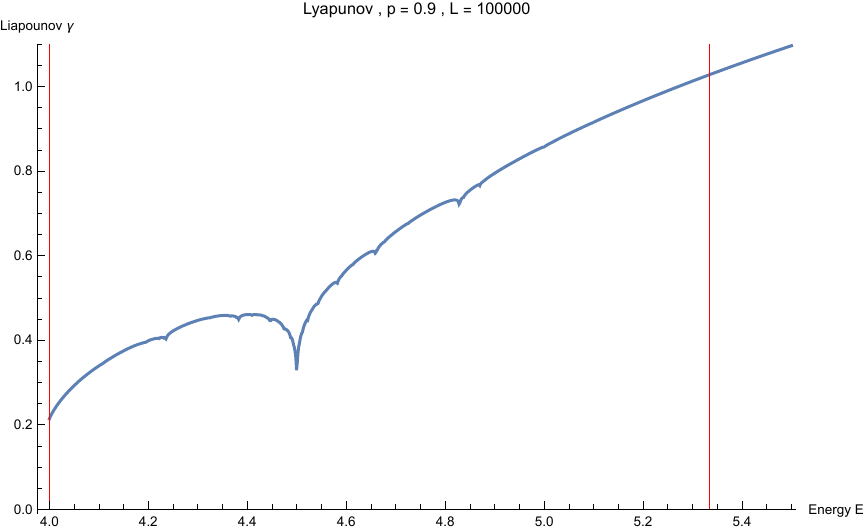}
\\
\includegraphics[width=3.8in]{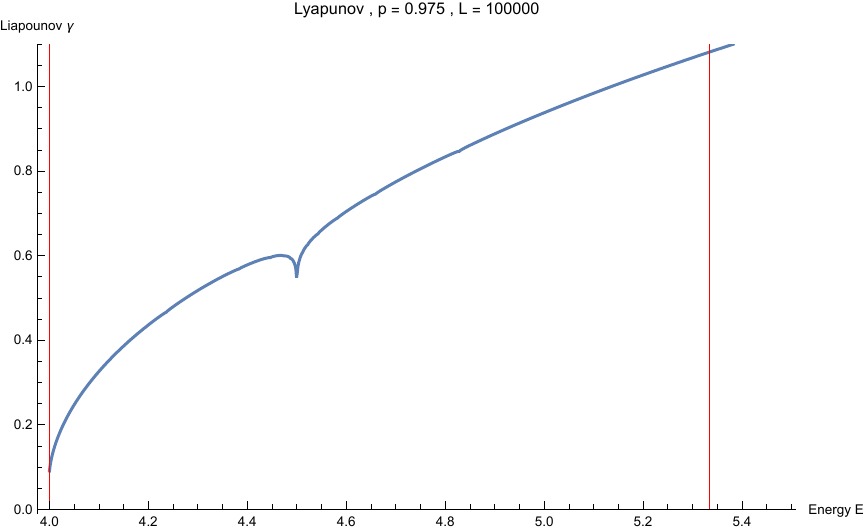}
\end{center}
\begin{figure}[h!]
\begin{center}
\caption{Lyapunov exponent $\gamma(E)$ as a function of the energy $E$ for $E>4$ states localized 
on the spine, for various values of the hole probability $p$. The curves are obtained 
using samples of size $\mathcal{N}=10^5$ ($\mathcal{N}$ is denoted $L$ in the figures).}
\label{lambdaE>4}
\end{center}
\end{figure}

When $p=0$ the Lyapunov exponent vanishes in the spectrum support
$$\gamma_{p=0}(E)= 0\quad,\quad\text{if}\ 4\le E\le 16/3$$
and the eigenstates are extended. This corresponds to the regular comb and the 
eigenfunctions are plane waves. This is in agreement with the results of 
\cite{David_2022}. When $E>16/3$, there are no  normalizable eigenstates, 
and the Riccati recurrence gives a Lyapunov exponent greater than zero.
As soon as $p>0$ the Lyapunov exponent is non-zero in the interval $4\le E\le 16/3$.

The last curve corresponds to $p=0.975$, close to 1. 
When $p=1$, the density of teeth is zero, and the comb reduces to 
the infinite linear chain. Then there are no $E>4$ eigenstates.  
This is consistent with the fact that for $p=1$ the Ricatti recurrence \ref{RecRelE>4} 
is deterministic, and has a fixed point $\psi=e^\sigma$, which gives the non-zero Lyapunov exponent
$$
\gamma_{p=1}(E)=\sigma=\mathrm{arcosh}((E-2)/2)
$$
close to the numerical curve obtained for $p=0.975$.   

For hole probability $p\in ]0,1[$, the Lyapunov exponent 
$\gamma(E)$ is found to be strictly positive, and is a continuous 
(but non-differentiable) function of the energy $E$.
It exhibits algebraic cusp-like singularities which are 
standard in this kind of disordered one-dimensional systems. 
These singularities were first discussed by Schmidt 
in \cite{PhysRev.105.425} for the harmonic chain, by Lifschitz \cite{Lifschitz1963} and notably by Halperin  
\cite{PhysRev.139.A104} for the quantum walk on a chain. They have been 
extensively studied in the literature (see e.g. \cite{LuckBook1992}). The singularities form a dense 
denumerable subset of the energy range. They are associated with specific 
values of the energy $E$ where some ``resonances'' occur for specific finite 
length subchains configurations of the random chain.

\subsubsection{Integrated density of states for $E>4$}
On Fig.~\ref{etaE>4} we show the integrated density of states $\eta(E)$, 
again obtained from \rf{Lambda+Eta} and the recursion relation \rf{RecRelE>4} for the $C_n$'s. 
The results for $\eta(E)$ are plotted as a function of the energy $E$, and for different 
values of the hole probability $p$ going from $0.1$ to almost 
one (the linear chain).
Again the curves are obtained from sample size $\mathcal{N}=10^5$.

\begin{center}
\includegraphics[width=3.8in]{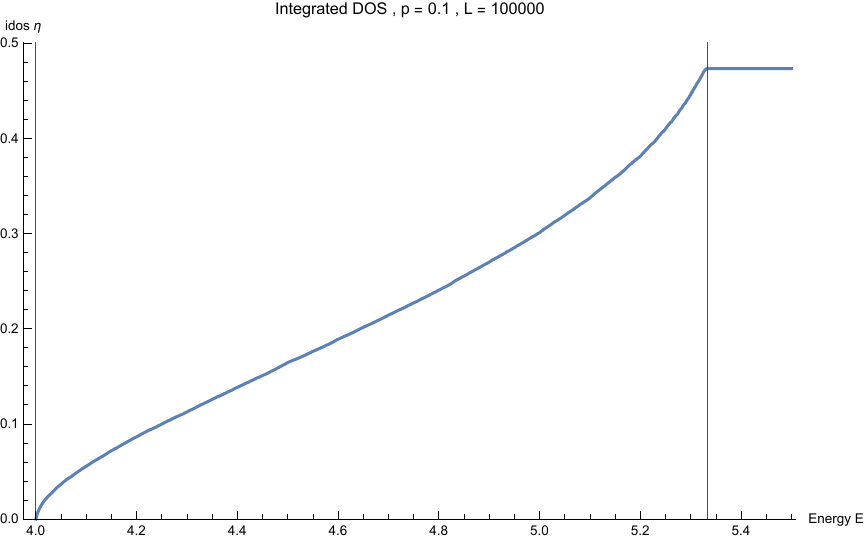}
\\
\includegraphics[width=3.8in]{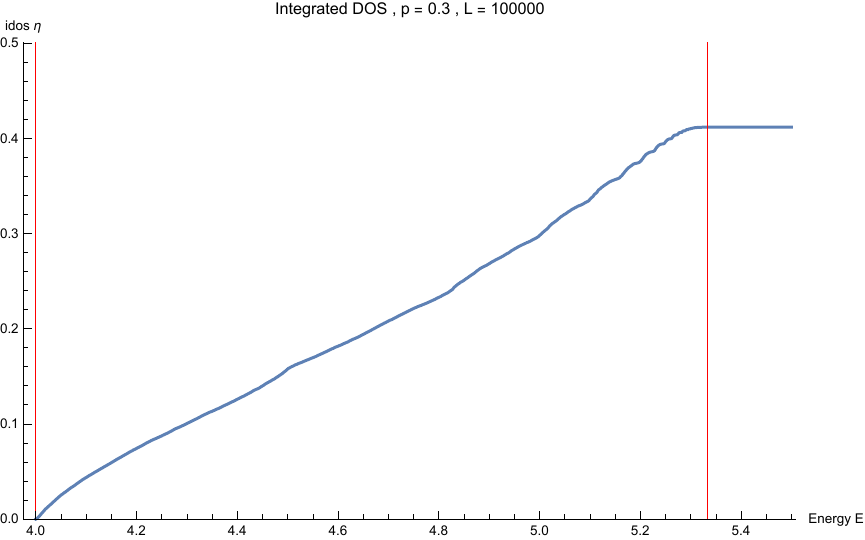}
\\
\includegraphics[width=3.8in]{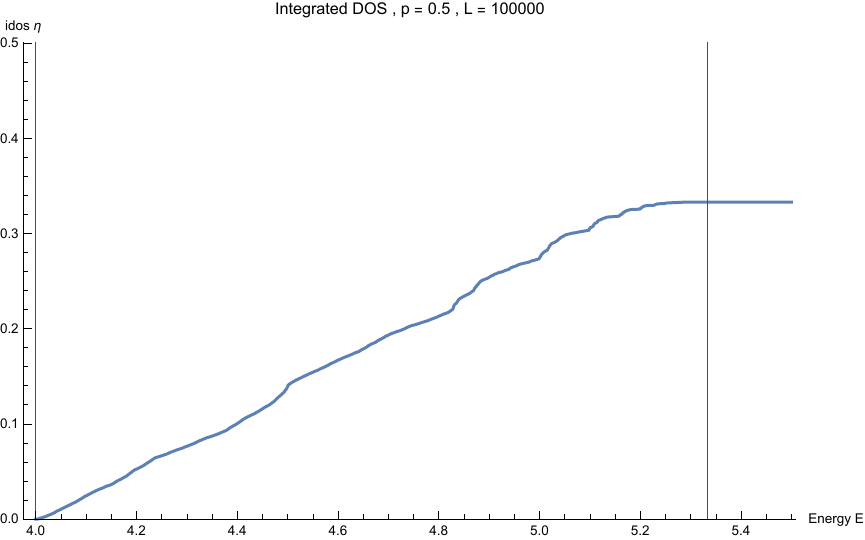}
\\
\includegraphics[width=3.8in]{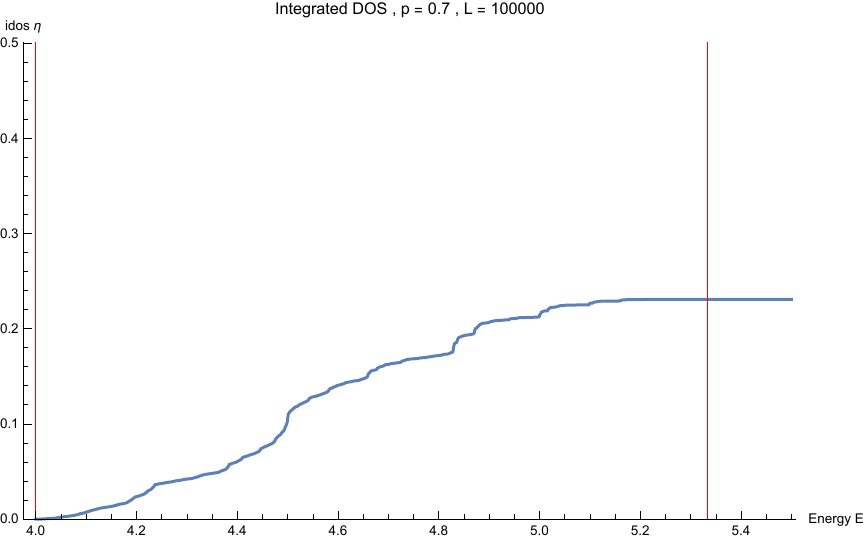}
\\
\includegraphics[width=3.8in]{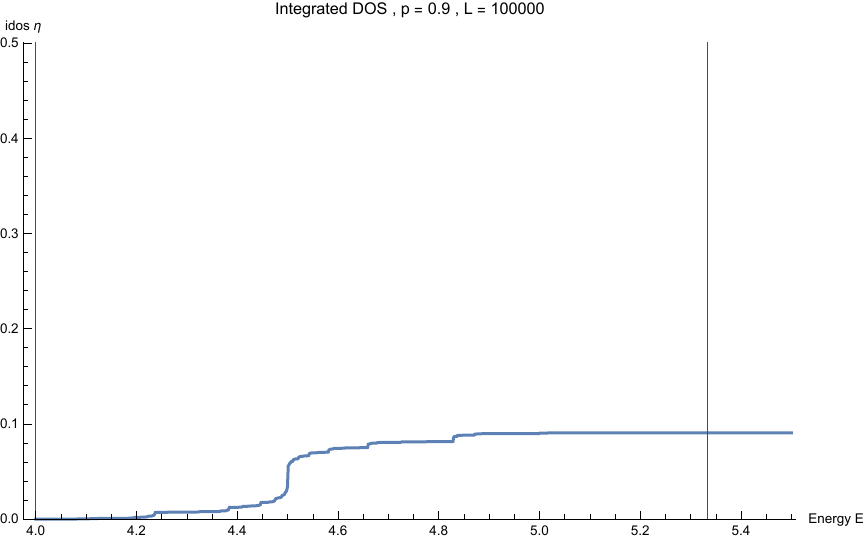}
\\
\includegraphics[width=3.8in]{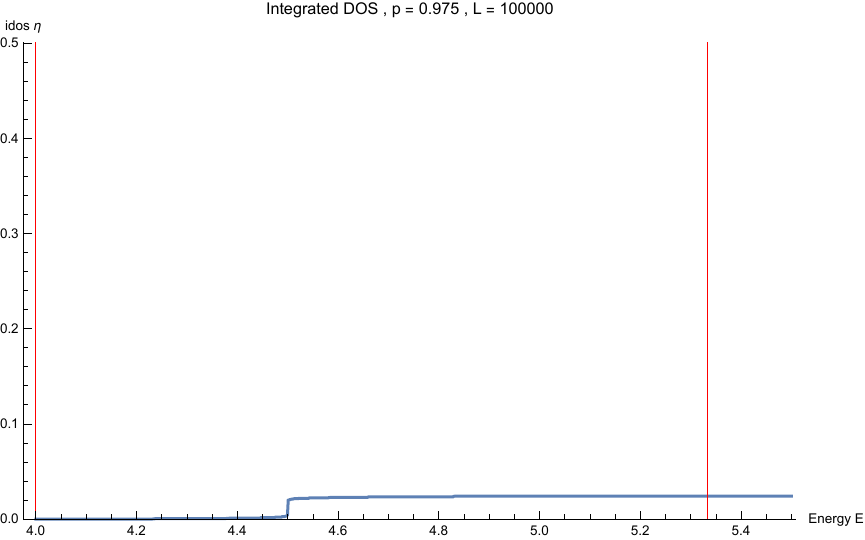}
\end{center}
\begin{figure}[h!]
\begin{center}
\caption{Integrated density of states $\eta(E)$ as a function of the energy $E$ 
for states localized on the spine, for various values of the hole probability $p$.}
\label{etaE>4}
\end{center}
\end{figure}
For $p=0$ which is the regular infinite comb with 
no disorder, we get a smooth curve which agrees with the analytical 
results of \cite{David_2022}.
For $p>0$, $\eta(E)$ is a continuous but non-differentiable function of the energy 
$E$, with singularities at the same dense set of points as the Lyapunov exponent 
$\gamma(E)$. Note also that at the maximal value $E_{\mathrm{max}}=16/3$, $\eta$ 
is the normalised number of $E>4$ states per 
unit of length for the spine, and agrees with the analytical result of Appendix \ref{AProbChain}
\begin{equation}
\label{NE>4bis}
\eta(E_{\mathrm{max}})\ =\ \overline N_{E>4}\ \mathop{=}\ {1-p\over 2-p}.
\end{equation}

\subsubsection{Minimal Lyapunov (maximal localization length)}
As stated above, for $0<p<1$, the Lyapunov exponent for energies $4\le E\le 16/3$ 
is always strictly positive.
On Fig.~\ref{minLiapE>4} we plot the minimal value for $\gamma(E)$ in this 
energy range, 
$$
\gamma_{\mathrm{min}}=\min_{4<E<16/3}\gamma(E)
$$
as a function of the hole probability $p$.
\begin{figure}[h!]
\begin{center}
\includegraphics[width=3.5in]{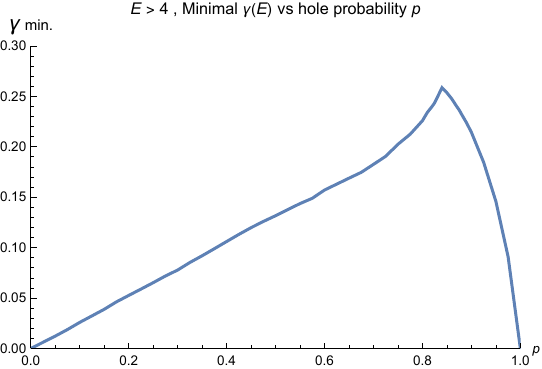}%
\caption{Minimal Lyapunov exponent for the
$E>4$ states, as a function of the hole probability $p$}
\label{minLiapE>4}
\end{center}
\end{figure}

The minimal Lyapunov exponent $\gamma_{\mathrm{min}}(p)$  is found to grow with $p$ in the range $0<p<p_{\scriptscriptstyle{\mathrm{max}}}\simeq.8$, and then to decrease sharply with $p$ when $p_{\scriptscriptstyle{\mathrm{max}}}<p<1$. The growth is approximately linear in the range $0<p<.7$.
If one looks back to the $\gamma(E)$ curves of fig.~\ref{lambdaE>4}, one numerically observes that for $p<p_{\scriptscriptstyle{\mathrm{max}}}$ the minimum of the curve $\gamma(E)$ occurs for a critical value 
of the energy $E_c\simeq 4.5$ which is strictly positive, and which corresponds to a cusp at $E=5/2$ 
in the curve $\gamma(E)$. 
When $p>p_{\scriptscriptstyle{\mathrm{max}}}$ the minimum of the curve 
$\gamma(E)$ occurs at $E=4$. This explains the abrupt change in the behaviour of 
$\gamma_{\mathrm{min}}(p)$ with $p$.

\subsection{The $E<4$ states}
\label{ssE<4Lyap}
\subsubsection{Lyapounov exponent for the S-matrix column states $|\Upsilon\kt$}
\label{sssE<4LyUp}
We now consider the localization effects for the $0<E<4$ states, which propagate 
along the teeth. When $p>0$, we expect localization 
along the spine, as discussed in Sec.~4.1.
First we consider, at a given energy $E=E(\theta)=2-2\cos \theta$, 
the states $|\Upsilon_{\theta,t}\kt$ 
defined in Sec.~\ref{sssScvec}, which are associated to the teeth $t$ of the 
comb, and form an orthonormal basis for the constant energy sub-Hilbert 
space $\mathcal{H}_\theta$ of the comb.
Each state $|\Upsilon_{\theta,t}\kt$ 
corresponds to an incoming particle with energy $E$ on the tooth $t$, scattered 
by the comb along the other teeth and the spine. The wave function for 
this state is of the form
\begin{equation}
\label{phit0form}
\begin{split}
\Upsilon_{\theta,t}(n,j)\ & 
= \ \delta_{n,t}\,e^{-\mathi\,\theta\,j} + A_n\,e^{-\mathi\,\theta\,j} \qquad \text{if $n$ is a tooth}
\\  
\Upsilon_{\theta,t}(n)\ & =\ A_n \qquad \text{if $n$ is a hole}
\end{split}
\end{equation}
with $A_n$ the probability amplitude for the outgoing particle on the tooth $n$,
see  Sec.~3.3.4.
We are interested in the effect of the disorder on the $A_n$'s, 
in particular on their large $n$ behaviour.

The equation for the $A_n$'s was obtained in Sec.~\ref{sssScvec} and for the 
$\tilde A_n= A_n-\delta_{n,t_0}$ it takes the explicit form
\rf{E<4columnA}. This can be rewritten as a recursion relation for the Riccati 
variables, which reads, as long as $n\neq t_0$,
\begin{equation}
\label{RecRelE<4a}
\tilde\psi_{n+1}=-{1\over \tilde \psi_{n}}+\begin{cases}
      1+e^{-\mathi\,\theta}& \text{if $n$ is a tooth}, \\
      e^{\mathi\theta}+e^{-\mathi\theta}& \text{if $n$ is a hole} \end{cases}
      \qquad\text{with}\qquad \tilde\psi_n=\tilde A_n/\tilde A_{n-1}.
\end{equation}
These are still $\mathrm{SL}(2,\mathbb{C})$ transformations, and the 
hypothesis of Furstenberg's \cite{Furstenberg1963} theorem is valid.
There is a unique invariant measure $W(\tilde\psi)$ for this random process, 
which is given by a solution of the fixed point integral equation analogous 
to \rf{WIntEqu}, but now with a probability distribution for a 
complex random potential $V$. The measure $W(\tilde\psi)$ is fractal, 
with support in $\mathbb{C}$.
One can define a Lyapunov exponent $\gamma$ for the amplitudes $A_n$
using the formula \rf{LyapAver}
\begin{equation}
\label{LyapAver2}
\gamma\ =\ \int_\mathbb{C} d^2\tilde\psi\ W(\tilde\psi)\, \log |\tilde\psi |
\ \mathop{=}\limits_{\mathrm{a.s.}}\ 
\lim_{L\to\infty} {1\over L} \sum_{i=1}^L \log |\tilde\psi_i | 
\end{equation}
so that one can bound the $A_n$ at large distance as
\begin{equation}
\label{ }
|A_n|\ \le \ \mathtt{cst.}\, e^{-\gamma\,|n-t  |}.
\end{equation}

We show in Fig.~\ref{lambdaE<4a} estimates for the Lyapunov exponent, as obtained by 
this random Riccati recursion (starting from some initial $\tilde\psi_0$, 
apply the random recursion \rf{RecRelE<4a}, and take the large $L$ average \rf{LyapAver2}).
The results for $\gamma$ are plotted as a function of the energy $E$, 
and for values of the hole probability $p$ going from $0.1$ 
to almost one (the linear chain).
The curves are obtained with $\mathcal{N}=10^5$ iterations.

\begin{center}
\includegraphics[width=3.8in]{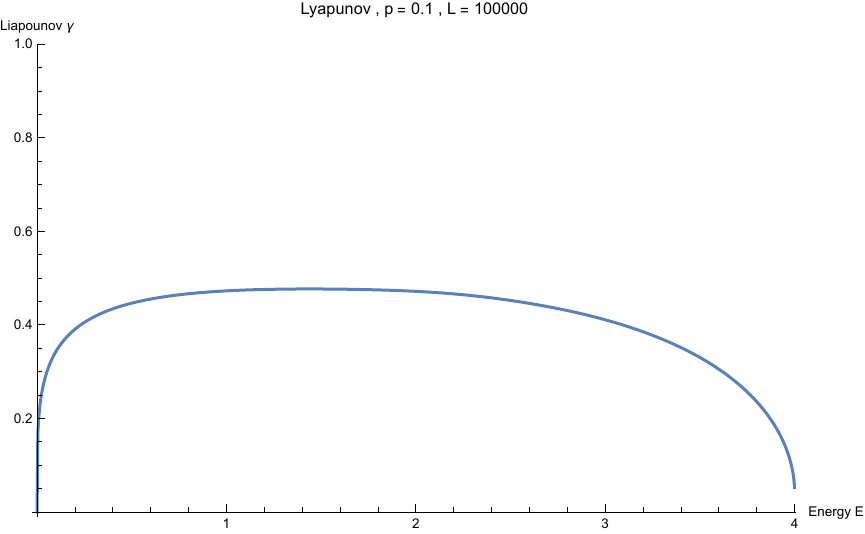}
\\
\includegraphics[width=3.8in]{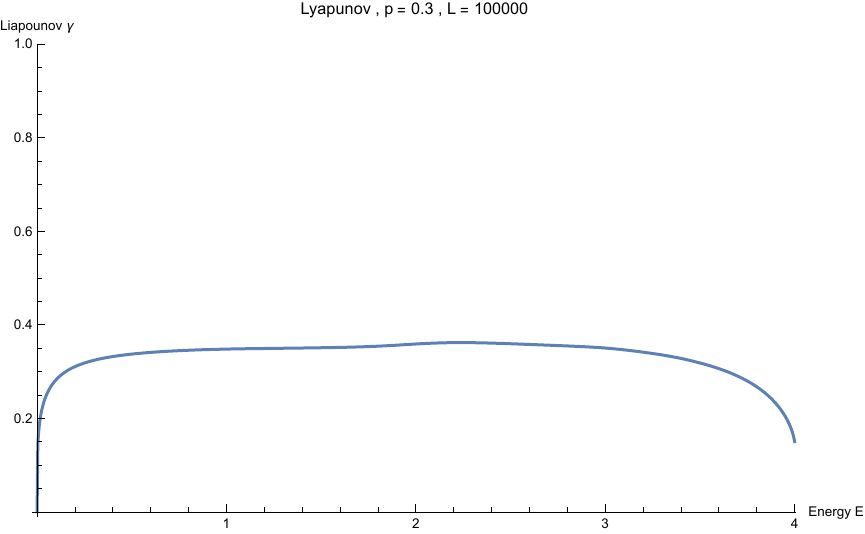}
\\
\includegraphics[width=3.8in]{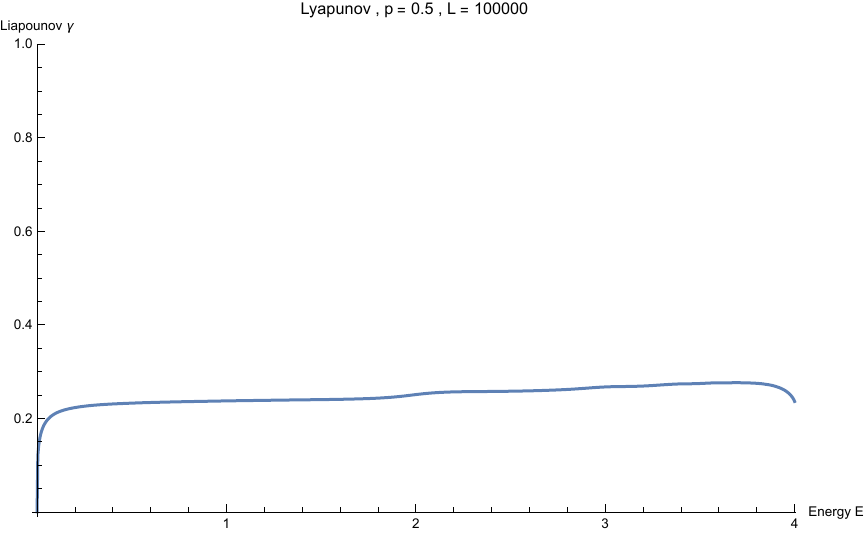}
\\
\includegraphics[width=3.8in]{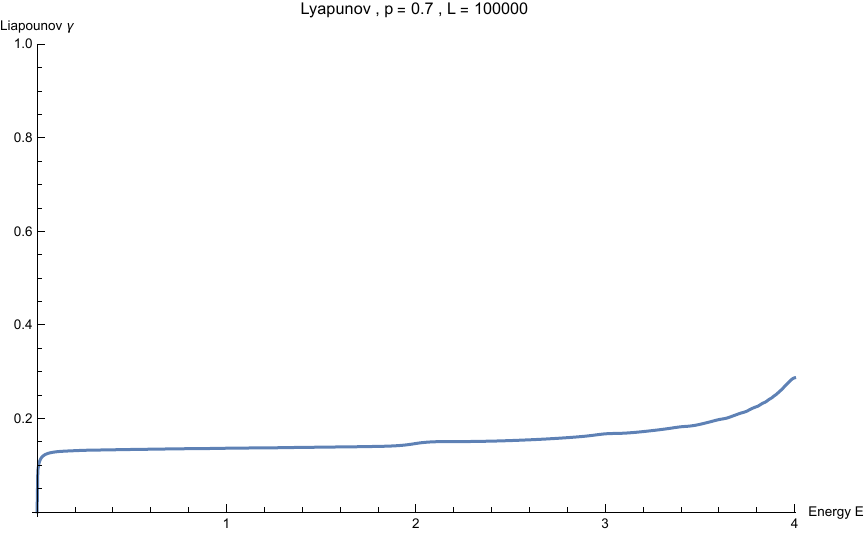}
\\
\includegraphics[width=3.8in]{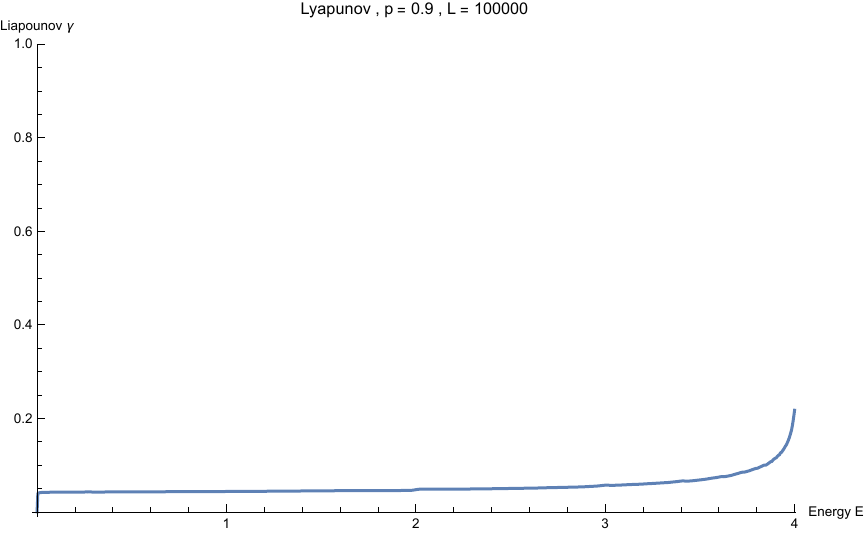}
\\
\end{center}
\begin{figure}[h!]
\begin{center}
\caption{The Lyapunov exponent $\gamma(E)$ as a function of the 
energy $0\le E\le 4$ for the S-matrix column vector states $|\Upsilon\kt$, for 
various values of the hole probability $p$. }
\label{lambdaE<4a}
\end{center}
\end{figure}
Already for $p=0$ the Lyapunov exponent is found to be non-zero when $E\neq 0,4$, 
and the amplitudes $A_n$ decay exponentially at infinity.
This result agrees with the analytical calculations of \cite{David_2022}, 
which implies that for $p=0$ one has
\begin{equation}
\label{LyapE<4p=0}
\gamma(E)= \log \left (\mathrm{Re}\left(w_-(e^{\mathi \theta})\right)\right) \quad,\qquad w_-(z)={1\over 2}\left(1+z-\sqrt{(z+3)(z-1)}\right)
\end{equation}
with the notation of Sec.~4.1 of \cite{David_2022}.

Fig.~\ref{lambdaE<4a} presents results for various values of $p$ in the  range $0<p<1$.
As the disorder parameter $p$ increases, the curve for the Lyapunov exponent $\gamma(E)$ deforms 
continuously with $p$. 
Contrary to the case for the $E>4$ $|\Phi\rangle$ states, for the $E<4$ $|\Upsilon\rangle$ states, 
we do not observe cusp-like singularities in the $\gamma(E)$ 
curves in the presence of disorder ($0<p<1$).

The last curve in fig.~\ref{lambdaE<4a} corresponds to $p$ close to the limit case $p=1$. 
It is consistent with the fact that for the linear chain ($p=1$) $\gamma(E)$ vanishes for all $0<E<4$.

At the maximal value for the energy of the $|\Upsilon\kt$ states, $E=4$, 
the Lyapunov exponent $\gamma$ coincides with the  Lyapunov exponent 
for the $E>4$  $|\Psi\kt$ states at their minimal energy $E=4$.
Indeed, the recursion relation \rf{RecRelE<4a} when $\theta=\pi$ is 
equivalent to the recursion relation \rf{RecRelE>4}  when $\sigma=0$ 
(they differ only by a change of sign of $\psi$).

\subsubsection{The Lyapunov exponent for $E<4$ S-matrix eigenstates}
For completeness, we present a few numerical estimates of the Lyapunov 
exponent for the eigenstates of the S-matrix $S(\theta)$, defined in 
Sec.~\ref{sssSmatrix}, which gives the $\texttt{in}\to\texttt{out}$ 
reflection amplitudes for states of fixed energy $E$ propagating 
along the teeth. For  a given energy
$E =2-2\,\cos\theta$
these eigenstates $| B_{\theta,\delta}\rangle$ are labelled by a phase shift 
$\delta$ and are solutions of the equation
$$
S(\theta)\,| B_{\theta,\delta}\rangle=\mathrm{e}^{\mathi\delta}| B_{\theta,\delta}\rangle .
$$
We refer to Appendix~\ref{ASmatrix} for details.
For the local coefficients $B_n$ of the $| B_{\theta,\delta}\rangle$ states, 
the above eigenvalue equation gives the recursion relation 
\rf{BthEq}, which is the same as the recursion relation for states of the binary 
chain for energy $E$ and disorder potential $V=V(E,\delta)$ given by \rf{VEd}.
The Lyapunov exponent can be evaluated as a function of $E$ and $\delta$ 
by the standard method reviewed in Sec.~\ref{sNumRes1D}.

On Fig.~\ref{lambdaSEdel}, we give numerical estimates of the 
Lyapunov exponent as a function of the phase shift $\delta$, 
for various values of the energy $E$, and for the specific value 
$p=1/4$ of the hole probability for the random comb.
\begin{center}
\includegraphics[width=2.5in]{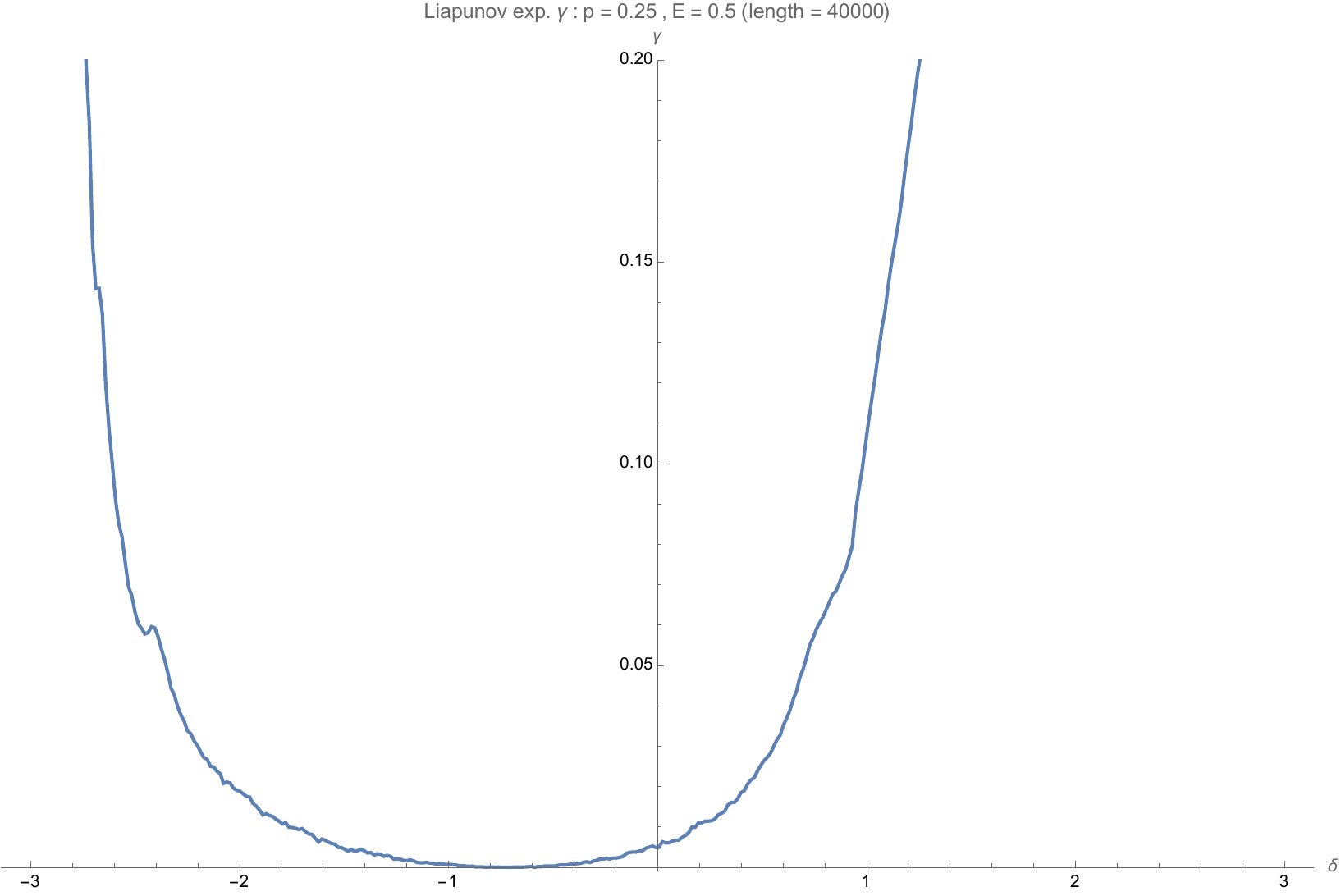}
\quad
\includegraphics[width=2.5in]{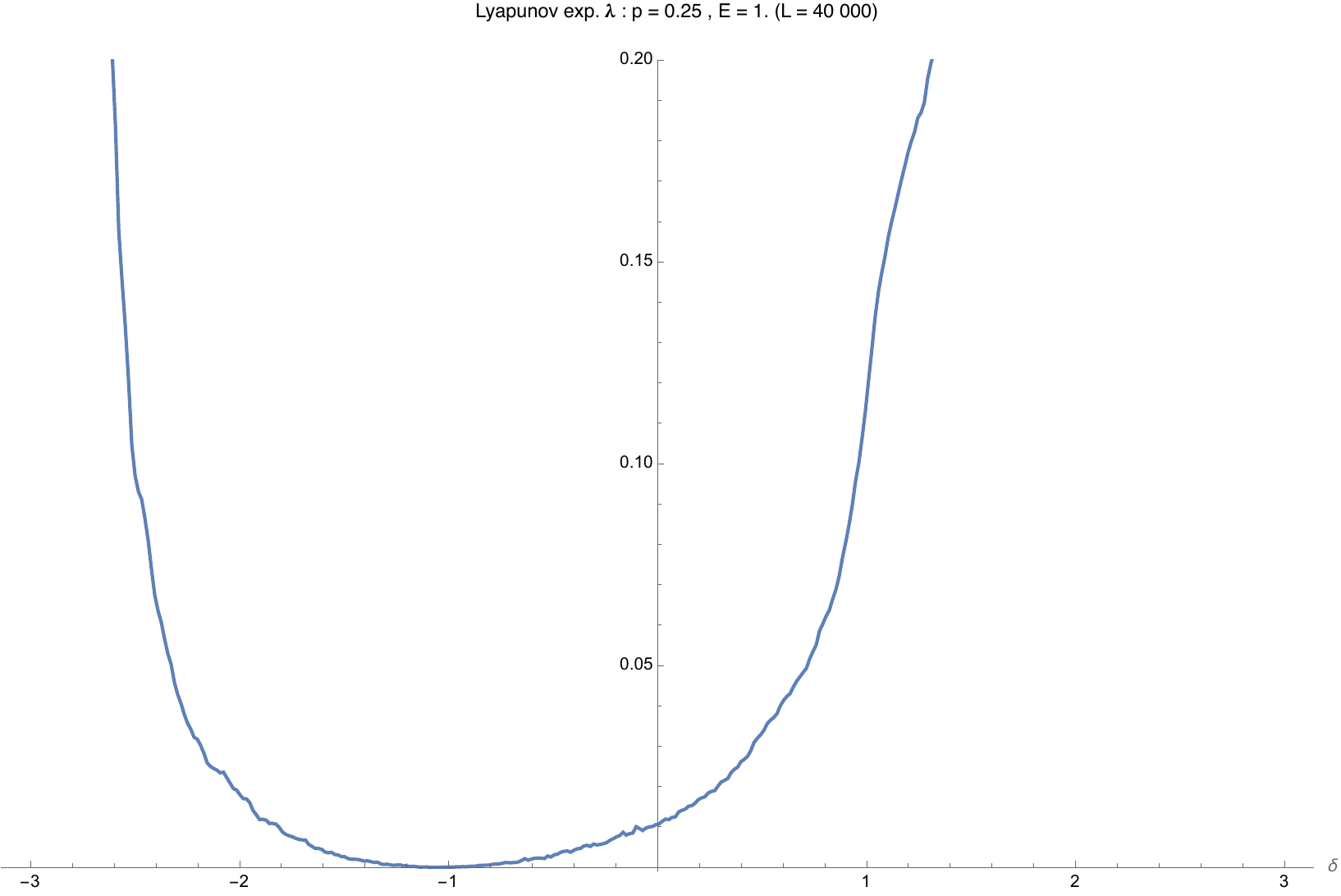}
\\
\includegraphics[width=2.5in]{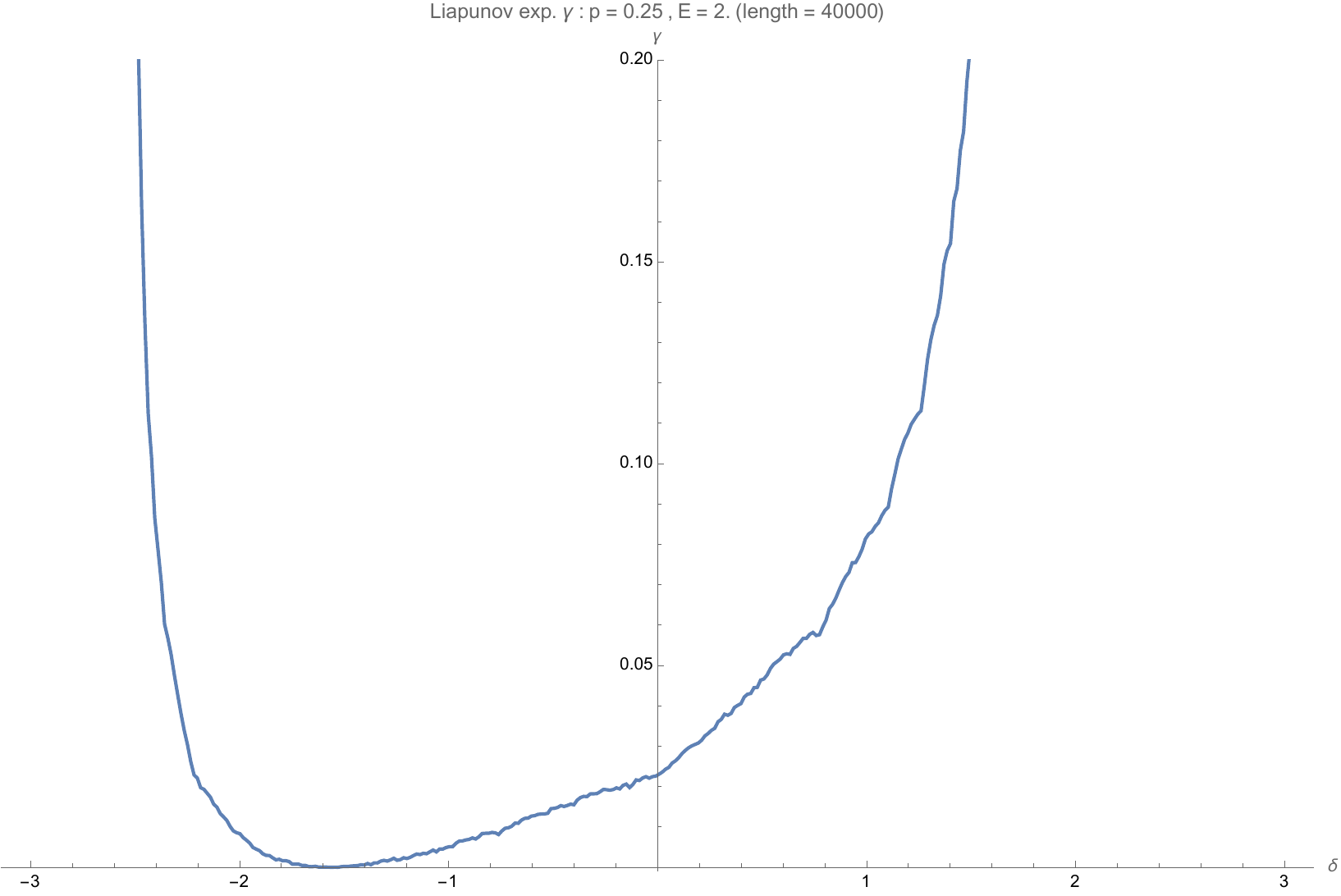}
\quad
\includegraphics[width=2.5in]{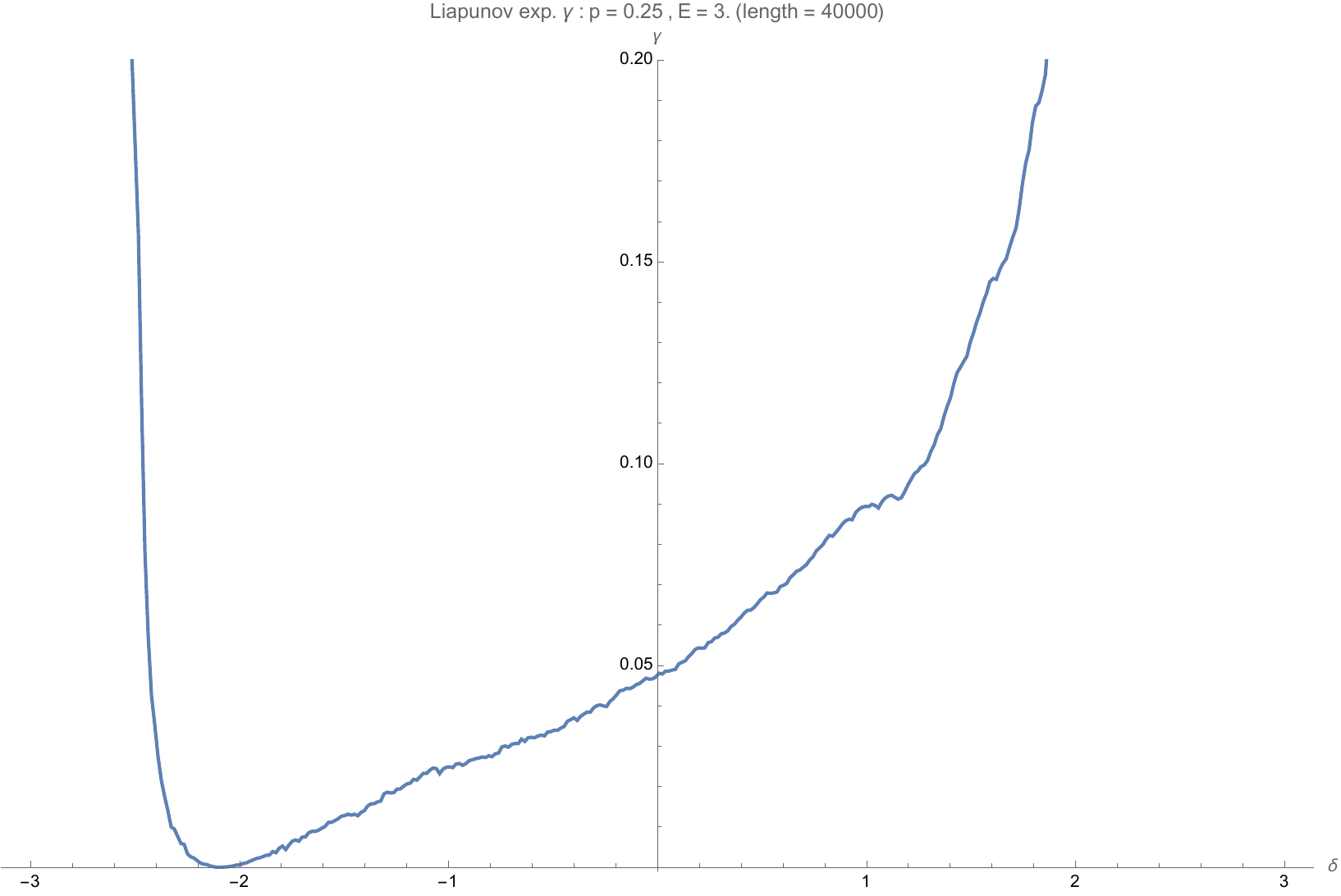}
\\
\includegraphics[width=2.5in]{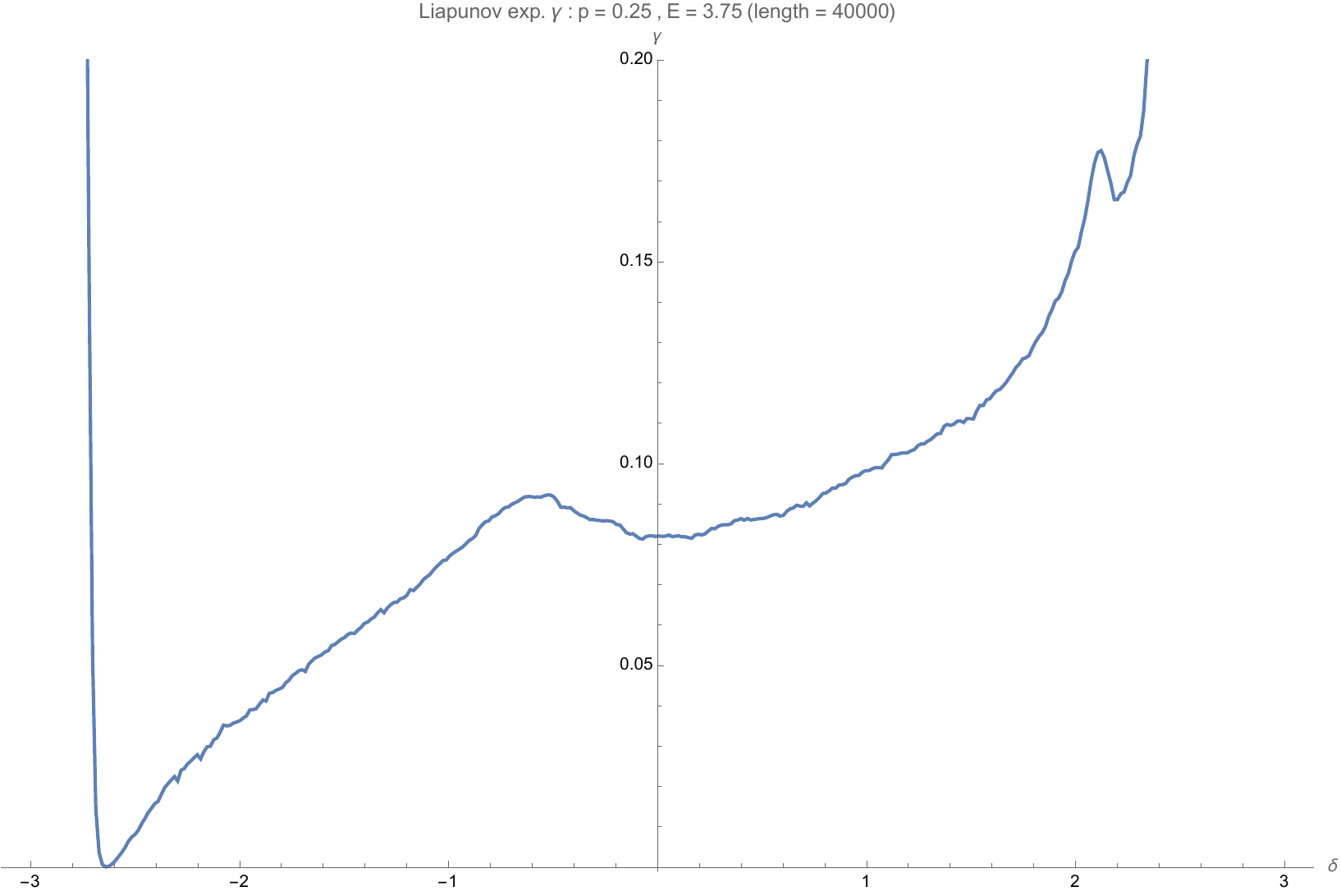}
\quad
\includegraphics[width=2.5in]{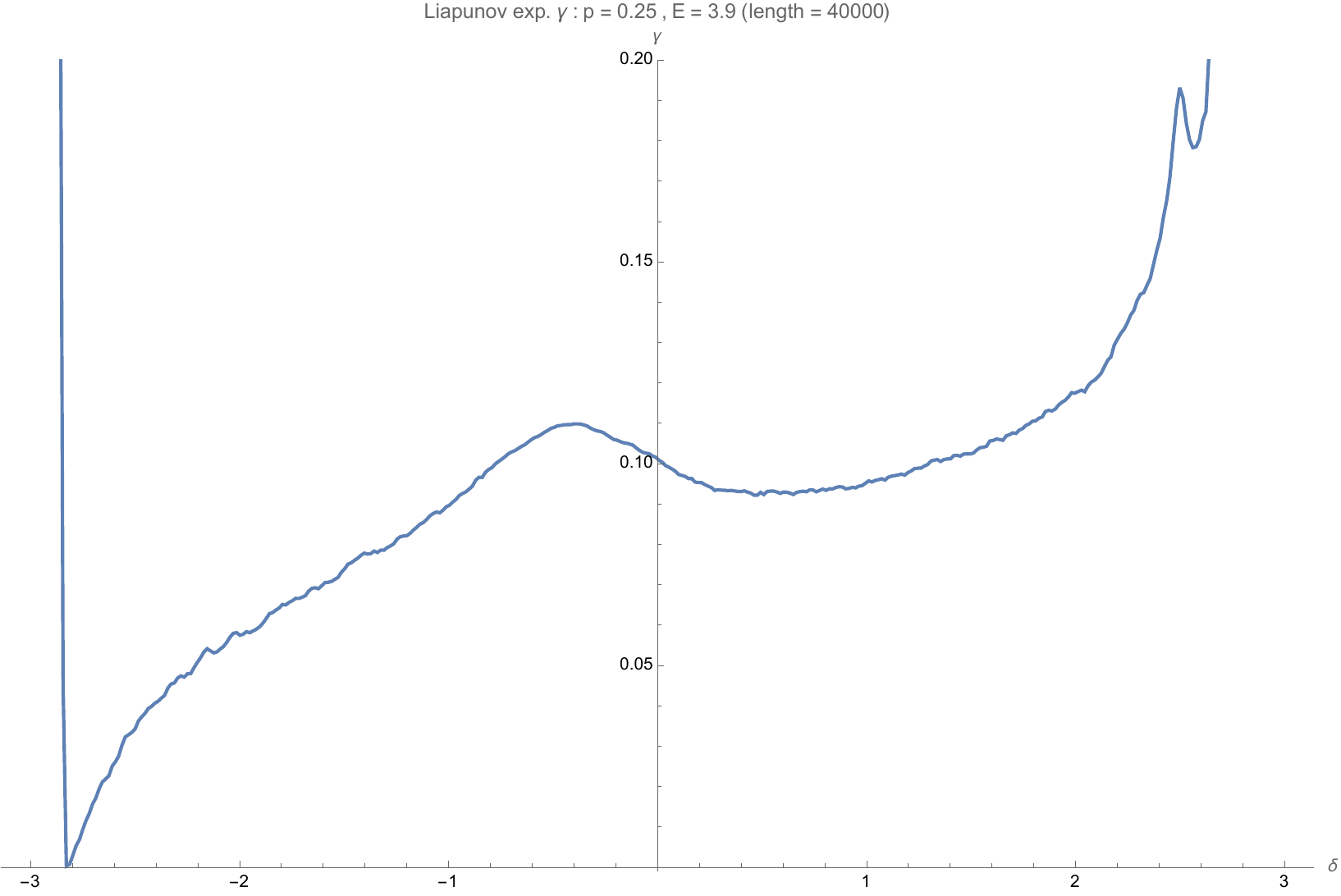}
\end{center}
\begin{figure}[h!]
\begin{center}
\caption{Lyapunov exponent $\gamma(\delta)$ as a function of the phase 
shift $-\pi<\delta<\pi$ for the S-matrix eigenstates, for various values 
of the energy $E$, and for hole probability $p=1/4$}
\label{lambdaSEdel}
\end{center}
\end{figure}
Again, the Lyapunov exponent is generically non-zero, and exhibits 
singularities. However, it vanishes for the specific value of the phase shift
\begin{equation}
\label{ }
\delta_0\ =\ -\,\theta\ =\ \arccos(2-E)/2
\end{equation}
since for this value of $\delta$, the potential $V(E,\delta)$  vanishes
independently of the hole probability $p$.


\section{Scaling for $\gamma$ and the $\Upsilon$ states in the $E\to 0$ limit }
\label{sssTheta2zero}

\subsection{Scaling for the Lyapunov exponent}
The result \rf{LyapE<4p=0} from \cite{David_2022} implies that for the regular 
comb ($p=0$) at small energies the Lyapunov exponent scales as
\begin{equation}
\label{LyapEto0}
\gamma_{\scriptscriptstyle{p=0}}\ \sim\ E^{1/4}\quad\text{when}\qquad E\to 0.
\end{equation}
We now show that this scaling is still valid for the random comb ($0<p<1$), 
and that the $|\Upsilon\kt$ eigenstates take a simple scaling form when $E\to 0$.

The recursion relation \rf{E<4columnA} for the components $\tilde A_{n,t_0}$ 
of the $|\Upsilon_{\theta_,t_0}\rangle$ column states gives the Riccati equation 
\rf{RecRelE<4a} for the ratios $\tilde\psi_n=\tilde A_n/ \tilde A_{n-1}$ 
(as long as $n\neq t_0$)
\begin{equation*}
\tilde\psi_{n+1}=-{1\over \tilde \psi_{n}}+\begin{cases}
      1+e^{-\mathi\,\theta}& \text{if $n$ is a tooth}, \\
      e^{\mathi\theta}+e^{-\mathi\theta}& \text{if $n$ is a hole}. \end{cases}
\end{equation*}
When $\theta\to 0$, we can neglect the terms of order $\theta^2$ and higher. 
The recursion becomes
\begin{equation*}
\tilde\psi_{n+1}=-{1\over \tilde \psi_{n}}+\begin{cases}
      2{-\mathi\,\theta}& \text{if $n$ is a tooth}, \\
      2& \text{if $n$ is a hole} .\end{cases}
\end{equation*}
The fixed points of this recurrence are (up to O($\theta$) terms)
$$
\tilde\psi_0^* = \begin{cases}
      1\pm \sqrt{-\mathi\,\theta}& \text{if all $n$ are teeth}, \\
      1& \text{if all $n$ are holes} 
      \end{cases}
$$
so it is natural to study the full random recurrence in the domain
$$
\tilde\psi-1 \sim \sqrt{\theta} .
$$
We rewrite
$$
\tilde\psi_{n}=1+\sqrt{\theta}\,\chi_n
$$
and the recurrence becomes (up to O($\theta$) terms)
\begin{equation}
\label{chirec}
\chi_{n+1}=\chi_n-\sqrt{\theta}\left(\,\chi_n^2 + \mathi \,\epsilon_n\right) \quad,\qquad \epsilon_n=\begin{cases}
      1& \text{if $n$ is a tooth}, \\
      0& \text{if $n$ is a hole}. \end{cases}
\end{equation}
When $\theta \ll 1$ one can view the variable $\tau$ defined as
\begin{equation}
\label{taudef}
\tau = n\,\sqrt{\theta}
\end{equation}
as a continuous time variable, and ${\chi}_n$ as a function of $\tau$. 
Then,  in the limit $\sqrt{\theta}\to 0$ and $\tau$ fixed, the stochastic 
recurrence equation \rf{chirec} becomes the deterministic differential flow equation
$$
{d\chi(\tau)\over d\tau} =-\chi(\tau)^2 - \mathi\ \mathbb{E}[\epsilon_n]
$$
with $\mathbb{E}[\epsilon_n]$ the expectation of the i.i.d.\ random variables 
$\epsilon_n$. Since
$\mathbb{E}[\epsilon_n] = 1-p$,
with $p$ the probability of holes in the chain,
the flow equation is simply
\begin{equation}
\label{chifloweq}
{d\chi(\tau)\over d\tau} =-\chi(x)^2 - \mathi\,(1-p)\ .
\end{equation}
The flow \rf{chifloweq} has two fixed points
$$
\chi^*_+ =\  e^{-\mathi\pi/4}\sqrt{1-p}\ ,\quad \chi^*_- =\  e^{3 \mathi\pi/4}\sqrt{1-p}
$$
and
$\chi^*_+$ is the attractive fixed point when $\tau\to +\infty$, and $\chi^*_-$ is the repulsive one.
Of course their roles are reversed in the $\tau\to -\infty$ limit.

Now, to compute the Lyapunov exponent at order $\sqrt{\theta}$, it is sufficient to take
\begin{equation}
\label{ }
\gamma(\theta)=\log (  | 1+\sqrt{\theta}\,\chi^*_+|) = 
\sqrt{\theta}\ \mathrm{Re}(\chi^*_+)+\mathrm{O}(\theta) .
\end{equation}
We get 
\begin{equation}
\label{gammasmalltheta}
\gamma(\theta)=\sqrt{\theta}\,\sqrt{(1-p)/2}+\mathrm{O}(\theta).
\end{equation}
Since $E=2-2\cos\theta=\theta^2+\mathrm{O}(\theta^4) $, the small $E$ scaling for $\gamma$ is
\begin{equation}
\label{LyapGenSmall}
\gamma_p\ =\ \sqrt{(1-p)/2}\ E^{1/4}\ +\ \mathrm{O}(E^{1/2})
\end{equation}
which generalizes the scaling \rf{LyapEto0} for the $p=0$ case.

\subsection{Fluctuations around the fixed point}
The effect of randomness on the recursion relation \rf{chirec} for the $\chi_n$'s is 
subdominant as $\theta\to 0$.
Its effect can be understood by linearizing \rf{chirec} near the attractive 
fixed point $\chi^*_+$. Let us write
\begin{equation}
\label{ }
\chi_n=\chi^*_+ \ +\ \theta^{1/4}\, \kappa_n\ .\end{equation}
We get at first order in $\kappa_n$ 
\begin{equation}
\label{linkaprec}
\kappa_{n+1}-\kappa_n = - \mathi\,\theta^{1/4}\, (\epsilon_n-(1-p)) \, -
\,\theta^{1/2}\, 2 \chi^*_+\,\kappa_n.
\end{equation}
In the scaling limit $\theta\to 0$, in terms of the continuous 
time $\tau=n\,\sqrt{\theta}$ defined by \rf{taudef}, the random variable 
$\theta^{-1/4}(\epsilon_n-(1-p))$ becomes a white noise $\eta(\tau)$
(the derivative of the Wiener process $W_\tau$)
\begin{equation}
\label{eps2eta}
\theta^{-1/4}(\epsilon_n-(1-p))\quad \longrightarrow \quad  \sqrt{p(1-p)}\ \eta(\tau)
\end{equation}
and the linear recursion \rf{linkaprec} becomes the stochastic equation 
\begin{equation}
\label{thestochkappa}
{d\kappa(\tau)\over d\tau}= -\,\mathi\, \sqrt{p(1-p)}\ \eta(\tau)\ -\ 2\,\sqrt{1-p}\ e^{-i\,\pi/4}\,\kappa(\tau).
\end{equation}
This Langevin equation defines a continuous time stochastic process $\kappa(\tau)$ in the complex plane. 
It can be viewed as a two-dimensional Ornstein–Uhlenbeck process, and it is obviously Markovian.

One can show that for this Markov process, any initial density distribution 
$P_0(\kappa)$ for $\kappa$ evolves with the time $\tau$ into  a distribution $P(\kappa;\tau)$.
Standard methods 
(see e.g.,  \cite{VanKampen-book}, \cite{Thygesen-book}) show that $P(\kappa;\tau)$ 
converges at large $\tau$ to a stable distribution $P_{\mathrm{stable}}(\kappa)$. 
Indeed, the affine stochastic equation \rf{thestochkappa} for $\kappa$ leads 
to a  Fokker-Planck equation for $P(\kappa;\tau)$
\begin{equation}
\label{stocheqP}
{\partial P\over\partial\tau} = 
2\sqrt{1-p} \left(e^{-\mathi\pi/4}\partial_\kappa( \kappa P) + 
e^{\mathi\pi/4}\partial_{\bar\kappa}( \bar\kappa P)\right) -
p(1-p)(\partial^2_\kappa+\partial^2_{\bar \kappa}-2\partial_\kappa\partial_{\bar \kappa})P.
\end{equation}
The first term with first derivatives is the advection term which comes from the 
second linear term in \rf{thestochkappa}. The second term with second derivatives 
is the diffusion term which comes from the first stochastic term in \rf{thestochkappa}.

The Fokker-Planck equation \rf{stocheqP} admits a unique attractive (positive and normalizable) 
stable solution 
$$
{\partial P_{\mathrm{stable}}\over\partial\tau} =0.
$$
This solution is simply a Gaussian distribution. 
Writing the complex variable $\kappa$ in terms of its real and imaginary part
\begin{equation}
\label{ }
\kappa=x+\mathi\,y\ ,\quad \bar\kappa=x-\mathi\,y
\end{equation}
$P_{\mathrm{stable}}$ is found to be
\begin{equation}
\label{ }
P_{\mathrm{stable}}(x,y)\ ={1\over\pi}\sqrt{2\over p}\ \exp\left({-(3 x^3 + 2 x y + y^2)/\sqrt{p}}\right).
\end{equation}
This is  a squeezed Gaussian centered at the origin, with standard deviation of 
order $p^{1/4}$, where $p$ is the probability of holes.

\subsection{Scaling for the $|\Upsilon\kt$ states}
\label{sssScalUpsilon}

We are now in a position to show that when $\theta\to 0$, the $|\Upsilon_{t,\theta}\kt$ 
eigenstates also take a deterministic scaling form which depends only on the strength of 
disorder, i.e., of the hole probability $p$, but not on the actual realization of the disorder.
This will be useful in the next section.

From the previous calculation, we know that the recurrence relation 
\rf{RecRelE<4a} for the components $\tilde A_{n,t}(\theta)$ of some 
$|\Upsilon_{\theta,t}\rangle$ column eigenstates has a fixed point characterized by 
a Lyapunov exponent $\gamma$ of order $\sqrt{\theta}$ given by \rf{gammasmalltheta} when $\theta$ is small.
The local effect of the disorder in the recurrence relation is of order $\theta$ and is subdominant.
It follows that the amplitudes $\tilde A_{n,t}(\theta)$  must take following the scaling 
form in the $\theta\to 0$ limit
\begin{equation}
\label{tAYscaling}
\tilde A_{n,t}(\theta)\ =\ \mathtt{C}\ \exp\left({-\sqrt{\theta (1-p)/2} \,|n-t|}\right)\,
\left(1+\mathrm{O}(\sqrt{\theta})\right).
\end{equation}
The local effect of the disorder is contained in the $\mathrm{O}(\sqrt{\theta})$ term, which is a random variable.
The constant $\mathtt{C}$ is given by Eq.~\rf{E<4columnA} at the tooth $t$, namely for $n=t$, 
and using the scaling ansatz \rf{tAYscaling}, keeping the terms of order at most 
$\mathrm{O}(\theta)$. The equation gives
$$
\mathtt{C}\ \left( 2\,\sqrt{\theta (1-p)/2} + \mathrm{O}(\theta) \right)\ =\ -2\,\mathi\,\theta+\mathrm{O}(\theta^2) .
$$
The randomness is contained in the $ \mathrm{O}(\theta)$ and $\mathrm{O}(\theta^2)$ terms.
This gives $\mathtt{C} = -\,\mathi\,\sqrt{2\theta/(1-p)}± +\ \mathrm{O}(\theta)$, and the final 
scaling form for the $A_{n,t}=\tilde A_{n,t}-\delta_{n,t}$ 
coefficients using \rf{ A2Atild} is
\begin{equation}     
\label{AYscaling}
A_{n,t}(\theta)\ =-\delta_{n,t}\ -\,\mathi\,\sqrt{2\theta/(1-p)}\,\exp\left({-\sqrt{\theta (1-p)/2} \,|n-t|}\right)\,\left(1+\mathrm{O}(\sqrt{\theta})\right).
\end{equation}


\section{Quantum diffusion from a point on the spine}
\label{sDiffusion}
\newcommand{\sloc}{{\scriptscriptstyle{\mathrm{loc}}}}
\newcommand{\nsz}{{n_{\scriptscriptstyle{0}}}}
\newcommand{\tsz}{{t_{\scriptscriptstyle{0}}}}
\newcommand{\Ploc}{P^{^{\scriptscriptstyle{\mathrm{loc}}}}}
\newcommand{\Philoc}{\Phi^{^{\scriptscriptstyle{\mathrm{loc}}}}}
\newcommand{\oPloc}{\overline{P}^{^{\scriptscriptstyle{\mathrm{loc}}}}}
\newcommand{\Pesc}{P^{^{\scriptscriptstyle{\mathrm{esc}}}}}
\newcommand{\Phiesc}{\Phi^{^{\scriptscriptstyle{\mathrm{esc}}}}}

We now come back to the diffusion problem discussed in \cite{David_2022} for the regular comb: 
What is the  large time behaviour of the wave function for a quantum particle initially located at a 
given vertex
on the spine at time zero ?
For a random comb $\mathcal{C}$ we start from the resolution of identity
\begin{equation}
\label{partition }
\mathbbold{1}=\sum_{\sigma} |\Gamma_\sigma\rangle \langle \Gamma_\sigma |\ +\ 
\int_0^\pi  {d\theta\over 2\pi}\ \sum_{t:\mathrm{tooth}} |\Upsilon_{\theta,t} \rangle 
\langle \Upsilon_{\theta,t} |
\end{equation}
in the Hilbert space $\cal{H}$.
Here $|\Gamma_\sigma\rangle$ are the normalized 
energy eigenstates with  $E=E_\sigma=2+2\cosh\sigma > 4$, 
defined in Sec.\ 3.4.  We have $\br n,j|\Gamma_\sigma \kt =C_ne^{-\sigma j}$
where the $C_n$'s are solutions to \rf{E>4equ}.
The $|\Upsilon_{\theta,t}\rangle$ are the energy eigenstates with  $E=2-2\cos\theta < 4$, 
labelled by the teeth $t$ of the comb.
Each state describes  an ingoing wave along the tooth $t$, and the outgoing waves 
along all the teeth. They are obtained from the columns of the reflection 
matrix $\mathbb{S}(\theta)$, defined by \rf{BigSdef} in Sec.\ \ref{ssSmatrix}.

The initial state at time $T=0$ is denoted $|\Phi_{n_0}\rangle$, localized at the site 
$\nsz$ on the spine, with a wave function
\begin{equation}
\label{localizedstate }
\br n,j|\Phi_{n_0}\kt =\phi_{\nsz}(n,j)=\delta_{\nsz,n}\delta_{j,0}.
\end{equation}
At time $T>0$ the state is
\begin{equation}
\label{Phi0t}
|\Phi_{n_0}(T)\rangle =
\sum_\sigma\,e^{-\mathi\,T\,E_\sigma} \, |\Gamma_\sigma\rangle\, \langle\Gamma_\sigma | \Phi_{n_0}\rangle
\ +\ \int_0^\pi {d\theta\over 2\pi}\,e^{-\mathi\,T\,E_\theta}\,\sum_{t:\mathrm{tooth}}\,|\Upsilon_{\theta,t}\rangle\,\langle \Upsilon_{\theta,t} | \Phi_{n_0}\rangle .
\end{equation}
The first term on the r.h.s.\ of \rf{Phi0t} is the $E>4$ part, which stays localized near the spine.
The second term on the r.h.s.\ of \rf{Phi0t} is the $E\le 4$ part, which propagates along the teeth 
and escapes to infinity as $T\to\infty$.

\subsection{Localization probabilities and profiles}
\label{ssLocPart}
\subsubsection{Probability $\Ploc(n)$ to stay localized at the spine}
Let us first discuss the $E>4$ part of the wave function, that we denote
\label{ssDifLoc}
\begin{equation}
\label{ PhiLocDef}
|\Philoc_{n_0}(T)\rangle =
\sum_\sigma\,e^{-\mathi\,T\,E_\sigma} \, |\Gamma_\sigma\rangle\, 
\langle\Gamma_\sigma | \Phi_{n_0}\rangle .
\end{equation}
Its squared norm is, of course, independent of time, and gives the probability
to stay localized near the spine at large $T$:
\begin{equation}
\label{PLocDef}
\Ploc(\nsz)\ =\ \langle\Philoc_{n_0}(T) | \Philoc_{n_0}(T)\rangle\ =\ \sum_\sigma\,\left | \langle\Gamma_\sigma 
| \Phi_{n_0}\rangle \right | ^2 .
\end{equation}
The quantity
$\Ploc(\nsz)$ depends on $\nsz$ and the comb $\mathcal{C}$.
Over the probability space of random combs with hole probability $p$, it is, of course, a random variable. 
For the periodic comb, the law for  $P^{\sloc}(\nsz)$ must be independent 
of $\nsz$ (by translation invariance).
In the infinite size limit $L\to\infty$, $P^{\sloc} (\nsz)$ must converge in 
law to a definite random variable, which depends only on $p$. 
However, it does not converge to a definite function of $p$, since it still depends on the 
random local environment of the initial position $\nsz$.

The average of $\Ploc\!(\nsz)$ over the ensemble of random combs can be 
bounded from above by the following simple argument.  We note that if the comb is periodic and
$\mathbb{E}[\cdot]$ denotes the average over the comb disorder, then
$\mathbb{E}[\Ploc(\nsz)]=\mathbb{E}[\Ploc]$ is independent of $n_0$. 
For a finite periodic comb $\mathcal{C}$ of length $L$, we can average \rf{PLocDef}
over the $n_0$ and get
\begin{equation}
\label{EPLoc}
\mathbb{E}[\Ploc]
  =  {1\over L}\, \sum_{n_0} \mathbb{E}\left[ \sum_\sigma
 \br \Gamma_\sigma |\Phi_{n_0}\kt\br\Phi_{n_0}|\Gamma_\sigma\kt \right] \ =\ 
{1\over L}\, \mathbb{E}\left[ \sum_\sigma \br \Gamma_\sigma |\mathbb{P}_{\scriptscriptstyle{\mathrm{spine}}}|\Gamma_\sigma\kt \right] 
\end{equation}
with $\mathbb{P}_{\scriptscriptstyle{\mathrm{spine}}}$ the projector onto the spine.
For any configuration, we have obviously
\begin{equation}
\label{ }
\sum_\sigma \br\Gamma_\sigma |\mathbb{P}_{\scriptscriptstyle{\mathrm{spine}}}|\Gamma_\sigma\kt \le 
\sum_\sigma \br\Gamma_\sigma |\Gamma_\sigma\kt  = N_{E>4}
\end{equation}
with $N_{E>4}$ the number of states with energy greater than 4. 
From Appendix \ref{AProbChain} we know that the density of $E>4$ states per unit
length converges almost surely when $L\to\infty$ towards a limit given by 
\rf{NE>4}.
We thus get for large chains the upper-bound
\beq{upperbound}
\mathbb{E}[\Ploc] \leq {1-p\over 2-p}
\eeq

\subsubsection{Numerical studies of $\Ploc(n)$}
\label{sssNumPloc}
We now present some illustrative numerical studies of $\Ploc$. 
First we compute the expectation of $\Ploc$ as a function of the hole probability $p$. 
Our calculation is based on \rf{EPLoc}, by constructing large samples of size $\mathcal{N}$ 
of random periodic combs of size $L$, for a chosen value of $p$, then by computing explicitly 
the $E>4$ eigenstates $\Gamma_\sigma$ for each chain, and finally by studying the statistics of $\Ploc$.

Some results are presented on Fig.~\ref{figEPloc}, and are obtained from samples of size 
$\mathcal{N}=10^3$ of chains of length $L=500$, for various values of $p$.
This size and number of samples are sufficient to get good enough statistics for 
our qualitative discussion.  The expectation
$\mathbb{E}[\Ploc](p)$ is found to decrease with $p$, starting from the exact value for $p=0$, 
$$
\mathbb{E}[\Ploc](0)= {1\over 2}-{2\over 3\pi}+{\sqrt{3}\over 9\pi}\log(2+\sqrt{3})=0.368\ldots
$$ 
calculated in our paper on the regular comb (see Eq.~107 in \cite{David_2022}), down to 
$\mathbb{E}[\Ploc](1)=0$ when $p=1$, where the comb reduces to a single line with no teeth. 
The numerical results are consistent with the upper-bound \rf{upperbound}.
\begin{figure}[h!]
\begin{center}
\includegraphics[width=3.5in]{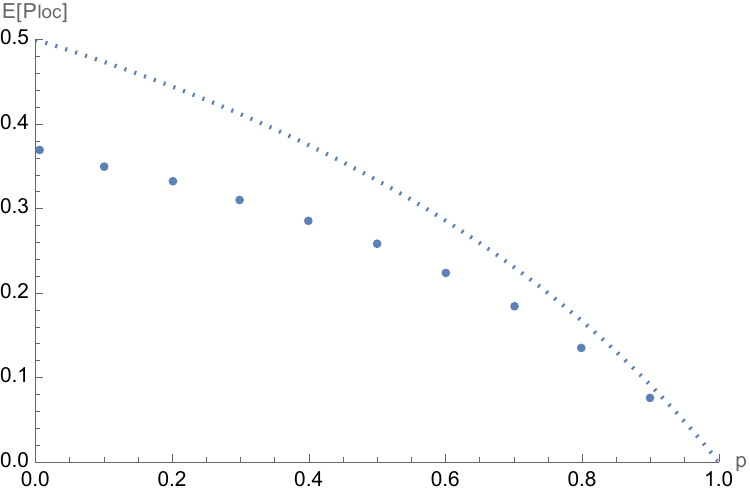}
\caption{The expectation $\mathbb{E}[\Ploc](p)$ of staying localized near the spine, when starting from a point on the spine, as a function of the hole probability $p$. The points represent the result of numerical estimates for samples of size $\mathcal{N}=10^3$ of combs of length $L=500$. The dashed line represents the upper bound \rf{upperbound}.}
\label{figEPloc}
\end{center}
\end{figure}

We have seen that $\Ploc(\nsz)$ is a random variable, since it depends on the 
environment of the initial site $\nsz$, and in particular on whether $\nsz$ is a tooth or a 
hole. A qualitative understanding of this effect can be obtained by plotting the probability 
distribution for this variable $\Ploc(\nsz)$, for the sample ensembles of combs constructed 
in the numerical study described above.

Let us first discuss this probability distribution in the hole/tooth symmetric case $p=1/2$, 
shown in Fig.~\ref{pdfPloc5}
\begin{figure}[h!]
\begin{center}
\includegraphics[width=3.5in]{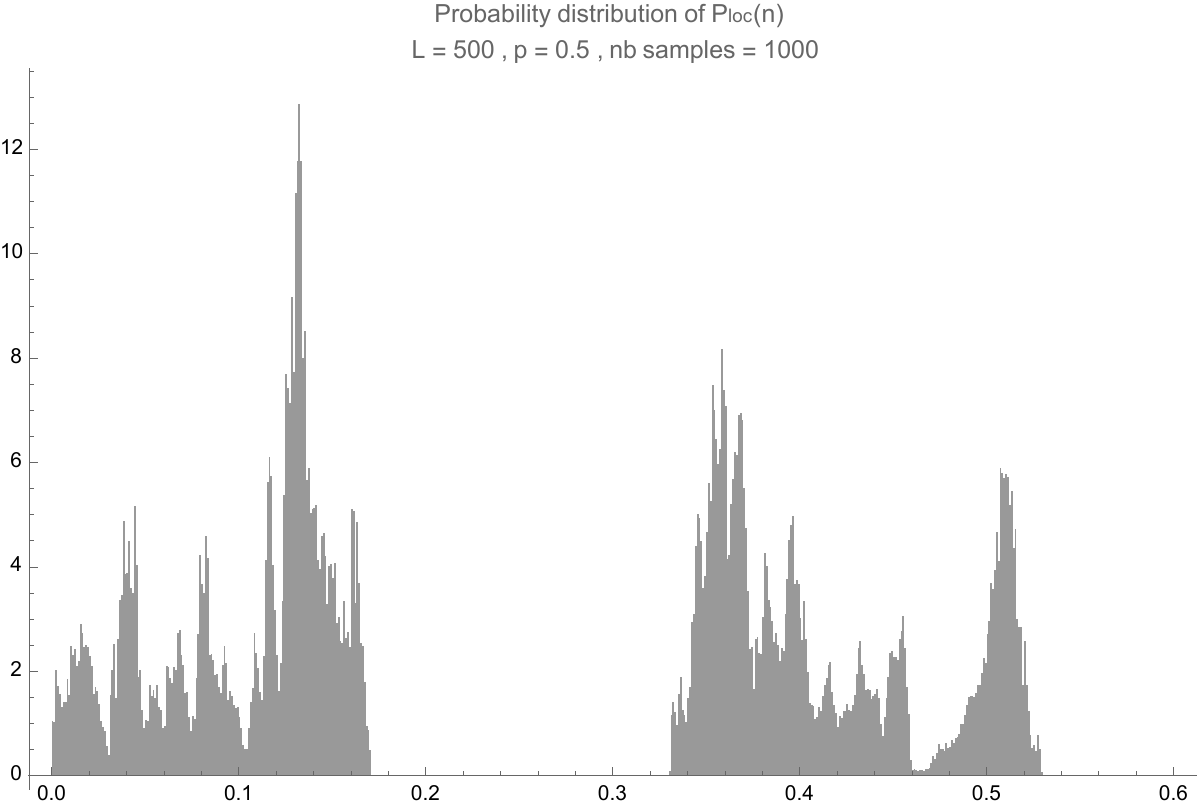}
\caption{Probability distribution for $\Ploc\!(n)$  for hole probability $p=1/2$.}
\label{pdfPloc5}
\end{center}
\end{figure}
We observe a rather spiky distribution function, separated into two distinct parts. The first 
part is for small values of $\Ploc$, typically $0\le\Ploc\le .17$, the second 
part is for higher values of $\Ploc$, typically  $0.33\le\Ploc\le .53$. These two distributions 
are separated by a large gap. The spiky character of the distribution may indicate a fractal nature.
Our study shows that the low $\Ploc$ distribution comes from comb configurations 
where the initial point $\nsz$ is a hole, while the large $\Ploc$ distribution comes from 
comb configurations where the initial point $\nsz$ is a tooth. 

This effect is corroborated when looking at the $\Ploc$ probability distributions 
for different values of the hole probability $p$.
The distribution for higher tooth probability $p=.2$ is depicted on Fig.~\ref{pdfPloc2}. 
There is still a gap between a small $\Ploc$ distribution and a large $\Ploc$ distribution, with 
a gap separating the two distributions, but the large $\Ploc$ distribution carries more weight 
than the small one.
\begin{figure}[h!]
\begin{center}
\includegraphics[width=3.5in]{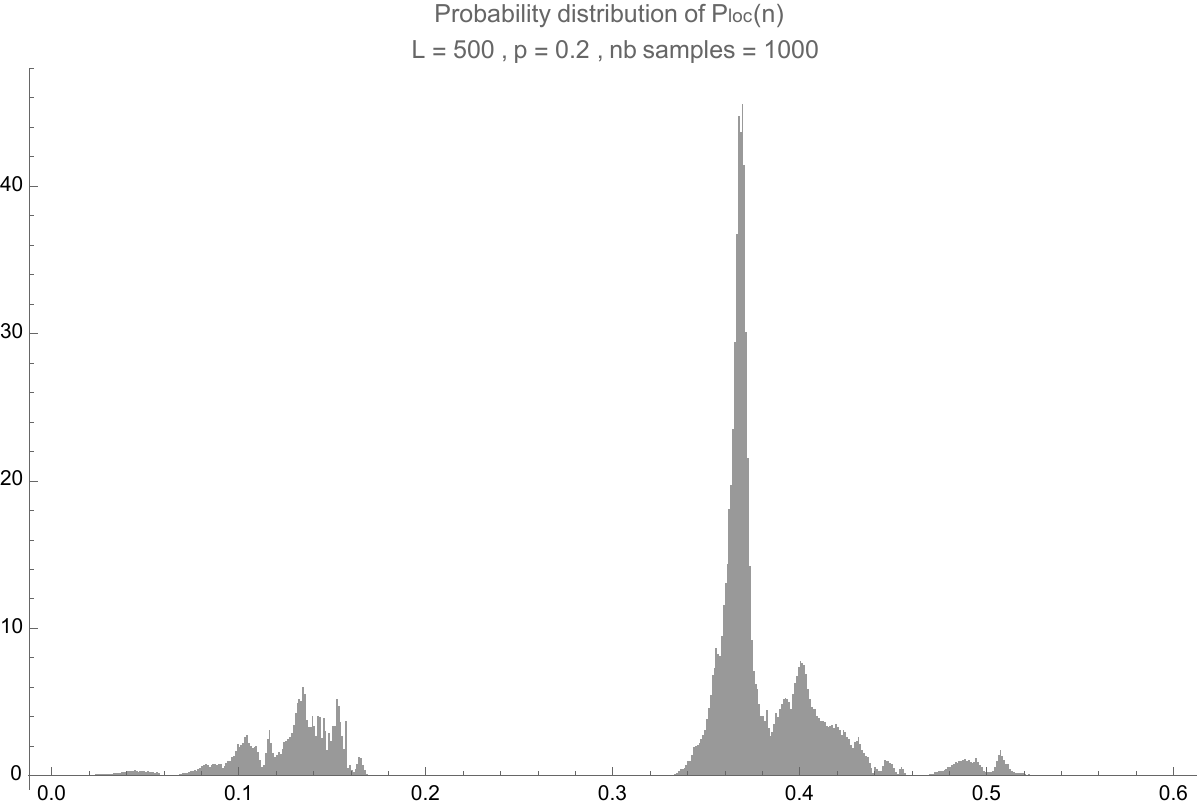}
\caption{Probability distribution for $\Ploc\!(n)$  for hole probability $p=.2$}
\label{pdfPloc2}
\end{center}
\end{figure}
In the reverse case of higher hole probability $p=.8$, depicted on Fig.~\ref{pdfPloc8}, 
one still has a distribution separated into two parts by a gap, but now 
the small $\Ploc$ distribution 
carries more weight than the high one.
\begin{figure}[h!]
\begin{center}
\includegraphics[width=3.5in]{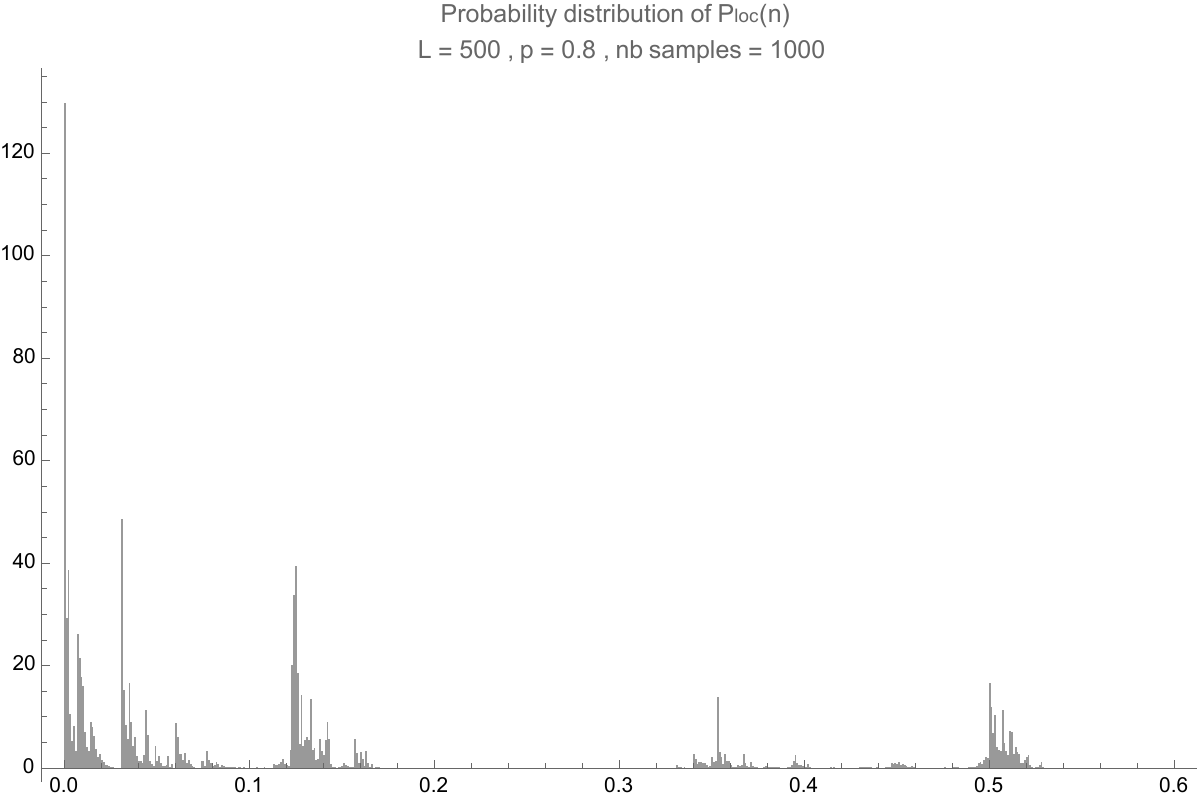}
\caption{Probability distribution for $\Ploc\!(n)$  for hole probability $p=.8$}
\label{pdfPloc8}
\end{center}
\end{figure}

This trend is also seen in calculations for other values of $p$ in the range $0<p<1$, but 
we will not present these here.  However, let us mention that the existence and the 
position of the gap seem unchanged by the hole probability $p$. It would be interesting to 
pursue an investigation of this point by analytical methods.

\subsubsection{Correlation profile $\Ploc\!(\nsz,n)$}
More information on the localization probabilities near the spine in the diffusion 
process is accessible by considering the probability 
$\Ploc(n_{\scriptscriptstyle{0}},n;T)$ to stay localized at  
a given site/tooth $n$ along the spine at time $T$, when starting from site $\nsz$ 
on the spine at time $T=0$.
We define the projector onto the tooth/hole $n$, $\mathbb{P}_n$, by
\begin{equation}
\label{Pndef}
\mathbb{P}_n \ =\ \begin{cases}
      \ \sum\limits_{j=0}^\infty | n,j\rangle \langle n,j|\quad&\text{if $n$ is a tooth}, \\
      \ | n,0\rangle \langle n,0|\quad&\text{if $n$ is a hole}.
\end{cases}
\end{equation}
The probability $\Ploc(n_{\scriptscriptstyle{0}},n;T)$ is given by
\begin{equation}
\label{ }
\Ploc(n_{\scriptscriptstyle{0}},n;T)\ =\ 
\langle \Philoc_{n_0}(T) | \mathbb{P}_n  | 
\Philoc_{n_0}(T) \rangle .
\end{equation}
For a given comb configuration $\mathcal{C}$ this probability
is a dynamical quantity which varies with $T$. Over the ensemble of combs it is, of course, a stochastic variable.
In any case there is no reason for  $\Ploc(\nsz,n;T)$ to have a $T\to\infty$ limit. 

However, the probability $\Ploc(\nsz,n;T)$ can be bounded uniformly, independently of $T$,  
by the following argument. Consider the function 
\begin{equation}
\label{envelope}
Q(\nsz,n)\ =\ 
\sum_{\sigma,\sigma'} 
\left\vert{\langle \Phi_\nsz | \Gamma_\sigma \rangle} 
{\langle  \Gamma_\sigma | \mathbb{P}_n | \Gamma_{\sigma'}\rangle } 
{\langle  \Gamma_{\sigma'} | \Phi_\nsz \rangle} \right\vert .
\end{equation}
It is easy to check that 
\begin{equation}
\label{ }
\Ploc(\nsz,n;T)\le Q(\nsz,n).
\end{equation}
Since the $\Gamma_\sigma$ eigenstates are localized, i.e.,
exponentially bounded at infinity with a minimal Lyapunov exponent 
which depends only on the hole probability $p$ (see the discussion in 
Sec.~\ref{sssGammaE>4}), one can argue that the function $Q(\nsz,n)$ has 
a.s.\ a large $L$ limit, which only depends on the distance $|n-\nsz |$, 
and is exponentially bounded when $|n-\nsz|\to\infty$. We will not elaborate further.

A simpler way to study the localization near the spine at large time in 
the diffusion process is to consider the time average of 
$\Ploc(n_{\scriptscriptstyle{0}},n;T)$ in the large $T$ limit.
This average is 
\begin{equation}
\label{PlocAvDef}
\overline{\Ploc}(\nsz,n)=\lim_{T\to\infty} {1\over T}\int_0^T 
dT'\,P^{\scriptscriptstyle{loc}}(n_{\scriptscriptstyle{0}},n;T').
\end{equation}
Here the $\overline{P}$ denotes the average over time $T$ of the process  
$P(T)$ for a given realization of the disorder, not its probabilistic expectation value.
For any finite comb configuration $\mathcal{C}$ this $T\to\infty$ limit exists. 
We have
\begin{equation}
\label{ }
\Ploc(n_{\scriptscriptstyle{0}},n;T')=\sum_{\sigma,\sigma'} e^{\mathi\,T'\,(E_\sigma-E_{\sigma'})}
{\langle \Phi_\nsz | \Gamma_\sigma \rangle} {\langle  \Gamma_\sigma | \mathbb{P}_n | \Gamma_{\sigma'}\rangle } 
{\langle  \Gamma_{\sigma'} | \Phi_\nsz \rangle} 
\end{equation}
and when averaging over $T'$, only the non-oscillating terms for which
$E_\sigma=E_{\sigma'}$ survive in the large $T$ limit. 
For an open comb there is no level crossing, so that only the diagonal terms 
$\sigma=\sigma'$ survive in the sum. For periodic combs, level crossing can occur 
but only for very specific comb configurations, whose measure goes to zero in the 
$L\to\infty$ limit.
Therefore, in the large $T$ limit we can write
\color{black}
\begin{equation}
\label{ }
\overline{\Ploc}(\nsz,n)=\sum_{\sigma} 
{\langle \Phi_\nsz | \Gamma_\sigma \rangle} {\langle  \Gamma_\sigma 
| \mathbb{P}_n | \Gamma_{\sigma}\rangle } 
{\langle  \Gamma_{\sigma} | \Phi_\nsz \rangle} .
\end{equation}
Using \rf{Pndef} and the fact that, with our notation, for 
any state $|\Gamma_\sigma\rangle$, $\langle\Gamma_\sigma | n,0\rangle= 
\langle\Gamma_\sigma |\Phi_n\rangle$, while
$\langle\Gamma_\sigma | n,j\rangle= e^{-j\sigma}\,\langle\Gamma_\sigma |\Phi_n\rangle$ if $n$ is a tooth, 
we have, summing over the $j$'s, if needed,
\begin{equation}
\label{ }
\langle  \Gamma_\sigma | \mathbb{P}_n | \Gamma_{\sigma}\rangle =\begin{cases}
\  \langle  \Gamma_\sigma | \Phi_n\rangle  \langle \Phi_n  | \Gamma_{\sigma} 
\rangle {(1- e^{-2\sigma})}^{-1}& \text{\ if $n$ is a tooth }, \\
\    \langle  \Gamma_\sigma | \Phi_n\rangle  \langle \Phi_n  | 
\Gamma_{\sigma} \rangle  & \text{\ if $n$ is a hole}.
\end{cases}
\end{equation}
Thus we get an explicit formula for $\overline{\Ploc}(\nsz,n)$ in terms 
of the $|\Gamma_\sigma\rangle$ states:
\begin{equation}
\label{PLocAvExpl}
\overline{\Ploc}(\nsz,n)=\sum_{\sigma} \left(\left|{\langle \Phi_\nsz | 
\Gamma_\sigma \rangle} {\langle  \Gamma_\sigma | \Phi_n\rangle}\right|^2\ \times\ \begin{cases}
  {(1- e^{-2\sigma})}^{-1}& \text{\ if $n$ is a tooth }, \\   
   \qquad 1   & \text{\ if $n$ is a hole}.
\end{cases}\right).
\end{equation}
\color{black}

For an ensemble of random combs, $\overline{\Ploc}(\nsz,n)$  is, of course, still a random variable. 
For a periodic comb, its distribution depends only on the distance $|n_0-n|$ by 
translation invariance. 
The infinite comb limit $L\to\infty$ is expected to exist, so that 
$\overline{\Ploc}(\nsz,n)$ must converge in law to a random variable, whose 
law depends only on $|\nsz-n|$. 
By the same argument as above for the function $Q(\nsz,n)$, 
since the $|\Gamma_\sigma\kt$ states are localized, we expect that the stochastic function 
 $\overline{\Ploc}(\nsz,n)$ for the infinite random comb is exponentially 
 decreasing as $|n-\nsz |\to\infty$, with a localization length which depends only on 
 the hole probability $p$. 

The interplay between the large $T$ limit of the average of  
$\Ploc(n_{\scriptscriptstyle{0}},n;T)$ (given by the r.h.s.\ of \rf{PlocAvDef}) 
and its large size limit $L\to\infty$ raises interesting questions, that 
we shall not discuss here. 
Finally, note that one can easily show that, as expected
\begin{equation}
\label{ }
\sum_n \overline{\Ploc}(\nsz,n)\ =\ \Ploc(\nsz).
\end{equation}

\subsubsection{Numerical results for $\overline{\Ploc}(\nsz,n)$}
To illustrate the above statements, we present the results of some numerical simulations. 
We use the same method as the one used previously in Sec.~\ref{sssNumPloc} to study $\Ploc(n)$.
We first construct large samples of random periodic comb configurations, 
for some hole probability $p$. 
The length of the combs is denoted $L$ and $\mathcal{N}$ is the size of the 
sample (the number of random comb configurations in the sample). 
For each configuration $\mathcal{C}$ we compute numerically its $E>4$ 
energy eigenstates $|\Gamma_\sigma\rangle$. 
Then we compute the corresponding time average $\overline{\Ploc}(\nsz,n;\mathcal{C})$ 
using the explicit formula \rf{PLocAvExpl}, for the chosen configuration $\mathcal{C}$.
Finally, we repeat the calculation over the sample of configurations $\mathcal{C}$, and 
study the resulting statistics of the $\overline{\Ploc}(\nsz,n;\mathcal{C})$'s.

We consider here only the expectation $\mathbb{E}\left[\overline{\Ploc}(\nsz,n)\right]$ 
of $\overline{\Ploc}(\nsz,n)$, given by  $\rf{PLocAvExpl}$, as a function of the 
initial and final points $\nsz$ and $n$.
\color{black}
By translation invariance, this expectation depends only on the distance 
$d=|n-\nsz|$ between the initial and final points.
To get a better statistics for this expectation
one may, for each comb configuration $\mathcal{C}$, take the average over all 
pairs of points $n,\nsz$, conditioned to be at a fixed distance $d$.

In practice, we construct a large sample $\frak{S}$ of 
$\mathcal{N}$ independent random combs $\mathcal{C}$, of length $L$, with hole probability
$p$, and compute the average 
\begin{equation}
\label{EPloc}
\mathbb{E}[\overline{\Ploc}](d,L)\ =\ 
{1\over \mathcal{N}} \sum_{\mathcal{C}\in\frak{S}} {1\over L} 
\sum_{\nsz} \overline{\Ploc}(\nsz,\nsz+d).
\end{equation}
For large comb sample size $\mathcal{N}=|\frak{S}|\to\infty$ this should converge 
a.s.\ towards a well-defined function $\mathbb{E}[\overline{\Ploc}](d,L)$
\begin{equation}
\label{EPloc2}
{1\over \mathcal{N}} \sum_{\mathcal{C}\in\frak{S}} {1\over L} \sum_{\nsz} 
\overline{\Ploc}(\nsz,\nsz+d)\ \mathop{\longrightarrow}\limits_{\mathrm{a.s.}}\ 
\mathbb{E}[\overline{\Ploc}](d,L).
\end{equation}
This function must be periodic in $d$ with period $L$, the length of the periodic random combs.
We also expect this function to have a limit as $L\to\infty$:
\begin{equation}
\label{EPloc3}
\mathbb{E}[\overline{\Ploc}](d,L)\ \mathop{\longrightarrow}\limits_{L\to\infty}\ 
\mathbb{E}[\overline{\Ploc}](d).
\end{equation} 

Since this function is constructed out of the $E>4$ eigenstates $|\Gamma_\sigma\rangle$ 
which are localized with non-zero Lyapunov exponents $\gamma(E)$  which 
are bounded from below by some $\gamma_{\mathrm{min}}$, we expect that 
$\mathbb{E}[\overline{\Ploc}](d)$ is bounded by an exponential function
\begin{equation}
\label{expboundEP}
\mathbb{E}[\overline{\Ploc}](d)\ <\ \mathtt{cst.}\,e^{-2\,\gamma_{\mathrm{min}}\,d}.
\end{equation}

\begin{figure}
\begin{center}
\includegraphics[width=3.5in]{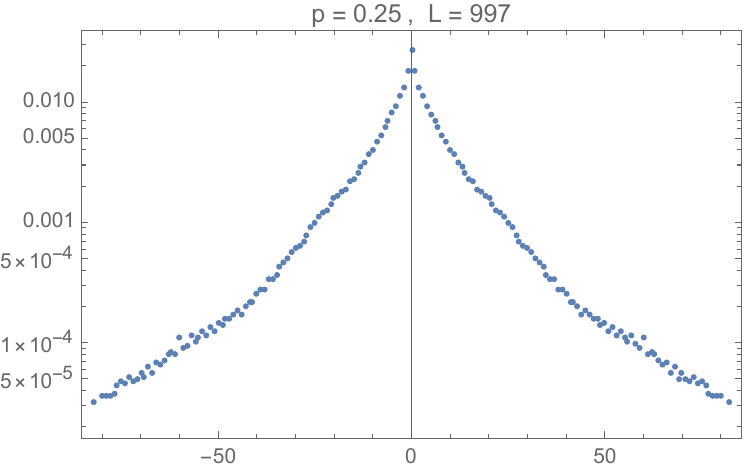}\\
\includegraphics[width=3.5in]{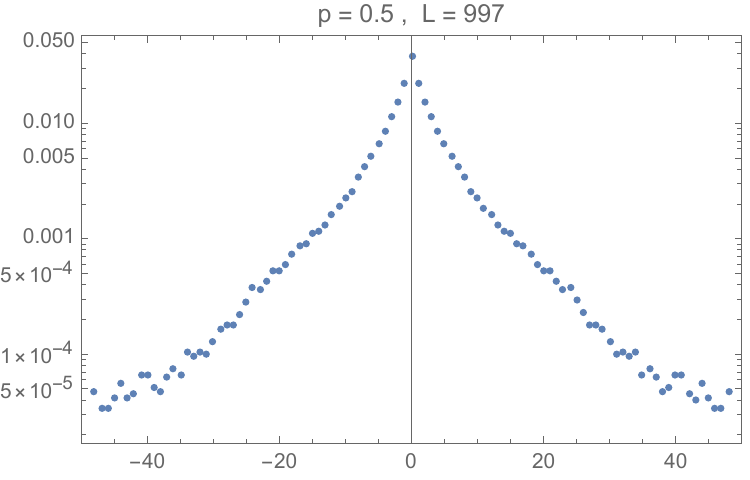}\\
\includegraphics[width=3.5in]{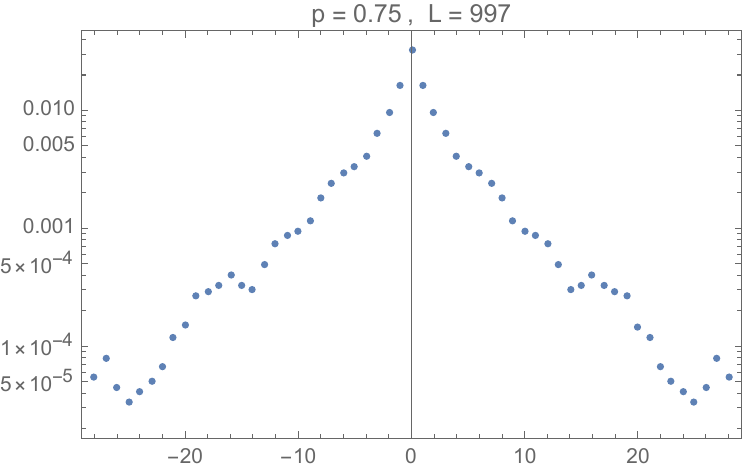}
\caption{Log plot of $\mathbb{E}[{\overline{\Ploc}}](\nsz,n)$ 
as a function of $n-\nsz$ for hole probabilities $p=.25$, $.50$ and $.75$ .
The estimates are obtained by using  a sample of $\mathcal{N}=10^3$ random periodic chains of length $L\simeq 10^3$.}
\label{fEPloc}
\end{center}
\end{figure}

Our numerical studies confirm these predictions. In the simulations we compute 
the average \rf{EPloc} for finite but large values of $L$ 
and sample size $\mathcal{N}$, and check the statistics.
We present in Fig.~\ref{fEPloc} the results for the profile of the function 
$\mathbb{E}[{\overline{\Ploc}}](d)$, in log plots, as a function of the distance $d$, 
for ensembles of random combs corresponding to three values of the hole probability $p$ 
namely for $p=.25$, $p=.5$  and $p=.75$.
In practice we consider combs of size $L$ of the order of $10^{3}$ and comb samples 
of similar size $\mathcal{N}=10^3$. This is a illustrative study and for 
this purpose, the statistic is good with these sizes of comb of samples.

The log plots show clearly decay with $d$ and are compatible with the expected bound 
\rf{expboundEP}. In particular, the decay characteristic length is found to decrease 
with the hole probability $p$, from $p=.25$ to $p=.75$, as expected from the $p$ 
dependence of $\gamma_{\mathrm{min}}(p)$ observed in Fig.~\ref{minLiapE>4}.

Similar curves can be constructed for the higher moments and cumulants of ${\overline{\Ploc}}$, 
such as $\mathbb{E}[ {\overline{\Ploc}}^2 ](d)$, but we shall not discuss this.
They could be useful to study whether the spatial distribution of 
$\overline{\Ploc}(d)$ exhibits multifractality.

\subsection{Escape probabilities and escape profiles}

\subsubsection{Total escape probability}

Now we discuss the second part of the wave function $|\Phi_\nsz(T)\rangle$, namely 
the second term on the r.h.s.\ of \rf{Phi0t}, which is the projection of $|\Phi_\nsz(T)\rangle$
onto the subspace of energy eigenstates with energy $E\le 4$.
These states describe a particle which propagates along the teeth and can escape to infinity.
This delocalized state is given by
\begin{equation}
\label{PhiEsc}
|\Phiesc_{n_0}(T)\rangle =
\int_0^\pi {d\theta\over 2\pi} \,e^{-\mathi\,T\,E_\theta} \sum_{t:\mathrm{tooth}}\, |\Upsilon_{\theta,t}\rangle\, \langle\Upsilon_{\theta,t} | \Phi_{n_0}\rangle .
\end{equation}
The norm squared of this state is, using \rf{OrthNUps},
\begin{equation}
\label{PEsc}
\Pesc\!(\nsz)=\int_0^\pi {d\theta\over 2\pi}\, \sum_{t:\mathrm{tooth}}\, 
{\left\vert{\langle\Upsilon_{\theta,t} | \Phi_{n_0}\rangle}\right\vert}^2
\end{equation}
and gives the probability for a particle initially at site $\nsz$ on the spine to escape 
to infinity along the teeth.
Obviously, we have $\Ploc\!(\nsz)+\Pesc\!(\nsz)=1$.

\subsubsection{Escape probability along a tooth}

A quantity of interest is the escape probability along a given tooth $t$, when 
starting from a site 
$\nsz$ on the spine.
We first define the escape probability to be on the tooth $t$ at time $T$
by
\begin{equation}
\label{PEsct1}
\Pesc\!(t,\nsz;T)\ =\
\langle \Phiesc_\nsz (T) | \mathbb{P}_{t} |
\Phiesc_{n_0}(T)\rangle ,
\end{equation}
where $\mathbb{P}_t=\sum_j|n,j\kt\br n,j|$ is the projection on the 
tooth $t$ rooted at the vertex
$(n,0)$ on the spine.
This probability is time dependent, but we show that it has an infinite time limit, 
\begin{equation}
\label{PEscLim}
\Pesc\!(t_1,\nsz)=\lim_{T\to\infty}\Pesc\!(t_1,\nsz;T)
\end{equation}
and we express this limit in terms of the columns of the S-matrix 
$| \Upsilon_{\theta,t} \rangle$, defined in Sec.\ 3.3, which are labelled by the teeth $t$.

Using the decomposition \rf{PhiEsc}, $\Pesc\!(t_1,\nsz;T)$ 
can be written
\begin{equation}
\label{PEsct2}
\sum_{t,t':\mathrm{teeth}}\int_0^\pi  {d\theta \over 2 \pi} \int_0^\pi 
{d\theta' \over 2 \pi}\, e^{\mathi\,T\,(E_\theta-E_{\theta'})}\,
\langle\Phi_\nsz | \Upsilon_{\theta,t} \rangle\,
\langle \Upsilon_{\theta,t} | \mathbb{P}_{t_1} | \Upsilon_{\theta',t'} \rangle
\,  \langle \Upsilon_{\theta',t'} | \Phi_\nsz \rangle .
\end{equation}
Now we use the form of the wave functions of the 
$| \Upsilon_{\theta,t}  \rangle$ states:
\begin{equation}
\label{phithetat}
\begin{split}
\Upsilon_{\theta,t}(n,j)\ & 
= \ \delta_{n,t}\,e^{-\mathi\,\theta\,j} + A_{\theta,t}(n)\,e^{\mathi\,\theta\,j},
\qquad \text{if $n$ is a tooth},
\\  
\Upsilon_{\theta,t}(n)\ & =\ A_{\theta,t}(n), \qquad \text{if $n$ is a hole},
\end{split}
\end{equation}
where the $A_{\theta,t}(n)$ are the solutions of the eigenvalue equation \rf{E<4column}. 
In terms of the $\mathbb{S}(\theta)$ matrix, defined 
in Sec\ 3.3, $A_{\theta,t}(n)= \mathbb{S}(\theta)_{n,t}$ where $t$ is a tooth.  
Using the notation
$\mathbbold{1}$ for the $N\times N$ identity matrix, \rf{phithetat} can be written
\begin{equation}
\label{ }
\begin{array}{lc}
\Upsilon_{\theta,t}(n,j)=\mathbb{S}(\theta)_{n,t}\,e^{\mathi\,\theta\,j}+ 
\mathbbold{1}_{n,t}\,e^{-\mathi\,\theta\,j}, & \qquad \text{if $n$ is a tooth}, \\
& \\
\Upsilon_{\theta,t}(n)\  =\mathbb{S}(\theta)_{n,t}, &\qquad \text{if $n$ is a hole}.
\end{array}
\end{equation}
In particular, one has
\begin{equation}
\label{P0Uexpl}
\langle \Phi_\nsz\vert \Upsilon_{\theta,t} \rangle =  A_{\theta,t}(\nsz)+\delta_{\nsz,t}
= \left(\mathbb{S}(\theta) +\mathbbold{1}\right)_{n_0,t}.
\end{equation}
We can write the term in the integral  in the r.h.s. of \rf{PEsct2} more explicitly as
\begin{equation}
\label{UPU}
\begin{split}
\langle \Upsilon_{\theta,t} | \mathbb{P}_{t_1} | \Upsilon_{\theta',t'} \rangle=\sum_{j=0}^\infty\ & 
\bar A_{\theta,t}(t_1)A_{\theta',t'}(t_1) e^{\mathi\,(\theta'-\theta) j}
+\delta_{t,t_1} A_{\theta',t'}(t_1)e^{\mathi\,(\theta'+\theta) j} \\
& + \delta_{t',t_1} \bar A_{\theta,t}(t_1)e^{-\mathi\,(\theta'+\theta) j}
+ \delta_{t,t_1}\delta_{t',t_1}e^{\mathi\,(\theta-\theta') j}.
\end{split}
\end{equation}
Summing over the $j$'s along the tooth $t_1$, the r.h.s.\ of (\rf{UPU}) becomes
\begin{equation}
\begin{split}
\label{UPUexpl}
\bar A_{\theta,t}(t_1)&A_{\theta',t'}(t_1)\left(1-e^{\mathi\,(\theta'-\theta)-\epsilon_+}\right)^{-1}
+ \delta_{t,t_1} A_{\theta',t'}(t_1)\left(1-e^{\mathi\,(\theta'+\theta)-\epsilon_+}\right)^{-1}
\\
&+ \delta_{t',t_1} \bar A_{\theta,t}(t_1)\left(1-e^{-\mathi\,(\theta'+\theta)-\epsilon_+}\right)^{-1}
+ \delta_{t,t_1}\delta_{t',t_1} \left(1-e^{\mathi\,(\theta-\theta')-\epsilon_+}\right)^{-1}
\end{split}
\end{equation}
where, as in Appendix~\ref{Appendix7}, a small positive parameter $\epsilon_+$ is 
introduced in order to make sense of the summation over the $j$'s, and has to be taken 
to zero at the end of the calculation.

We know from Appendix B.3 that for a finite comb, the $A_{\theta,t}(n)$ are 
meromorphic functions in the variable $z=e^{\mathi\theta}$ and they are analytic in the 
variable $\theta$ on the real axis.
Therefore the integrand in \rf{PEsct2} is meromorphic in the integration 
variables $\theta$ and 
$\theta'$, and analytic for real $\theta$ and $\theta'$, as long as the parameter 
$\epsilon_+$ is non-zero.
We can study the large $T$ asymptotics of $P^{^{\scriptscriptstyle{esc}}}\!(t_1,\nsz;T)$ 
by deformations of the integration paths in the complex plane over the variables $\theta$ 
and $\theta'$ in \rf{PEsct2}, and use Cauchy residue formula and complex 
saddle point methods.

Let us first fix $\theta\in (0,\pi)$ and look at the integration in 
$\theta'$ over the interval $[0,\pi]$.
We see from \rf{UPUexpl} that
$\langle \Upsilon_{\theta,t} | \mathbb{P}_{t_1} | \Upsilon_{\theta',t'} 
\rangle$ has only two simple poles near the real axis. The first one is at 
$\theta'=\theta-\mathi\epsilon_+$ and comes from the first term of \rf{UPUexpl}. 
The second one is at $\theta'=\theta+\mathi\epsilon_+$ and comes from the fourth 
term of \rf{UPUexpl}.
Thus, 
\begin{equation}
\begin{split}
\label{UPUres}
\langle \Upsilon_{\theta,t} | \mathbb{P}_{t_1} | \Upsilon_{\theta',t'} \rangle\ =\  
\bar A_{\theta,t}(t_1)A_{\theta,t'}(t_1){\mathi\over \theta'-\theta+\mathi\epsilon_+}
+\delta_{t,t_1}\delta_{t',t_1}{\mathi\over \theta'-\theta-\mathi\epsilon_+} \ +\ \cdots
\end{split}
\end{equation}
where $\cdots$ denotes a term regular at $\theta'=\theta$.

The time-dependent term under the integral in \rf{PEsct2} 
behaves near $\theta'=\theta$ as
$$
e^{\mathi T(E_\theta-E_\theta')}=e^{-\mathi\, 2 T \,\sin\theta\, (\theta'-\theta) 
+O((\theta-\theta')^2)}.
$$
The $T\to\infty$ limit is obtained by deforming the integration contour over $\theta'$ 
into the lower 
complex half plane $\mathrm{Im}(\theta')<0$, as depicted on Fig.~\ref{fThetaInt}, 
thus picking up 
the residue of the first pole at $\theta'=\theta-\mathi\epsilon_+$, 
but not the residue of the second pole at $\theta'=\theta+\mathi\epsilon_+$.
In the limit $\epsilon_+\to 0$ the residue becomes
$\bar{A}_{\theta, t}(t_1) A_{\theta',t'}(t_1)$ which is, of course, $T$ independent.
The rest of the contour lies in the lower half plane and 
decays exponentially when $T$ goes to infinity.   

\begin{figure}[h!]
\begin{center}
\includegraphics[width=4in]{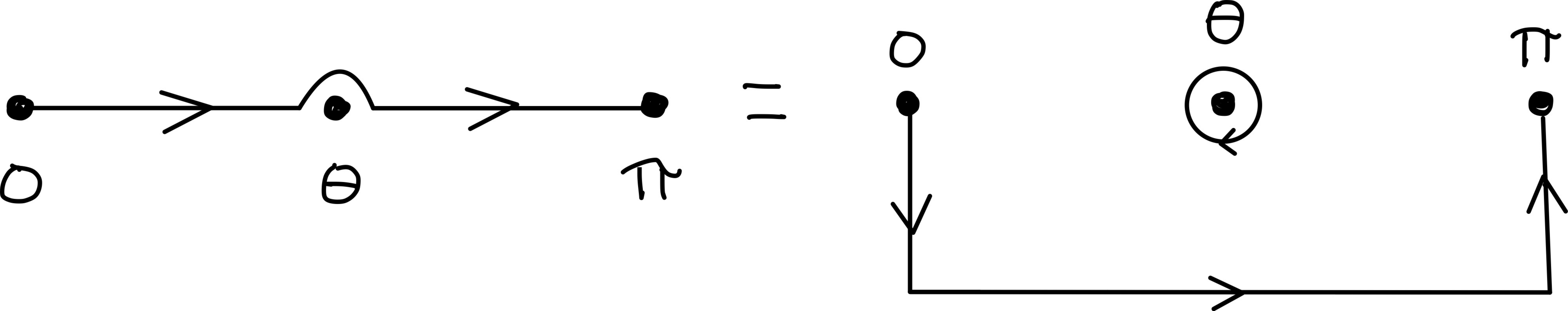}
\caption{ 
Deformation of the integration contour over $\theta'$}
\label{fThetaInt}
\end{center}
\end{figure}
We end up with 
the following expression for the large time escape probability 
\begin{equation}
\label{PEsct4}
\Pesc\!(t_1,\nsz)=\sum_{t,t'\mathrm{:teeth}} \int_0^\pi 
{d\theta\over 2\pi}\, 
\langle\Phi_\nsz | \Upsilon_{\theta,t} \rangle
\bar A_{\theta,t}(t_1)A_{\theta,t'}(t_1) 
 \langle \Upsilon_{\theta,t'} | \Phi_\nsz \rangle .
\end{equation}
This can be reexpressed in terms of the matrix $\mathbb{S}$.
Since $t_1$, $t$ and $t'$ are teeth, using the fact that the $N_t\times N_t$ S-matrix 
$S(\theta)$ 
is both symmetric and unitary, and that the $N\times N$ matrix $\mathbb{S}(\theta)$ 
is such that $\mathbb{S}(\theta)^{-1}=\overline{\mathbb{S}(\theta)}$ 
(see Appendix~\ref{App5}) we have
$$
\overline{A_{\theta,t}}(t_1)=\overline{S(\theta)}_{t_1,t}=S(\theta)^{-1}_{t,t_1}=
\mathbb{S}(\theta)^{-1}_{t,t_1}=\overline{\mathbb{S}(\theta)}_{t,t_1}.
$$
Hence, the sum over the teeth $t$ can be written 
\begin{equation}
\label{ }
\begin{split}
\sum_{t\mathrm{:teeth}}\langle\Phi_\nsz | \Upsilon_{\theta,t} \rangle\,\bar A_{\theta,t}(t_1)&=
\sum_{t\mathrm{:teeth}} (\mathbb{S}(\theta)+\mathbbold{1})_{\nsz,t}\, \mathbb{S}(\theta)^{-1}_{t,t_1}
=\left(\mathbb{S}(\theta)^{-1}+\mathbbold{1}\right)_{n_0,t_1}\\
&= \left(\overline{\mathbb{S}(\theta)}+\mathbbold{1}\right)_{n_0,t_1}=
\overline{\langle\Phi_\nsz | \Upsilon_{\theta,t_1}\rangle} .
\end{split}
\end{equation}
The sum over the teeth $t'$ gives the complex conjugate
\begin{equation}
\label{ }
\sum_{t'\mathrm{:teeth}}A_{\theta,t'}(t_1) 
 \langle \Upsilon_{\theta,t'} | \Phi_\nsz \rangle = \langle\Phi_\nsz | \Upsilon_{\theta,t_1}\rangle .
\end{equation}
We end up with the simple  formula for the total probability to escape along the tooth $t_1$ 
as $T\to\infty$:
\begin{equation}
\label{Pescnt}
\Pesc\!(t_1,\nsz)
\ =\ \int_0^\pi {d\theta\over 2\pi}\ \left\vert \langle \Phi_\nsz \vert \Upsilon_{\theta,t_1}\rangle 
\right\vert ^2 .
\end{equation}
This result does not depend on whether $n_0$ is a tooth or a hole on the comb $\mathcal{C}$.
It involves simply the overlap between the initial state $|\Phi_{n_0}\rangle$ and 
the final states $|\Upsilon_{\theta,t_1}\rangle$ corresponding to a 
particle moving only on the outgoing tooth $t_1$. 
The result \rf{Pescnt} is natural, considering the time 
reversal invariance of the problem.

From \rf{Pescnt}, the total probability to escape, starting from site $\nsz$, 
agrees with \rf{PEsc}, as expected
\begin{equation}
\label{ }
\sum_{t_1\mathrm{:teeth}}\Pesc\!(t_1,\nsz) =\Pesc\!(n_0)= 1-\Ploc\!(n_0)\ .
\end{equation}

\subsection{Large distance asymptotics of $P^{^{\mathrm{esc}}}\!(t,n)$}
Using the results of Sec.~\ref{sssTheta2zero} , we now can 
derive the asymptotic form for the probability to escape along a tooth 
$t$, starting  from an initial point $n_{\scriptscriptstyle{0}}$ on the spine, 
in the case when the distance between the tooth $t$ and the initial 
point $n_{\scriptscriptstyle{0}}$ is large. 
This escape probability $P^{^{\mathrm{esc}}}\!(t,n_{\scriptscriptstyle{0}})$ 
is given by \rf{Pescnt}
and is, of course, a random variable since the eigenstates 
$|\Upsilon_{\theta,t}\kt$ are stochastic variables.
Using \rf{P0Uexpl}, for $t\neq\ n_{\scriptscriptstyle{0}}$, this quantity is
\begin{equation}
\label{inttheta}
 \int_0^\pi {d\theta\over 2\pi}\ \left\vert A_{n_0,t}(\theta)
\right\vert ^2 .
\end{equation}
As long as the distance between the initial point $n_{\scriptscriptstyle{0}}$ 
and the tooth $t$ is large enough, the dominant contribution to 
the integral  \rf{inttheta} is given by the values of $\theta$ such that the 
overlap between the states $|\Phi_{n_{\scriptscriptstyle{0}}}\kt$ 
and $|\Upsilon_{\theta,t}\kt$ is large.
This occurs for values of $\theta$ such that the localization length 
$\xi(\theta)$ of $|\Upsilon_{\theta,t}\kt$ is large and of the same order as 
the distance $|t-n_{\scriptscriptstyle{0}}|$. This is equivalent to 
the Lyapunov exponent $\gamma(\theta)$ being small and of order 
$|t-n_{\scriptscriptstyle{0}}|^{-1}$.
In this domain, we can use the small $\theta$ scaling form 
\rf{AYscaling} for $A_{n,t}(\theta)$ obtained in Sec.~\rf{sssScalUpsilon}. 
This gives the dominant contribution
\begin{equation*}
\label{ }
P^{^{\scriptscriptstyle{\mathrm{esc}}}}\!(t,n_{\scriptscriptstyle{0}})\ 
\simeq\ \int_0^\infty {d\theta\over 2\pi}\  2\,\theta (1-p)^{-1}
\,\exp\left({-\sqrt{2\theta (1-p)} \,|n_{\scriptscriptstyle{0}}-t|}\right).
\end{equation*}
The integral can be evaluated and gives the final result
\begin{equation}
\label{EscProbAsym}
P^{^{\scriptscriptstyle{\mathrm{esc}}}}\!(t,n_{\scriptscriptstyle{0}})\ 
=\ {6\over (1-p)^3}\,{1\over |t-n_{\scriptscriptstyle{0}}|^4}\ +\ \cdots 
\quad\text{when}\quad  |t-n_{\scriptscriptstyle{0}}|\ \to\ \infty ,
\end{equation}
where $\cdots$ denotes terms that vanish faster than $|t-n_{\scriptscriptstyle{0}}|^{-4}$
as $|t-n_{\scriptscriptstyle{0}}|\ \to\ \infty$.

Let us comment on this result.
\begin{itemize}
  \item Firstly, the escape probability along a tooth, starting from an 
  initial point on the spine, takes a universal and deterministic form when the 
  distance between the initial point and the tooth is large. 
  \item The algebraic decay exponent $-\,4$ with the distance $d=|t-n_{\scriptscriptstyle{0}}|$ is universal, 
  and independent of the strength of the disorder (hole probability) $p$.
  \item The algebraic $(1-p)^{-3}$ dependence of the coefficient means that at large distance $d$ 
  the escape probability takes the form of a continuous algebraic distribution probability
  $$dP^{^{\scriptscriptstyle{\mathrm{esc}}}}=6\, dx\,x^{-4},\  x=(1-p)|t-n_{\scriptscriptstyle{0}}|$$
  which depends on the rescaled universal distance $x$.
  
\item The subdominant terms represented by the $\cdots$ in 
\rf{EscProbAsym} are stochastic terms depending on the actual 
realization of the disorder in the infinite random chain.
\end{itemize}

\section{Conclusion}
In this paper we have extended the study of the continuous time quantum
walk on a regular infinite comb of \cite{David_2022} to the case of the infinite random comb. 
As for the regular comb, the energy eigenstates behave differently at low energy  
$0\le E\le 4$ and 
at high energy $E>4$.
For $E<4$ the eigenstates describe particles propagating in and out along the teeth of the comb. 
For $E>4$ the eigenstates correspond to a particle localized in the vicinity of the infinite spine 
of the comb. 

We have shown that the introduction of geometric randomness (by assigning a probability 
$p$ for having a tooth missing along the spine) changes qualitatively some 
features of the quantum walk, in both energy regimes.
The problem of quantum diffusion from a point on the comb also changes and the walk can
be trapped in a finite region with nonzero probability.

Most results are obtained by mapping the different regimes to the classical 
one-dimensional Anderson model of strong localization and to the special case 
of the random binary chain, sometimes for real energy and disorder strength, 
and sometimes for complex ones. We have presented both analytical results using 
these mappings, and numerical results based on the application of standard methods: 
exact diagonalization of finite dimensional Hamiltonians and Riccati recurrences.

We have shown that the $E>4$ energy eigenstates states are also localized along the 
spine, with a localization length bounded from above uniformly w.r.t.\ the energy $E>4$. 
For the $0\le E\le 4$ states, we have shown that the S-matrix $S(E)$ 
(describing the scattering of in- and out-states) also exhibits 
localization along the spine, both for the elements of the S-matrix 
(columns states) and the eigenstates of the S-matrix (phase shift eigenstates).

We have studied analytically the low energy limit ($0<E\ll 4$) of the S-matrix 
at finite disorder ($0<p<1$) and shown that it takes a scaling 
form controlled by a fixed-point equation for a Riccati recursion relation.
Finally, these results enable us to study quantitatively the large time limit of 
the quantum diffusion of a particle starting from an initial point on the spine. 
Furthermore, we can estimate the asymptotic escape probability profile 
along the teeth, as well as the 
localization length of the average probability profile for a particle to stay 
localized near the initial point as time becomes large.

We believe that the methods developed in this paper can be applied to more complicated
quantum walk problems, e.g., quantum walk on a random tree.  For a generic random tree
such a problem is essentially one-dimensional and one would expect to find localization.

\section*{Acknowledgements}
FD thanks Jean-Marc Luck for useful discussions.
TJ is grateful for hospitality at IPhT, Saclay.
FD is grateful for hospitality at Science Institute, U. of Iceland, Reykjavik.
FD was partly funded by the ERC-SyG project, Recursive and Exact
New Quantum Theory(ReNewQuantum) which received funding from the European Research
Council (ERC) under the European Union’s Horizon2020 research and innovation programme
under Grant Agreement No. 810573.


\appendix

\newtheorem{lemma}{Lemma}
\newtheorem{proof}{Proof}
\newcommand{\sotimes}{\raisebox{.27ex}{${\scriptscriptstyle{\otimes}}$}}

\section{Some results on the spectrum of the binary chain}

\subsection{$k$-chains}
We start from the eigenvalue equation for a binary chain 
\rf{bcgeneric} corresponding to a finite comb $\mathcal{C}$ with $N_t$ 
teeth and $N_h$ holes:
\begin{equation}
\label{bcgeneric1}
\begin{split}
H\,\phi(n)&=-\phi({n+1})-\phi({n-1}) + V(n)\,\phi(n)=E\,\phi(n) 
\\
V(n) &= \begin{cases}
0& \text{if $n$ site is a hole $\circ$}, \\
V& \text{if $n$ site is a tooth $\bullet$}
\end{cases}
\end{split}
\end{equation}
with open or periodic boundary conditions. For fixed $V$ 
the $N=N_t+N_h$ eigenvalues of $H$ are denoted $E_\alpha(V)$, $\alpha=1,\cdots N$.
We know from Section 3.6.3 that 
$$
0\le {dE_\alpha(V)\over dV} \le 1.
$$
Any finite chain can be written as an alternating string of holes $\circ$  
and teeth $\bullet$.  If
$k$ is a positive integer, we define a \textbf{$k$-hole chain} as 
a chain such that: 
\begin{itemize}
  \item[(i)] Its length $L$ is a integer multiple of $k$, namely
  $$L\equiv 0  \pmod{k}$$
  \item[(ii)] The distance between holes are multiples of $k$ 
  (treating the endpoints $n=0$ and $L=N+1$ as holes if the chain is open). 
  This is equivalent to the positions of holes satisfying
  $$n_\circ\equiv \begin{cases}
      0\pmod{k}& \text{ if the chain $\mathcal{C}$ is open with Dirichlet b.c.}, \\
      \mathrm{const}\pmod{k}& \text{ if the chain $\mathcal{C}$ is closed 
      with periodic b.c.}.
\end{cases}
$$
\end{itemize}
Here is an example of a 3-hole open chain of length $L=24$ where the
endpoints are denoted by $\sotimes$'s:
$$\raisebox{.2ex}{${\scriptscriptstyle{\otimes}}$}\bullet\bullet\circ\bullet\bullet\bullet\bullet\bullet\circ\bullet\bullet\circ\bullet\bullet\bullet\bullet\bullet\circ\bullet\bullet\circ\bullet\bullet\sotimes$$
and a corresponding 3-hole closed chain of length $L=24$
$$\cdots\bullet\circ\bullet\bullet\circ\bullet\bullet\bullet\bullet\bullet\circ\bullet\bullet\circ\bullet\bullet\bullet\bullet\bullet\circ\bullet\bullet\circ\bullet\cdots$$
Similarly we define a \textbf{$k$-tooth chain} by interchanging 
holes $\circ$ and teeth $\bullet$, and treating the end points of open 
chains as teeth.
Here is an example of 3-tooth open chain:
$$\raisebox{.2ex}{${\scriptscriptstyle{\otimes}}$}\circ\circ \bullet \circ\circ\circ\circ\circ \bullet \circ\circ \bullet \circ\circ\circ\circ\circ \bullet \circ\circ \bullet \circ\circ\sotimes$$
and a corresponding 3-tooth closed chain of length $L=24$
$$\cdots\circ \bullet \circ\circ \bullet \circ\circ\circ\circ\circ \bullet \circ\circ \bullet \circ\circ\circ\circ\circ \bullet \circ\circ \bullet \circ\cdots$$

Note that if $k$ can be factorized into $k=k_1\,k_2$ 
any $k$-chain (hole of tooth) is also a $k_1$-chain and a $k_2$-chain. 
So it suffices to consider only prime $k$-chains.

\subsection{Some no-go results}
\begin{lemma}
\label{lemma1}
Consider a chain $\mathcal{C}$ and its energy spectrum $\{ E_\alpha(V)\}$. The following statements are equivalent:
\begin{enumerate}
\item There is an energy level $E_\alpha$ and a value $V_0$ for the potential 
where the first derivative of $E_\alpha$ w.r.t. $V$ vanishes
$$E'_\alpha(V_0)={dE_\alpha\over dV}(V_0)=0$$.
\item 
There is an integer $k_0>1$ such that the chain $\mathcal{C}$ is a $k_0$-tooth chain, and
 $E_p= -2\,\cos(\pi\, p/k_0)$ is an eigenvalue of $H$ with $p$ an integer such that $1\le p<k_0$.
\end{enumerate}
\end{lemma}
\paragraph{Proof:}\par\noindent
Let us assume that $dE_\alpha/dV=0$. For the corresponding normalized eigenstate $|\phi_\alpha\rangle$ we have
$$dE_\alpha/dV=\sum_{\mathrm{teeth}\, t} |\phi_\alpha(t)|^2$$
This implies that on every tooth $t$ of $\mathcal{C}$, $\phi_\alpha(t)=0$. Now let us consider in 
$\mathcal{C}$ a string ${s}$ consisting of $m_s>0$ holes (there is at least one), which is bounded 
by two teeth of respective positions $n_1$ and $n_2=n_1+m_s+1$ (note that for an open chain 
it can happen that $n_1=0$ or $n_2=L$). For the holes in this string, $n_1<n<n_2$, the wave equation is 
$- \phi(n-1) - \phi(n+1)=E\, \phi(n)$ with b.c. $\phi(n_1)=\phi(n_2)=0$. The solution must be proportional to
$\phi(n)=\sin(\vartheta (n-n_1))$ with $\vartheta=\pi\, r_s/(m_s+1)$ for some positive integer $r_s$. 
This must be satisfied for all the strings of holes in $\mathcal{C}$.  We connect these local solutions 
through the teeth and note that every tooth must be isolated with two neighbouring holes
(otherwise the solution is identically zero), while all the distances between 
successive teeth $\ell_s=m_s+1$ must be multiples of $k_0$, the greatest common divisor of the $\ell_s$'s, 
so that $\vartheta=\pi\,p/k_0$ for some $p$ and the same for all strings of holes.  
Note that $k_0>1$, otherwise some teeth are not isolated.

Conversely, if $\mathcal{C}$ is a $k_0$-tooth chain, then $E_p=-2 \cos(\pi p/k_0)$ are eigenvalues of $H$
which is proven by explicitly constructing eigenfunctions of the same form as $\phi$ above.
$E_p$ is clearly independent of $V$.
The lemma follows.
\hfill$\square$

Using the hole-tooth symmetry, a similar result holds for saturating the 
upper-bound $dE/dV\le 1$, with a similar proof, and we find the following result:

\begin{lemma}
\label{lemma1}
Consider a chain $\mathcal{C}$ and its energy spectrum $\{ E_\alpha(V)\}$. The following 
statements are equivalent:
\begin{enumerate}
\item There is an energy level $E_\alpha$ and a value $V_0$ where the first derivative of 
$E_\alpha$ w.r.t. $V$ is given by
$$
E'_\alpha(V_0)={dE_\alpha\over dV}(V_0)=1.
$$
\item 
There is an integer $k_0>1$ such that the chain $\mathcal{C}$ is a $k_0$-hole chain, and, moreover,
 $E_\alpha= -2\,\cos(\pi\, p/k_0)+V$ with $p$ an integer such that $1\le p<k_0$.
\end{enumerate}
\end{lemma}

A finite chain $\mathcal{C}$ which is not a $k$-chain for any $k$ will be called a \textbf{generic chain}.
In the random chain model where each site has a probability $p$ to be a hole and $1-p$ to be a tooth, a finite length random chain has a finite probability $P_{\text{n-g}}$ to be non-generic. It is easy to bound this probability 
$P_{\text{n-g}}(p,L)$ exponentially as a function of the length $L$ of the chain, as
\begin{equation}
\label{ }
P_{\text{n-g}}(p,L) \le C(p)^L
\ \text{with $0<C(p)<1$ some function of $p$}
\end{equation}
so that an infinite random chain is a.s.\ generic.

One can get some more specific results on the spectral flow. Here is a useful one.

\begin{lemma}
\label{lemma3}
Let $\mathcal{C}$ be a finite chain with at least one tooth and one hole.
If there is a value of the disorder parameter $V>2$ such that an eigenstate  
(solution of \rf{bcgeneric1}) has eigenvalue zero
$$
E_\alpha(V)=0\quad\text{and}\quad 2<V<\infty ,
$$
then $\mathcal{C}$ is a 2-tooth chain. Moreover, if $\mathcal{C}$ is a 
periodic chain its length $L$ must be a multiple of 4 ($L\equiv 0\mod{4}$).
If this is satisfied, $E=0$ is an eigenvalue of $H$ for any value of $V$.

Symmetrically, if there is an eigenstate such that
$$
E_\alpha(V)=V\quad\text{and}\quad 2<V<\infty ,
$$
then $\mathcal{C}$ is a 2-hole chain, and if $\mathcal{C}$ is periodic then $L\equiv 0\mod{4}$.
In this case $E=V$ is an eigenvalue of $H$ for any value of $V$.
\end{lemma}

\paragraph{Proof}\par\noindent
For $E=0$ the eigenvalue equation becomes
\begin{equation}
\label{E=0equation}
\phi({n+1})+\phi({n-1})= \begin{cases}
0& \text{if $n$ is a hole $\circ$}, \\
V\,\phi(n)& \text{if $n$ is a tooth $\bullet$}
\end{cases}
\end{equation}
Let us first consider an open chain $\mathcal{C}$ of length $L$ with Dirichlet b.c.\
$\phi(0)=\phi(L)=0$, with $N_h>0$ holes and $N_t>0$ teeth ($L=N_h+N_t+1$).
For a real solution of \rf{E=0equation}, we note that for any consecutive pair of holes, e.g.,
$n_0$ and $n_0+1$, \rf{E=0equation} implies that 
$\phi(n_0-1)=-\phi(n_0+1)$ and $\phi(n_0)=-\phi(n_0+2)$, so we can forget the pair of holes, 
and reduce the chain $\mathcal{C}$ to the chain $\mathcal{C}'$ of length $L-2$
and the wave function $\phi$ to $\phi'$ such that
$\phi'(n)=\phi(n)$ if $n< n_0$, $\phi'(n)=-\phi(n+2)$ if $n\ge n_0$.  
The function $\phi'$  is a solution of  \rf{E=0equation} on $\mathcal{C}'$. 
$$
\cdots\circ\bullet\circ\circ \color{red} \circ\circ\color{black}\bullet\bullet\circ\cdots\quad \equiv\quad \cdots\circ\bullet\circ\circ\bullet\bullet\circ\cdots
$$
We can thus reduce the problem to chains which have only \textbf{isolated holes} 
between strings of teeth.
$$
\cdots\bullet\bullet\circ\bullet\bullet\bullet\circ\bullet\circ\bullet\bullet\circ\bullet\bullet\bullet\bullet\bullet\circ\bullet\cdots
$$
Let $\phi$ be a non-trivial solution of \rf{E=0equation} with $V>2$ on such a finite 
reduced open chain. Assume that $|\phi |$ has one of its maxima at $n_0$. We can
assume that $\phi(n_0)=\phi_0>0$.
Then, if $n_0$ is a tooth,
$$
\phi(n_0-1)+\phi(n_0+1)= V \phi_0> 2\, \phi_0 \quad
\implies \quad \phi(n_0-1)\, \text{or}\, \phi(n_0+1) > \phi_0
$$
so $n_0$ is not a maximum for $\phi$.
Therefore $n_0$ has to be a hole, and  $n_0-1$ and $n_0+1$ are teeth. This implies that
$$
\phi(n_0-1)+\phi(n_0+1)=0\quad \text{and}\quad \phi(n_0\pm 2)=V\phi(n_0\pm 1)-\phi_0.
$$
Let us assume that $\phi(n_0+1)<0$ and $\phi(n_0-1)>0$.  Then 
$\phi(n_0+2)<-\phi_0$ and therefore $n_0$ is not a maximum of $|\phi|$ as assumed. 
The same is true if $\phi(n_0-1)<0$. Hence,
$$
\phi(n_0-1)=\phi(n_0+1)=0 \quad\implies\quad \phi(n_0-2)=\phi(n_0+2)=-\phi_0
$$
so that $n_0-2$ and $n_0+2$ are vertices where $|\phi|$ has a maximum.
We can repeat the argument, $n_0-2$ and $n_0+2$ are holes, $n_0-3$ and $n_0+3$ are teeth, 
and $\phi(n_0-3)=\phi(n_0+3)=0$, etc., up to the endpoints of the chain, where $\phi$ must vanish.
Therefore $\mathcal{C}$ must be an alternating chain of holes and teeth, such as the one 
depicted here
$$
\sotimes\circ\bullet\circ\bullet\circ\bullet\circ\bullet\circ\bullet\circ\bullet\cdots\cdots\cdots\circ\bullet\circ\bullet\circ\bullet\circ\bullet\circ\bullet\circ\sotimes
$$
with $\phi=0$ on teeth, and $\phi=\pm\,\phi_0$ alternatively on holes, i.e., 
$\phi(n)=\phi_0 \sin(\pi n/2)$ and the length $L$ of the comb is necessarily even.

We can go back to the case of a chain whose holes are not necessarily isolated.
The above argument shows that the strings of holes must be separated by 
isolated teeth
such has the one depicted here:
$$
\sotimes\circ\circ\bullet\circ\circ\circ\bullet\circ
\bullet\circ\circ\circ\bullet\circ\cdots\cdots\cdots\circ\bullet\circ\bullet
\circ\bullet\circ\circ\circ\bullet\sotimes
$$
This is the definition of a 2-tooth open chain.

The same argument applies if the chain is closed. However, one must  check that the 
wave function $\phi(n)=\phi_0\,\sin(\pi\, n/2)$ is periodic, which requires $L$ to be an 
integer multiple of $4$.

Finally, using the $\mathtt{hole}\leftrightarrow\mathtt{tooth}$ symmetry 
\rf{ToothHoleSymmetry}, the symmetric part of the lemma is proved along the same lines.
\hfill$\square$

Of course Lemma \ref{lemma3} implies that if a chain is generic, there are no 
eigenvalues of $H$ such that $E=0$, $|V|>2$, as well as for $E=V$, $|E|>2$. 
Moreover, for generic chains all eigenvalues satisfy
$$
0<{dE_\alpha\over dV} <1
$$
as shown in Lemmas 1 and 2.

\subsection{An explicit formula for $N_{E>4}$}
There is an explicit formula for the number $N_{E>4}$ of states with energy $E>4$ 
for a comb $\mathcal{C}$ which depends in a simple way on the structure of the comb. 
\begin{figure}[h!]
\begin{center}
\includegraphics[width=5 in]{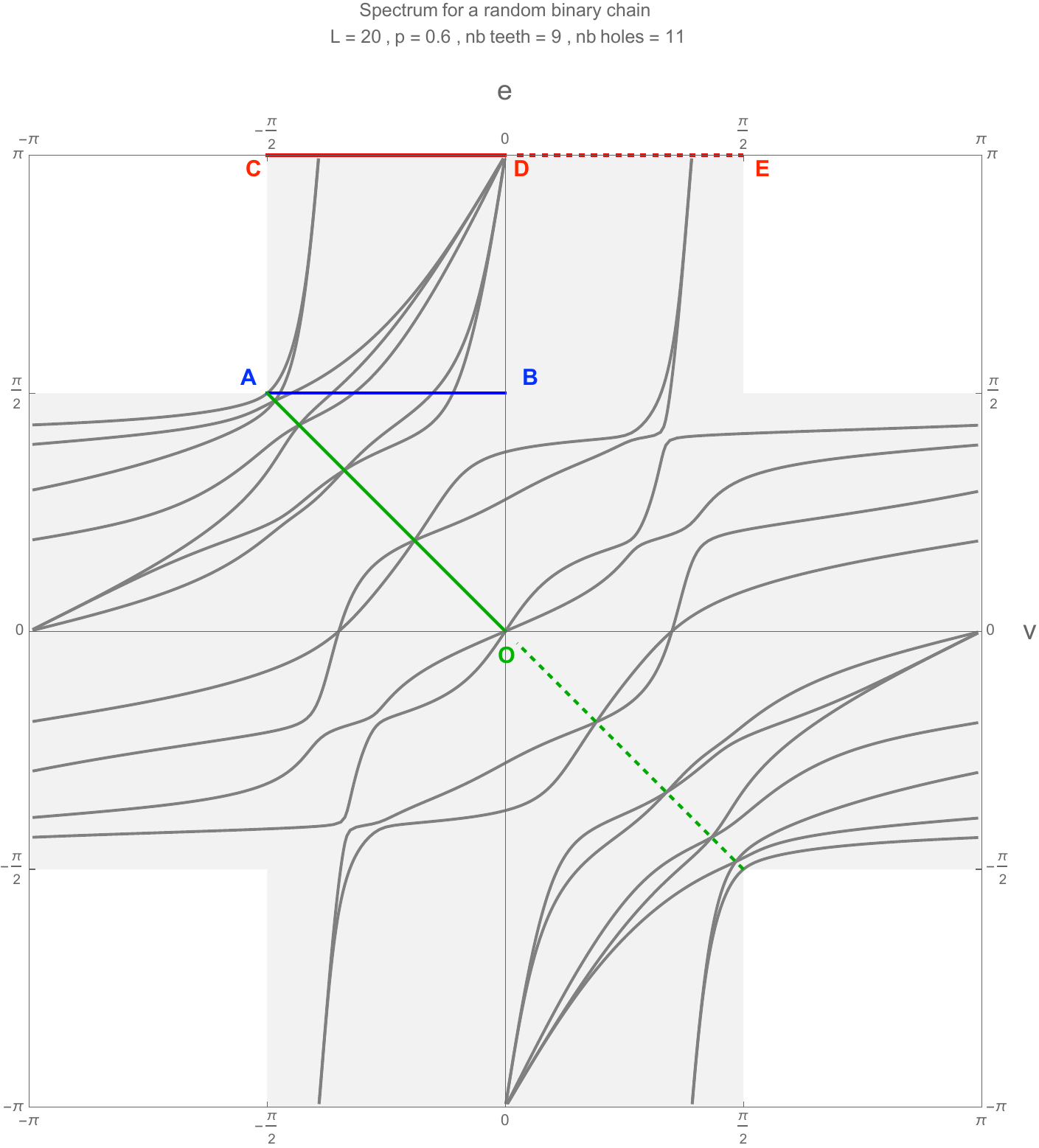}
\caption{Examples of a spectral flow in the $\mathbf{v},\mathbf{e}$ coordinates.
The blue segment 
is $[AB[$, the red one is $[CD]$ and the green one is $[AO[$.}
\label{Fveflows2}
\end{center}
\end{figure}
We use the 
coordinates of \rf{vevar}, $\mathbf{e}=2\,\arctan({E/2})$ and 
$\mathbf{v}=2\,\arctan({(V-E)/2})$, to describe the spectral flow.
The number of eigenstates of the random comb model localized near the spine 
is equal to the number of eigenstates of the binary chain with $E=2$, $0\le V < 2$, 
namely $\mathbf{e}=\pi/2$, $-\pi/2\le \mathbf{v} <0$. 
This is the number $N_{[\mathtt{AB}[}$ of flow lines that cross the segment 
$[\texttt{AB}[$ with endpoints $\texttt{A}=(-\pi/2,\pi/2)$ and $\texttt{B}=(0,\pi/2)$ 
in Fig.\ \ref{Fveflows2}.
Let us compare this with the number of eigenstates at $\mathbf{e}=\pi$.
For a generic chain (in fact for any chain which is not a $2$-teeth chain), 
we know from Lemma\ \ref{lemma3} that no flow lines starting from 
$\mathbf{e}=\pi/2$, $0<\mathbf{v}\le\pi/2$ can cross the vertical segment 
$[\texttt{D},\texttt{B}[$, where $\texttt{D}=(0,\pi )$. 
Therefore, $N_{[\texttt{AB}[}$ is equal to the number $N_{[\texttt{CD}]}$ of flow lines 
that crosses the segment $[\texttt{C},\texttt{D}]$, where
$\texttt{C}= (-\pi/2,\pi)$, i.e., flow lines which satisfy 
$\mathbf{e=\pi}$, $\pi/2<\mathbf{v}\le 0$.

If $\texttt{E}=(\pi/2,\pi)$, $N_{[\texttt{CE}]}$ the number of flow lines that cross 
$[\texttt{C},\texttt{E}]$ and $N_{[\texttt{DD}]}$ the number of flow lines that pass 
through the point $\texttt{D}$, we know that the total number of teeth of $\mathcal{C}$ is 
$$
N_t=N_{[\texttt{CE}]}=N_{[\texttt{CD}[}+N_{[\texttt{DD}]}+N_{]\texttt{DE}]}.
$$
By the symmetry of the spectrum described in Sec.\ 3.6.2, 
we know that that $N_{[\texttt{CD}[}=N_{]\texttt{DE}]}$.
Hence, using $N_{[\texttt{CD}]} = N_{[\texttt{CD}[} +N_{[\texttt{DD}]} $,
we obtain
$$
N_{[\texttt{AB}[}=(N_t+N_{[\texttt{DD}]})/2\,.
$$

Now $N_{[\texttt{DD}]}$ is the number of independent eigenstates at 
$\mathbf{e}=\pi$, $\mathbf{v}=0$, namely $E\to\infty$, $V=E$ for the binary chain. 
We know that the states at $\mathbf{e}=\pi$ are localized on the strings of teeth of 
$\mathcal{C}$, and vanish on the holes. 
Thus, to a single string of teeth $s$ of length $\ell_s$  are associated 
$\ell_s$ eigenvectors of the Hamiltonian restricted to $s$ 
with eigenvalues
$2 \tan(\mathbf{v}_k/2)=2\cos(\pi k/ (\ell_s+1))$, $k=1,\cdots,\ell_s$. 
We see that 
$0$ will be an eigenvalue if and only if
$\ell_s+1$ is even and $k=(\ell_s+1)/2$. Therefore, $N_{[\texttt{DD}]}$ 
is equal to the number of strings of teeth with an odd number of teeth:
$$N_{[\texttt{DD}]} = N_{t,\mathtt{odd}}=\text{number of strings of 
teeth with an odd number of teeth}$$
Hence, we have an explicit formula for any given generic chain
\begin{equation}
\label{NE<4expl}
N_{E>4}=(N_t+N_{t,\mathtt{odd}})/2\,.
\end{equation}
This formula is in fact valid for chains which are not $2$-chains, 
and it can be shown along the same lines as for the special 
case of $2$-teeth chains one has
\begin{equation}
\label{NE<4expl}
N_{E>4}=(N_t+N_{t,\mathtt{odd}})/2-1.
\end{equation}

\subsection{Probabilities for chains}
\label{AProbChain}
For a long random binary chain, the probability $P_h$  (resp.\ $P_t$) for a site to be a hole 
(resp.\ tooth) inside a string of holes (resp.\ teeth) of length $\ell$ is asymptotically
given by
$$
P_h(\ell) = \ell (1-p)^2 p^\ell\qquad\text{resp.}\qquad P_t(\ell)= \ell\, p^2 (1-p)^\ell .
$$
To prove this we note that given a random site, it has a probability $p$ to be a hole, 
and when moving left, the conditional probability  to meet $n_1$ holes before meeting a tooth 
is $p^{n_1}(1-p)$. 
The same holds when moving right, and the result follows. Note that the above formulas 
are exact for a 
finite periodic chain if its length $L$ is larger than $\ell+2$.

Therefore the density of strings of holes (resp.\ teeth) 
with length $\ell$ (number of strings divided by the length 
$L$ of the chain) is, when $L\to\infty$,
$$
D_h(\ell)=(1-p)^2 p^\ell\qquad\text{resp.}\qquad D_t(\ell)= p^2 (1-p)^\ell 
$$
Therefore the density of strings of holes with an odd (resp.\ even) length $\ell$ are
\begin{equation}
\label{Dholes}
D_{h,\mathtt{odd}}=p{(1-p)\over (1+p)}\qquad,\qquad D_{h,\mathtt{even}}= 
p^2{(1-p)\over (1+p)}
\end{equation}
\begin{equation}
\label{Dteeth}
D_{t,\mathtt{odd}}=p{(1-p)\over (2-p)}\qquad\,\qquad D_{t,\mathtt{even}}=
p{(1-p)^2\over (2-p)}\ .
\end{equation}
From \rf{NE<4expl}, 
the density of $E>4$ states, $\overline N_{E>4}$, defined as 
\begin{equation}
\label{DE>4}
\overline N_{E>4}=\lim_{L\to\infty}N_{E>4}/L
\ =\ 
 D_{t,\mathtt{even}}+ 2\,D_{t,\mathtt{odd}}
\end{equation}
Using \rf{Dteeth} we get the final result for $\overline N_{E>4}$ as a function of the hole probability $p$
\begin{equation}
\label{NE>4}
\overline N_{E>4}\ =\ 
{1-p\over 2-p}
\end{equation}
Note that this density of $E>4$ states is always smaller than the density of teeth $\overline N_t=1-p$.

\section{Existence and unitarity of the S-matrix}
\label{ASmatrix}

\subsection{A first result on the unitarity of $S$}
\label{App5}
We consider the $N_t\times N_t$ S-matrix $S$ and the $N\times N$ matrix 
$\mathbb{S}$ defined in Sec.~\ref{ssSmatrix}.
The eigenvalue equation \rf{E<4tfull} for an eigenstate with energy $0<E<4$ 
with $E=2-2\cos\theta$ is a linear equation \ref{XAYBeq} relating the 
column vectors $A=\{A_n\}$ (outgoing waves) and $B=\{B_n\}$ (incoming waves)
of seize $N$:
\begin{equation}
\label{XAYBeq2}
 \mathbb{X}(\theta)A+\mathbb{Y}(\theta)B=0.
\end{equation}
The matrices $\mathbb{X}$ and $\mathbb{Y}$ are $N\times N$ symmetric complex 
matrices which depend on $\theta$.
They satisfy (for real $\theta$)
\begin{equation}
\label{XYT2}
\mathbb{X}(\theta)={\overline{\mathbb{Y}}}(\theta)\ ,\quad \mathbb{X}^\top (\theta)=\mathbb{X}(\theta)\ ,\quad \mathbb{Y}^\top (\theta)=\mathbb{Y}(\theta)\ ,\quad \mathbb{X}(\theta)-\mathbb{Y}(\theta)= 2\,\mathi\,\sin(\theta)\,\mathbb{T}
\end{equation}
with $\mathbb{T}$ the projector on the teeth of the comb.

If the vectors $A$ and $B$ satisfy \rf{XAYBeq2}, then one has
$$ \bar A{\cdot}\mathbb{X}(\theta)A+\bar A{\cdot}\mathbb{Y}B(\theta)=0\ ,
\quad \bar B{\cdot}\mathbb{X}(\theta)A+\bar B{\cdot}\mathbb{Y}(\theta)B=0$$
and taking the complex conjugate
$$ 
\bar A{\cdot}\mathbb{Y}(\theta)A+\bar B{\cdot}\mathbb{X}(\theta)A=0\ ,
\quad \bar A{\cdot}\mathbb{Y}(\theta)B+\bar B{\cdot}\mathbb{X}(\theta)B=0 .
$$
Combining these four identities, we get
$$
\bar A{\cdot}(\mathbb{X}(\theta)-\mathbb{Y}(\theta))A=
\bar B{\cdot}\mathbb{X}(\theta)A-\bar A{\cdot}\mathbb{Y}(\theta)B=
\bar B{\cdot}(\mathbb{X}(\theta)-\mathbb{Y}(\theta))B .
$$
Using the last equation of \rf{XYT2} we get
$$
2\,\mathi\sin\theta\  \bar A{\cdot}\mathbb{T} A=2\,\mathi\sin\theta\  
\bar B{\cdot}\mathbb{T} B.
$$
As long as  $0<\theta<\pi$ (i.e.\ $0<E<4$), $\sin\theta\neq 0$, and therefore
\rf{XAYBeq2} implies
\begin{equation}
\label{UnitarityS1}
 \bar A{\cdot}\mathbb{T} A=\bar B{\cdot}\mathbb{T} B
\end{equation}
which reads explicitly
\begin{equation}
\label{barAA=barBB}
\sum_{t\text{:tooth}} \bar A_t A_t \ =\ \sum_{t\text{:tooth}} \bar B_t B_t.
\end{equation}

The matrix $\mathbb{S}$ defined by 
Eq.~\rf{BigSdef} which allows us to calculate the $A$'s from the $B$'s  
$$ 
\mathbb{S}(\theta)=-\mathbb{X}^{-1}(\theta)\mathbb{Y}(\theta)\qquad\text{so that}
\qquad A=\mathbb{S}(\theta) B
$$
apparently only makes sense if $\mathbb{X}(\theta)$ is invertible, 
but it is shown below in Appendix~\ref{App6} that $\mathbb{S}(\theta)$ is 
indeed well defined for all $\theta\in[0,\pi]$.

The  matrix $\mathbb{S}$ can be written as a block matrix, where the blocks 
correspond to the tooth and the hole subspaces as
\begin{equation}
\label{ }
\mathbb{S}(\theta)=\begin{pmatrix}
     S(\theta) &  0  \\
      C(\theta)&  -\mathbb{I}.
\end{pmatrix}
\end{equation}
Here $S(\theta)$ is an $N_t\times N_t$ matrix, the restriction of 
$\mathbb{S}(\theta$) to the teeth, $C(\theta)$ 
a $N_t\times N_h$ matrix, and $\mathbb{Id}$ the $N_h\times N_h$ identity matrix.
The matrix $S(\theta)$ is the S-matrix relating the $B_{\mathrm{teeth}}$ to 
the $A_{\mathrm{teeth}}$ where the subscript "teeth" indicates the restriction to the 
tooth subspace.
Eq.\ \rf{barAA=barBB} implies that the S-matrix $S(\theta)$ is unitary, 
provided it is well defined.

Finally, the fact that the matrix $\mathbb{Y}$ is the complex conjugate of the matrix $\mathbb{X}$ (if $\theta$ is real) implies that the inverse of the matrix $\mathbb{S}$ is simply its complex conjugate
\begin{equation}
\label{ }
\mathbb{S}^{-1}= -\mathbb{Y}^{-1}\mathbb{X}= -\bar{\mathbb{X}}^{-1}\bar{\mathbb{Y}}=
-\overline{\mathbb{X}^{-1}\mathbb{Y}}=\bar{\mathbb{S}}.
\end{equation}
Since on general grounds
$$\mathbb{S}^{^{-1}=}\begin{pmatrix}
     {S}^{-1} &  0  \\
      C\, S^{-1}&  -\mathbb{I}
\end{pmatrix}
$$
this implies that
\begin{equation}
\label{ }
S^{-1}(\theta)=\overline{S(\theta)}= S(\theta)^\top ,
\end{equation}
i.e.\ $S(\theta)$ is a unitary and symmetric matrix,
and
\begin{equation}
\label{ }
C(\theta)=\overline{C(\theta)}\,{S}(\theta).
\end{equation}

\subsection{The special cases $\theta=0$ or $\pi$}
\label{App5b}
For the special cases $\theta=0$ and $\theta=\pi$, $\mathbb{X}=\mathbb{Y}$ is a 
real symmetric matrix, independent of the number and configuration of teeth 
in the comb $\mathcal{C}$. In fact, it corresponds to the discrete Laplacian on 
a linear chain.  A general solution of \rf{XAYBeq} is obviously $A=-B$ so that 
one can take for the S-matrix $S=-1$, which is also unitary. 
However, $A+B=0$ corresponds to trivial eigenfunctions $\phi(n,j)=0$ if $n$ is a tooth.

It remains to look for a possible zero mode 
$A_0$  of $\mathbb{X}$, such that $\mathbb{X}A_0=0$. It corresponds to the 
eigenstate with $E=0$ (respectively $E=4$) on a linear chain, hence
${A_0}_n=1$ if $\theta=0$ and ${A_0}_n=(-1)^n$ if $\theta=\pi$. Such a zero 
mode does not exist for open combs. It exists for periodic combs of arbitrary length $L$ 
for $\theta=0$, and only for periodic combs with even length $L=2M$ if $\theta=\pi$. 
In any case, it does not correspond to a propagating particle along the teeth, 
and no $S$-matrix is associated to such zero modes.

\subsection{Proof of the existence and analyticity of ${S}$}
\label{App6}
When arguing from Eq.~\ref{XAYBeq} that there is an $N\times N$ matrix $\mathbb{S}$ 
such that $A=\mathbb{S}B$, we assumed that the matrix $\mathbb{X}$ is invertible. 
It turns out that this might not be the case,  but only for some specific values of $\theta$.

The $N\times N$ matrix  $\mathbb{X}$ (hence $\mathbb{Y}=\overline{\mathbb{X}}$) 
depends on the geometry of the comb $\mathcal{C}$ and on the parameter 
$\theta\in [-\pi,\pi]$.
Let us consider $\mathbb{X}$ and $\mathbb{Y}$ as functions of the variable 
$z=e^{i\theta}$ and analytically continue $\mathbb{X}$ from the unit circle 
to the full complex plane
 through the extension of \rf{E<4tfull}
\begin{equation}
\label{E<4tfullb}
\begin{split}
-A_{n-1}-A_{n+1}+A_n(1+z^{-1})&=B_{n-1}+B_{n+1}- B_n(1+z)
\ \text{if $n$ is a tooth}\\
-A_{n-1}-A_{n+1}+A_n(z+z^{-1})&= B_{n-1}+B_{n+1}-B_n(z+z^{-1})\ \text{if $n$ is a hole}
\end{split}
\end{equation}
which in matrix-vector notations reads
\begin{equation}
\label{XA+YBz}
\mathbb{X}(z) A+\mathbb{Y}(z)B=0\quad\text{with}\quad \mathbb{Y}(z)=\mathbb{X}(z^{-1}).
\end{equation}
The matrix $\mathbb{X}(z)$ is of the general form
$$\mathbb{X}(z)= z^{-1} \mathbb{X}_{\scriptscriptstyle{(\text{-}1)}} +\mathbb{X}_{\scriptscriptstyle{(0)}}+ z \mathbb{X}_{\scriptscriptstyle{(1)}}$$
with $\mathbb{X}_{\scriptscriptstyle{(\text{-}1)}}$ and $\mathbb{X}_{\scriptscriptstyle{(\text{}1)}}$  real diagonal matrices, and $\mathbb{X}_{\scriptscriptstyle{(\text{}0)}} $ a real tridiagonal matrix.
These three matrices have integer coefficients, in fact the coefficients are only $-1$, $0$ or $1$. 
Therefore, its  determinant is of the form
\begin{equation}
\label{detXz}
\det\mathbb{X}(z)= z^{-N}{d(z)}\quad\implies\quad \det\mathbb{Y}(z)= z^N d(z^{-1}) 
\end{equation}
where $N$ the length of the comb, and $d(z)$ a polynomial of degree $2N$, with real 
integer coefficients.
Thus, $\det\mathbb{X}(z)$ is a rational function of $z$ on the Riemann plane, 
with $2N$ zeros (possibly coinciding), and a pole of degree at most $N$ at $z=\infty$.
This implies that the matrix $\mathbb{X}(z)$ is invertible away from the zeros of $d(z)$, 
and that the matrix
$$
\mathbb{S}(z)=-\mathbb{X}(z)^{-1}\, \mathbb{Y}(z)=-\mathbb{X}(z)^{-1}\, \mathbb{X}(z^{-1})
$$ 
exists as a meromorphic matrix values function of $z$.
It can be written as
$$ 
\mathbb{S}(z) = {\tilde{\mathbb{N}}(z)\over d(z)}\ ,
$$
where $\tilde{\mathbb{N}}(z)$ is a matrix which is a polynomial of degree $2N$ in $z$.
It follows that $\mathbb{S}(z)$ has at most $2N$ poles in the complex plane, which 
correspond to the zeroes of $d(z)$.
Away from the zeroes of $d(z)$  (and the corresponding zeroes of $d(z^{-1})$), 
Eq.\ \rf{XA+YBz} makes perfect sense, and given $B\neq 0$, it has a  unique solution
$$A(z)=\mathbb{S}(z)\,B$$
which is meromorphic in $z$.

An obvious solution to \rf{XA+YBz} is $A+B=0$ with $B_n=0$ for all sites $n$ on the spine
where there is a tooth, namely taking  the vector $B\neq 0$ but localized on the holes, i.e.,
$\mathbb{T}\,B=0$.
Separating the matrix $\mathbb{S}$ into blocs corresponding to the tooth subspace 
and the node subspaces, this means that it can be written as
\begin{equation}
\label{ }
\mathbb{S}(z)=\begin{pmatrix}
     S(z) &  0  \\
      C(z)&  -\mathbb{I}
\end{pmatrix}
\end{equation}
with $S(z)=\mathbb{T}\,\mathbb{S}(z)\mathbb{T}$ the $N_t\times N_t$ restriction of $\mathbb{S}$ to the teeth, $C(z)$ a $N_t\times N_h$ matrix, and $\mathbb{I}$ the $N_h\times N_h$ identity matrix.
Both $S(z)$ and $C(z)$ are meromorphic.

The results of Appendix \rf{App5} show that on the unit circle, as long as
$$|z|=1\quad\text{and}\quad z\neq\pm1$$
$S(z)$ is a unitary matrix if it is nonsingular. Hence, it is a bounded matrix 
of norm $\parallel\! S(z)\!\parallel =1$. Since $S(z)$ is meromorphic, this 
implies that $S(z)$ has no poles on the unit circle.
Therefore, $S$ is unitary and analytic on the whole unit circle, 
hence for all $\theta\in [0,2\pi)$.

Note that the matrix $\mathbb{S}$ is not unitary for $|z|=1$, unless $C(z)=0$, 
which occurs for $z=\pm 1$, i.e., for the trivial cases $\theta=0$ and $\theta=\pi$.

\subsection{Derivation of the scalar product formula \rf{SPPhiAB} }
\label{Appendix7}
We now derive Eq.\ \rf{SPPhiAB} which gives the scalar product of energy eigenstates 
with energies $E\le 4$.
We consider two such states $| \Phi_\theta\rangle$ and $| \Phi'_{\theta'}\rangle$, 
of the form given by \rf{phiEle4bis}, for respective values $\theta$ and $\theta'$ of 
the phase factor along the teeth
\begin{equation*}
\label{phiEle4ter}
\begin{split}
\Phi (n,j)&=A_n\,e^{\mathi\theta j}+B_n\,e^{-\mathi\theta j}
\ ,\quad
\Phi' (n,j)=A'_n\,e^{\mathi\theta' j}+B'_n\,e^{-\mathi\theta' j}
\quad\text{if $n$ is a tooth,}\\
\Phi (n,0)&=A_n+B_n=C_n
\ ,\quad
\Phi' (n,0)=A'_n+B'_n=C'_n\quad\text{if $n$ is a hole},
\end{split}
\end{equation*} 
with the $A_n,B_n$ and the $A'_n,B'_n$ obeying Eq.\ \rf{E<4tfull} for $\theta$ and $\theta'$
respectively.

Their scalar product is formally
\begin{equation}
\label{SPE<4expl}
\begin{split}
\langle\Phi_\theta\vert\Phi'_{\theta'}\rangle = & \sum_{t:\text{tooth}} 
\sum_{j=0}^\infty \left(\bar A_t\,e^{-\mathi\theta j}+\bar B_t\,e^{\mathi\theta j}\right)
\left(A'_t\,e^{\mathi\theta' j}+B'_t\,e^{-\mathi\theta' j}\right)
+  \sum_{h:\text{hole}} \bar C_h C'_h .
\end{split}
\end{equation}
However, as for the scalar product over a single line, the sum over 
the $j$'s is not absolutely convergent. 
In order to make it well-defined, one adds a 
small exponential factor to the sums
\begin{equation}
\label{ }
\sum_{j=0}^\infty e^{\mathi(\pm\theta\pm\theta')j}\to 
\sum_{j=0}^\infty e^{-\epsilon_+ j}e^{\mathi(\pm\theta\pm\theta')j}
\end{equation}
with $\epsilon_+$ a small positive parameter, and let $\epsilon_+$ go to
$0$ at the end of the calculation.
We get for the sum over the teeth in \rf{SPE<4expl}
\begin{equation}
\label{SPttexpl}
\begin{split}
\sum_{t:\text{tooth}} 
{\bar A_t A'_t \over 1-e^{-\epsilon_+ +\mathi (\theta'-\theta)}}
+
{\bar A_t B'_t \over 1-e^{-\epsilon_+ -\mathi (\theta+\theta')}}
+
{\bar B_t A'_t \over 1-e^{-\epsilon_+ +\mathi (\theta+\theta')}}
+
{\bar B_t B'_t \over 1-e^{-\epsilon_+ +\mathi (\theta-\theta')}}\,.
\end{split}
\end{equation}
If $\theta$ and $\theta'$ are in the interval $]0,\pi[$ 
and $|\theta-\theta'|\neq 0$, then clearly the limit $\epsilon_+\to 0$ 
exists and we get for $\langle\Phi_\theta\vert\Phi'_{\theta'}\rangle$ 
a finite result, which is zero since their respective energy 
eigenvalues are different $E_\theta\neq E_{\theta'}$, and
$
\langle\Phi_\theta\vert H \vert \Phi'_{\theta'}\rangle=E_\theta \langle\Phi_\theta\vert\Phi'_{\theta'}\rangle= E_{\theta'} \langle\Phi_\theta\vert\Phi'_{\theta'}\rangle
$.

It remains to calculate the contribution of the $\theta-\theta'\to 0$ 
domain. For this one considers the region
$$
\epsilon_+ \ll |\theta-\theta'|\ll 1.
$$
Then the most singular contribution to 
$\langle\Phi_\theta\vert\Phi'_{\theta'}\rangle$ in 
\rf{SPE<4expl} is given by the first and fourth terms in 
\rf{SPttexpl}, which give
\begin{equation}
\label{SPttsing}
\sum_{t:\text{tooth}} {\bar A_t A'_t \over \epsilon_+ 
-\mathi (\theta'-\theta)}
+
\sum_{t:\text{tooth}} {\bar B_t B'_t \over \epsilon_+ 
+\mathi (\theta'-\theta)}\,.
\end{equation}
In the limit $\epsilon_+\to 0$, the fractions converge in the 
sense of distributions
\begin{equation}
\label{ }
\lim_{\epsilon_+\to 0_+} {1\over \epsilon_+\mp (\theta'-\theta)}\ 
= \ \mathrm{p.v.}{\pm\,\mathi\over \theta'-\theta}\ 
+\ \pi\ \delta(\theta'-\theta)
\end{equation}
where $\mathrm{p.v.}\,1/x$ the Cauchy principal value of $1/x$, 
and $\delta(x)$ the Dirac point distribution at $x=0$.
We get for the $\epsilon_+\to 0$ limit of  \rf{SPttsing}
\begin{equation}
\label{SPtts2}
\mathrm{p.v.}\, {i\over \theta'-\theta}\,\sum_{t:\text{tooth}}(\bar A_t A'_t-\bar B_t B'_t )
\ +\  \pi\ \delta(\theta'-\theta) \sum_{t:\text{tooth}} (\bar A_t A'_t+\bar B_t B'_t)
\end{equation}
This is the dominant contribution when $|\theta-\theta'|\ll 1$.
The first term in \rf{SPtts2} seems singular. 
However, when $\theta'\to\theta$, the $(A,B)$'s and $(A',B')$'s 
obey the same equation \rf{phiEle4bis} with the same 
parameter $\theta=\theta'$. In other words, one has in general
$
A=S(\theta)B$, $A'=S(\theta') B'$ and as 
$\theta'\to\theta$, $A'=S(\theta) B$. From the unitarity 
of the $S(\theta)$ matrices, one has in full generality
$$
\sum_{t:\text{tooth}}\bar A_t A'_t=\sum_{t:\text{tooth}}\bar B_t B'_t=
(A|A')=(B,B')
$$
when $ \theta'\to\theta$
so the first term in \rf{SPtts2} disappears and we are left with
$$
2\pi\ \delta(\theta-\theta')\  (A|A')\ =\ 2\pi\ \delta(\theta-\theta')\ (B|B').
$$
Since the finite term when $\theta\neq\theta'$ is zero, we are left with 
$$
\langle\Phi_\theta\vert\Phi'_{\theta'}\rangle = 
2\pi\ \delta(\theta-\theta')\  (A|A')\ =\ 2\pi\ \delta(\theta-\theta')\ (B|B')
$$
This ends the proof of \rf{SPPhiAB}. \hfill$\square$

\end{document}